\documentclass[aps,prd,reprint,superscriptaddress,nofootinbib,%
  longbibliography]{revtex4-2}

\usepackage{lmodern}
\usepackage{amsmath,amssymb}
\usepackage{booktabs}
\IfFileExists{siunitx.sty}{%
  \ifdefined\IfFormatAtLeastTF
    \IfFormatAtLeastTF{2020/10/01}{\usepackage{array}[=2016-10-06]}{}%
  \fi
  \usepackage{siunitx}}{}
\usepackage{graphicx}
\usepackage{xcolor}
\usepackage{hyperref}
\hypersetup{hidelinks}

\graphicspath{{figures/}}

\newcommand{\NRHJ}{\textsc{NRHJSur3dq8}}
\newcommand{\NRAAS}{\NRHJ}
\newcommand{\AAf}{\textsc{AA}}
\newcommand{\AdA}{\textsc{AdA}}
\newcommand{\AAint}{\textsc{AAint}}
\newcommand{\AAintPN}{\textsc{AAintPN}}
\newcommand{\Jphi}{\ensuremath{J_\phi}}
\newcommand{\mm}[1]{\ensuremath{#1}}

\newcommand{\AlModelModeCountWord}{seven}
\newcommand{\AlModelModeList}{$(2,1)$, $(2,2)$, $(3,2)$, $(3,3)$, $(4,3)$, $(4,4)$, $(5,5)$}
\newcommand{\AlTrainingSetSize}{260}
\newcommand{\AlLooTrainingSetSize}{259}
\newcommand{\AlWindowEndAfterPeak}{124.5}

\newcommand{\AlTimeShiftBound}{20}
\newcommand{\AlNRExchangeFixedSimCount}{18}
\newcommand{\AlNRExchangeFixedOddAmp}{8.2^{+278}_{-7.6}\times10^{-7}}
\newcommand{\AlNRExchangeFixedOddAmpCount}{72}
\newcommand{\AlNRMirrorPairCountWord}{six}
\newcommand{\AlNRMirrorOverlapDeficit}{2.5^{+217}_{-2.5}\times10^{-6}}
\newcommand{\AlNRMirrorOverlapCount}{42}
\newcommand{\AlLooMismatchTwoTwo}{1.3^{+32}_{-1.2}\times10^{-6}}
\newcommand{\AlLooMismatchAllModes}{2.3^{+59}_{-2.2}\times10^{-6}}

\newcommand{\AlInSampleMismatchTwoTwo}{7.5^{+106}_{-6.8}\times10^{-7}}
\newcommand{\AlInSampleMismatchAllModes}{1.3^{+16}_{-1.2}\times10^{-6}}
\newcommand{\AlBlendedLooMismatchTwoOne}{1.3^{+49}_{-1.2}\times10^{-5}}
\newcommand{\AlBlendedLooMismatchThreeThree}{5.5^{+224}_{-5.2}\times10^{-5}}
\newcommand{\AlBlendedLooMismatchThreeTwo}{1.6^{+33}_{-1.4}\times10^{-5}}
\newcommand{\AlBlendedLooMismatchFourFour}{4.6^{+374}_{-4.4}\times10^{-5}}
\newcommand{\AlBlendedLooMismatchFourThree}{9.8^{+262}_{-9.2}\times10^{-5}}
\newcommand{\AlBlendedLooMismatchFiveFive}{3.6^{+80}_{-3.4}\times10^{-4}}

\newcommand{\AlDirectInSampleMismatchTwoTwo}{6.3^{+74}_{-5.8}\times10^{-7}}
\newcommand{\AlDirectInSampleMismatchAllModes}{1.0^{+14}_{-0.96}\times10^{-6}}
\newcommand{\AlBlendedLooMismatchTwoTwo}{1.6^{+62}_{-1.5}\times10^{-6}}
\newcommand{\AlBlendedLooMismatchAllModes}{2.8^{+84}_{-2.7}\times10^{-6}}
\newcommand{\AlLooOverInSampleMedianRatioBlendedTwoTwo}{2.1}
\newcommand{\AlLooOverInSampleMedianRatioBlendedAllModes}{2.2}

\newcommand{\AlBlendedBetterThanDirectLooPctTwoTwo}{36}

\newcommand{\AlBlendedBetterThanDirectLooPctAllModes}{34}
\newcommand{\AlLooInspiralMismatchTwoTwo}{8.4^{+188}_{-8.0}\times10^{-7}}
\newcommand{\AlLooInspiralMismatchAllModes}{9.8^{+203}_{-9.2}\times10^{-7}}
\newcommand{\AlLooMergerRingdownMismatchTwoTwo}{2.9^{+509}_{-2.6}\times10^{-6}}
\newcommand{\AlLooMergerRingdownMismatchAllModes}{1.2^{+88}_{-1.1}\times10^{-5}}
\newcommand{\AlLooMergerRingdownOverInspiralFactorTwoTwo}{3}
\newcommand{\AlLooMergerRingdownOverInspiralFactorAllModes}{12}

\newcommand{\AlLooRatioAAOverReleasedAdATwoTwo}{1.1^{+3.8}_{-0.67}}

\newcommand{\AlLooAAAtOrBelowReleasedAdAPctTwoTwo}{36}
\newcommand{\AlLooRatioAAOverReleasedAdAAllModes}{1.1^{+2.3}_{-0.56}}

\newcommand{\AlLooAAAtOrBelowReleasedAdAPctAllModes}{34}
\newcommand{\AlLooRatioAAOverReleasedAdAHigherModesMin}{1.01}
\newcommand{\AlLooRatioAAOverReleasedAdAHigherModesMax}{1.03}

\newcommand{\AlInSampleRatioAAOverReleasedAdAMedianTwoTwo}{1.16}

\newcommand{\AlInSampleRatioAAOverReleasedAdAMedianAllModes}{1.11}

\newcommand{\AlInSampleRatioAAOverReleasedAdAHigherModesMin}{1.01}
\newcommand{\AlInSampleRatioAAOverReleasedAdAHigherModesMax}{1.04}

\newcommand{\AlLooWorstSimID}{SXS:BBH:4121}
\newcommand{\AlLooSecondWorstSimID}{SXS:BBH:4123}
\newcommand{\AlLooThirdWorstSimID}{SXS:BBH:2757}
\newcommand{\AlLooWorstPairMassRatio}{4}
\newcommand{\AlLooWorstPairChiOne}{+0.90}
\newcommand{\AlLooWorstSimChiTwo}{+0.50}
\newcommand{\AlLooSecondWorstSimChiTwo}{-0.50}
\newcommand{\AlLooWorstSimMismatch}{8.2\times10^{-4}}
\newcommand{\AlLooSecondWorstSimMismatch}{7.8\times10^{-4}}
\newcommand{\AlLooThirdWorstSimMismatch}{4.4\times10^{-4}}
\newcommand{\AlLooThirdWorstSimMassRatio}{8}
\newcommand{\AlLooThirdWorstSimChiOne}{-0.40}
\newcommand{\AlLooThirdWorstSimChiTwo}{+0.80}
\newcommand{\AlWorstSimInSampleMismatch}{2.0\times10^{-4}}
\newcommand{\AlSecondWorstSimInSampleMismatch}{8.3\times10^{-5}}
\newcommand{\AlWorstSimInSampleAfterPeakMismatch}{2.1\times10^{-3}}
\newcommand{\AlSecondWorstSimInSampleAfterPeakMismatch}{9.1\times10^{-4}}
\newcommand{\AlInSampleAfterPeakMedianMismatch}{3.7\times10^{-6}}
\newcommand{\AlWorstPairOwnCoefficientMismatch}{10^{-10}}
\newcommand{\AlInSampleThirdWorstLooSimMismatch}{5.1\times10^{-6}}
\newcommand{\AlResolutionModernQEightMassRatio}{8}
\newcommand{\AlResolutionModernQEightMin}{3\times10^{-7}}
\newcommand{\AlResolutionModernQEightMax}{1.3\times10^{-5}}
\newcommand{\AlHomPlateauMassRatioMin}{6}

\newcommand{\AlHomPlateauOverResolutionMin}{30}
\newcommand{\AlHomPlateauOverResolutionMax}{60}
\newcommand{\AlResolutionMismatchTwoTwo}{8.4^{+162}_{-8.3}\times10^{-6}}
\newcommand{\AlResolutionMismatchTwoOne}{5.5^{+201}_{-5.3}\times10^{-6}}
\newcommand{\AlResolutionMismatchThreeThree}{2.5^{+62}_{-2.3}\times10^{-5}}
\newcommand{\AlResolutionMismatchThreeTwo}{6.2^{+159}_{-6.1}\times10^{-6}}
\newcommand{\AlResolutionMismatchFourFour}{2.8^{+140}_{-2.8}\times10^{-5}}
\newcommand{\AlResolutionMismatchFourThree}{8.5^{+428}_{-8.3}\times10^{-6}}
\newcommand{\AlResolutionMismatchFiveFive}{3.4^{+234}_{-3.1}\times10^{-5}}

\newcommand{\AlAAOverNRHybSurFactorTwoTwo}{5.2^{+57}_{-5.0}}
\newcommand{\AlAAOverNRHybSurFactorAllModes}{3.6^{+50}_{-3.4}}
\newcommand{\AlAABetterThanNRHybSurPct}{80}
\newcommand{\AlNRHybSurMedianThreeThree}{4.4\times10^{-5}}
\newcommand{\AlAAMedianThreeThreeNRHybSurSet}{5.6\times10^{-5}}
\newcommand{\AlAABetterThanNRHybSurThreeThreePct}{58}
\newcommand{\AlNRHybSurMismatchTwoTwo}{6.9^{+38}_{-5.1}\times10^{-6}}
\newcommand{\AlNRHybSurMismatchAllModes}{8.7^{+39}_{-6.5}\times10^{-6}}
\newcommand{\AlNRHybSurUncoveredSimCountWord}{Twelve}
\newcommand{\AlNRHybSurCoverageSpinLimit}{0.9496}
\newcommand{\AlNRHybSurScoredEvenSimCount}{248}
\newcommand{\AlNRHybSurScoredOddSimCount}{237}
\newcommand{\AlNRHybSurScoredExchangeSymmetricCountWord}{eleven}
\newcommand{\AlModeCompSimCount}{30}
\newcommand{\AlModeCompTotalMass}{60}
\newcommand{\AlModeCompBandLowPct}{10}
\newcommand{\AlModeCompBandHighPct}{90}
\newcommand{\AlModeCompEdgeOnSevenModeFloor}{1.7\times10^{-4}}
\newcommand{\AlModeCompFiveToSevenFloorGain}{2.9}
\newcommand{\AlModeCompEdgeOnFiveModeFloor}{4.8\times10^{-4}}
\newcommand{\AlModeCompEdgeOnLooModel}{1.9\times10^{-4}}
\newcommand{\AlModeCompEdgeOnModelOverFloor}{1.1}
\newcommand{\AlModeCompFaceOnSevenModeFloor}{5\times10^{-7}}
\newcommand{\AlModeCompLooFitLevel}{1.2\times10^{-5}}
\newcommand{\AlModeCompCrossingInclination}{42}
\newcommand{\FrontTrainingMaxMassRatio}{8}
\newcommand{\FrontSpanLowerPercentile}{5th}
\newcommand{\FrontSpanUpperPercentile}{95th}

\newcommand{\FrontSkyAvgOrientations}{32}
\newcommand{\FrontSkyAvgMismatchLow}{8.9\times10^{-7}}
\newcommand{\FrontSkyAvgMismatchHigh}{2.8\times10^{-5}}
\newcommand{\FrontSkyAvgMassLow}{40}
\newcommand{\FrontSkyAvgMassHigh}{120}
\newcommand{\FrontSingleCallMs}{12.5}
\newcommand{\FrontBatchedMsPerWaveform}{0.43}
\newcommand{\FrontBatchSize}{4096}
\newcommand{\FrontBatchedStartFrequencyHz}{25}
\newcommand{\GradNumTrainingSims}{260}
\newcommand{\GradNumSurrogatedVariables}{19}

\newcommand{\GradNumPhysicalParamsWord}{eight}
\newcommand{\GradFigPoint}{(3.5,0.30,-0.20)}
\newcommand{\GradFigSpanM}{300}
\newcommand{\GradFigStep}{10^{-4}}

\newcommand{\GradErrBandQ}{4}
\newcommand{\GradErrBandChiOne}{0.30}
\newcommand{\GradErrBandChiTwo}{-0.20}
\newcommand{\GradErrBandRingdownFrameStartM}{60}
\newcommand{\GradErrBandInsetWidthM}{2}
\newcommand{\GradErrBandExtremumTimeM}{-99}
\newcommand{\GradErrBandInsetBandPercent}{0.06}
\newcommand{\GradErrBandFinalBandFraction}{0.545}
\newcommand{\GradErrBandClockSigmaMedianM}{0.238}
\newcommand{\GradCalRefOneSigmaPercent}{63}
\newcommand{\GradCalRefTwoSigmaPercent}{98}
\newcommand{\GradCalRefThreeSigmaPercent}{100}
\newcommand{\GradEnvelopePercentile}{90}
\newcommand{\GradEnvelopeZScore}{1.2816}

\newcommand{\GradPEInjMassMsun}{60}
\newcommand{\GradPEInjSNR}{40}
\newcommand{\GradFisherSigmaMass}{1.4}
\newcommand{\GradFisherSigmaQ}{0.35}
\newcommand{\GradFisherSigmaChiOne}{0.25}
\newcommand{\GradFisherSigmaChiTwo}{1.0}
\newcommand{\GradFisherSigmaDistMpc}{43}
\newcommand{\GradFisherSigmaTcMs}{0.26}
\newcommand{\GradFisherSigmaPhiRef}{0.13}
\newcommand{\GradFisherSigmaIota}{0.12}
\newcommand{\GradFisherSigmaIotaThreeFig}{0.124}
\newcommand{\GradHMCMaxBiasSigma}{0.04}
\newcommand{\GradNestedMaxBiasSigma}{0.08}
\newcommand{\GradHMCAcceptPercent}{86}
\newcommand{\GradHMCKeptSamples}{7.7\times10^{4}}
\newcommand{\GradNestedSigmaDistMpc}{42}
\newcommand{\GradNestedSigmaIota}{0.116}

\newcommand{\DiscSwapInteriorMaxDev}{4\times10^{-7}}

\newcommand{\DiscSwapWorstDev}{4\times10^{-6}}

\newcommand{\DiscSwapMaxMismatch}{5\times10^{-12}}

\newcommand{\DiscOddResidualHybLow}{7\times10^{-5}}
\newcommand{\DiscOddResidualHybHigh}{2\times10^{-4}}
\newcommand{\DiscOddResidualSurSevenLow}{7\times10^{-4}}
\newcommand{\DiscOddResidualSurSevenHigh}{2\times10^{-3}}
\newcommand{\DiscSwapThreeTwoHyb}{1.2\times10^{-3}}
\newcommand{\DiscSwapFourFourHyb}{1.1\times10^{-3}}
\newcommand{\DiscSwapThreeTwoSurSeven}{2.4\times10^{-3}}
\newcommand{\DiscSwapFourFourSurSeven}{1.7\times10^{-3}}
\newcommand{\DiscSwapThreeThreeSurSeven}{6\times10^{-2}}
\newcommand{\DiscDuHTwoTwoSurSeven}{3\times10^{-3}}
\newcommand{\DiscDuHTwoTwoHybLargestStep}{-4\times10^{-3}}
\newcommand{\DiscDuHTwoTwoHybSmallestStep}{2\times10^{-4}}
\newcommand{\GradCalCoverOneSigmaLowPercent}{59}
\newcommand{\GradCalCoverOneSigmaHighPercent}{89}
\newcommand{\GradCalCoverTwoSigmaLowPercent}{82}
\newcommand{\GradCalCoverTwoSigmaHighPercent}{97}
\newcommand{\GradCalCoverThreeSigmaLowPercent}{89}
\newcommand{\GradCalCoverThreeSigmaHighPercent}{99}
\newcommand{\GradCalRatioMedianLow}{0.15}
\newcommand{\GradCalRatioMedianHigh}{1.00}
\newcommand{\GradRhoCritCI}{421^{+1709}_{-345}}
\newcommand{\GradRhoCritMedian}{421}
\newcommand{\GradRhoCritMedianRounded}{420}
\newcommand{\GradRhoCritMin}{25}
\newcommand{\GradBiasToNoiseAtThirtyPercent}{8}
\newcommand{\GradNoiseWeightedHarshness}{2.3}
\newcommand{\GradNoiseWeightedRhoScale}{0.66}
\newcommand{\RemMfAbsErrLOOCI}{4.5^{+21}_{-4.1}\times10^{-5}}
\newcommand{\RemChifAbsErrLOOCI}{1.6^{+9.9}_{-1.5}\times10^{-5}}

\newcommand{\RemOutlierMfMiss}{3.4\times10^{-3}}
\newcommand{\DiscMetricEigMedianSoft}{6.1\times10^{-4}}
\newcommand{\DiscMetricEigMedianMid}{2.3}
\newcommand{\DiscMetricEigMedianStiff}{152}
\newcommand{\DiscMetricConditionMedian}{2.3\times10^{5}}
\newcommand{\DiscMetricSpearmanPath}{0.97}
\newcommand{\DiscMetricSpearmanQuad}{0.87}
\newcommand{\DiscMetricQuadDepartMM}{0.6}
\newcommand{\DiscMetricMeasuredMMMax}{0.91}
\newcommand{\DiscMetricNumPositiveDefinite}{26\,028}
\newcommand{\DiscMetricNumPathIntegrals}{58\,146}
\newcommand{\DiscCoverageGridSide}{128}
\newcommand{\IotaVariationAALow}{1.5}
\newcommand{\IotaVariationAAHigh}{4.9}
\newcommand{\IotaVariationHybLow}{1.8}
\newcommand{\IotaVariationHybHigh}{6.8}
\newcommand{\IotaVariationXPHMLow}{15}
\newcommand{\IotaVariationXPHMHigh}{53}
\newcommand{\IotaVariationTPHMLow}{16}
\newcommand{\IotaVariationTPHMHigh}{55}
\newcommand{\IotaVariationSEOBLow}{19}
\newcommand{\IotaVariationSEOBHigh}{36}
\newcommand{\IotaRankWorstFirst}{12}
\newcommand{\IotaRankWorstSecond}{19}
\newcommand{\IotaRankWorstThird}{36}
\newcommand{\IotaNumCommonSims}{248}
\newcommand{\IotaNumInclinations}{13}
\newcommand{\IotaTotalMassMsun}{110}
\newcommand{\IotaNumPoints}{39}
\newcommand{\IotaAAWinsPoints}{23}
\newcommand{\IotaAAWinsPercent}{59}
\newcommand{\IotaHybWinsPoints}{16}
\newcommand{\IotaHybWinsPercent}{41}
\newcommand{\IotaAAWinsTwoSevenHundred}{13}
\newcommand{\IotaAAWinsOneFourFourZero}{7}
\newcommand{\IotaAAWinsOneFourThreeNine}{3}
\newcommand{\IotaRatioAAHybLow}{0.05}
\newcommand{\IotaRatioAAHybHigh}{3.8}
\newcommand{\IotaCatalogAAWinPercent}{80}
\newcommand{\IotaFaceOnAAOneFourThreeNine}{5.2\times10^{-5}}

\newcommand{\IotaFaceOnHybOneFourFourZero}{4.1\times10^{-5}}
\newcommand{\IotaFaceOnAATwoSevenHundred}{3.3\times10^{-5}}
\newcommand{\IotaFaceOnHybTwoSevenHundred}{2.0\times10^{-4}}
\newcommand{\IotaFaceOnHybOneFourThreeNineMantissa}{1.8}
\newcommand{\IotaFaceOnAAOneFourFourZeroMantissa}{3.7}
\newcommand{\IotaFaceOnHybOverAATwoSevenHundred}{6}
\newcommand{\PerfNativeSpanOrbits}{18}
\newcommand{\PerfMinMassLow}{43}
\newcommand{\PerfMinMassMedian}{56}
\newcommand{\PerfMinMassHigh}{96}
\newcommand{\PerfShorteningMedianOrbits}{5}
\newcommand{\PerfShorteningMaxOrbits}{129}
\newcommand{\PerfNativeDurationSec}{1.04}
\newcommand{\PerfNativeStartFreqHz}{23.2}
\newcommand{\PerfLongDurationSec}{40}
\newcommand{\PerfModeCountWord}{seven}
\newcommand{\PerfBenchMassRatio}{4}
\newcommand{\PerfBenchSpinOne}{0.6}
\newcommand{\PerfBenchSpinTwo}{0.2}
\newcommand{\PerfBenchTotalMass}{60}
\newcommand{\PerfSampleRate}{4096}
\newcommand{\PerfHybCallFminHz}{10}
\newcommand{\PerfLongFminHz}{10}
\newcommand{\PerfBatchedFminHz}{25}
\newcommand{\PerfLongMassRatio}{1.5}
\newcommand{\PerfLongTotalMass}{20}
\newcommand{\PerfMassRangeFlowHz}{20}

\newcommand{\HybMultiStartMismatchOrder}{10^{-12}}

\newcommand{\HybAdiabaticFactor}{2\times10^{3}}
\newcommand{\HybClosurePredictedMedian}{4.6\times10^{-4}}
\newcommand{\HybClosureRandomPoints}{260}
\newcommand{\PerfNRSurWindowM}{4300}
\newcommand{\PerfNRSurOrbitsLow}{15}
\newcommand{\PerfNRSurOrbitsHigh}{22}
\newcommand{\PerfNRSurMinMassLow}{48}
\newcommand{\PerfNRSurMinMassHigh}{66}
\newcommand{\SoftSWSHLALTolerance}{10^{-12}}
\newcommand{\PerfNativeAACompiledMs}{12.5}
\newcommand{\PerfNativeAdAMs}{2.4}
\newcommand{\PerfHybCallAACompiledMs}{30}
\newcommand{\PerfHybCallNRHybSurMs}{74}
\newcommand{\PerfNativeAANumpyMs}{43}
\newcommand{\PerfHybCallAANumpyMs}{93}
\newcommand{\PerfNativeAAGradMs}{43}
\newcommand{\PerfNativeAdAGradMs}{44}
\newcommand{\PerfNativeAAGradNumpyMs}{240}
\newcommand{\PerfLongAACompiledMs}{62}
\newcommand{\PerfLongNRHybSurMs}{121}
\newcommand{\PerfLongAANumpyMs}{226}
\newcommand{\PerfBatchedSingleCallMs}{15.1}
\newcommand{\PerfHybCallSpeedup}{2.5}
\newcommand{\PerfLongSpeedup}{2}
\newcommand{\PerfNativeNRHybSurLowMs}{100}
\newcommand{\PerfNativeNRHybSurHighMs}{110}
\newcommand{\PerfBatchedThreads}{128}
\newcommand{\PerfBatchedBatchOneMs}{7.8}
\newcommand{\PerfBatchedAAMs}{0.43}
\newcommand{\PerfBatchedAABatch}{4096}
\newcommand{\PerfBatchedAmortization}{35}
\newcommand{\PerfBatchedBatchingGain}{18}
\newcommand{\PerfBatchedAdAMs}{0.14}
\newcommand{\PerfBatchedAdABatch}{4096}
\newcommand{\PerfDriftLimitPct}{10}

\newcommand{\HybEndpointAddedOrbits}{200}

\newcommand{\HybHoldbackMaxDephasingRad}{0.27}
\newcommand{\HybHoldbackOrbits}{74}
\newcommand{\HybHoldbackStartX}{0.039}
\newcommand{\HybSpliceMaxModeMismatch}{4.5\times10^{-3}}
\newcommand{\HybSpliceAllModeMismatch}{4.1\times10^{-7}}
\newcommand{\HybActionAllModeMismatch}{2.3\times10^{-10}}
\newcommand{\HybSpliceFourThreeMismatch}{4.5\times10^{-3}}
\newcommand{\HybActionFourThreeMismatch}{5.8\times10^{-5}}
\newcommand{\HybSpliceGainTwoOneFiveFive}{500}
\newcommand{\HybActionCleanAllModeMismatch}{5.9\times10^{-12}}
\newcommand{\HybFirstLawPNSideMedian}{2.1\times10^{-10}}
\newcommand{\HybLongSimCountWord}{eight}
\newcommand{\HybExtOrbitsLow}{46}
\newcommand{\HybExtOrbitsHigh}{129}
\newcommand{\HybNativeFracLowPct}{4}
\newcommand{\HybNativeFracHighPct}{16}
\newcommand{\HybExtDephasingMaxRad}{1.4}
\newcommand{\HybExtDephasingLowRad}{0.17}
\newcommand{\HybExtDephasingHighRad}{1.41}
\newcommand{\HybDephasingPerOrbitLowMrad}{4}
\newcommand{\HybDephasingPerOrbitHighMrad}{11}

\newcommand{\HybAmpRatioDeviationBound}{10^{-2}}
\newcommand{\HybFirstLawNRSideMedian}{2.9\times10^{-6}}
\newcommand{\LevDephasingLongFiveLowRad}{1.5}
\newcommand{\LevDephasingLongFiveHighRad}{2.1}
\newcommand{\LevDephasingLongFivePerOrbitLowMrad}{12}
\newcommand{\LevDephasingLongFivePerOrbitHighMrad}{16}
\newcommand{\LevDephasingHighQLowRad}{0.098}
\newcommand{\LevDephasingHighQHighRad}{0.12}
\newcommand{\LevDephasingHighQPerOrbitLowMrad}{1.6}
\newcommand{\LevDephasingHighQPerOrbitHighMrad}{2.6}
\newcommand{\HybFirstLawBandPPM}{8.1^{+16}_{-6.1}}
\newcommand{\HybClosureExcessRatio}{160}
\newcommand{\MethATrainingSimCount}{260}
\newcommand{\MethAOddRatioTrainingCount}{242}
\newcommand{\MethAExchangeSymmetricCount}{18}
\newcommand{\MethAMassRatioMin}{1}
\newcommand{\MethAMassRatioMax}{8}
\newcommand{\MethASpinMin}{-0.95}
\newcommand{\MethASpinMax}{0.998}
\newcommand{\MethAOddRatioSpinEdge}{0.95}
\newcommand{\MethAEvenRatioSpinEdge}{0.998}
\newcommand{\MethAPhaseSign}{-1}
\newcommand{\MethABlendStart}{-20}
\newcommand{\MethABlendEnd}{-5}
\newcommand{\MethAInspiralElementCount}{14}
\newcommand{\MethAInspiralOrder}{24}
\newcommand{\MethAInspiralCoeffsPerElement}{25}
\newcommand{\MethASpectralRowCount}{24472}
\newcommand{\MethAModeTruncReleasedMedian}{1.9\times10^{-6}}
\newcommand{\MethAModeTruncReleasedFracBelowTol}{91.7}
\newcommand{\MethAInPlaneSpinTol}{10^{-3}}
\newcommand{\MethAEccentricityTol}{10^{-3}}
\newcommand{\MethAMassRatioCut}{8}
\newcommand{\MethAMinOrbitsCut}{18}
\newcommand{\MethARadialActionBound}{5\times10^{-7}}
\newcommand{\MethAModeWorkingTolerance}{10^{-4}}
\newcommand{\MethAClockWorkingTolerance}{10^{-7}}
\newcommand{\MethAClockToleranceFactor}{10^3}
\newcommand{\MethAMinElementWidth}{2\pi}
\newcommand{\MethACandidateSimCount}{309}
\newcommand{\MethADistinctPointCount}{260}
\newcommand{\MethAReplacementCount}{Three}
\newcommand{\MethAReplacementMassRatioHigh}{7}
\newcommand{\MethAReplacementMassRatioLow}{1}
\newcommand{\MethAOrbitsMin}{18}
\newcommand{\MethAOrbitsMax}{147}
\newcommand{\MethAOrbitsMedian}{23}
\newcommand{\MethAFirstLawResidualInspiral}{9\times10^{-6}}
\newcommand{\MethAFirstLawGrowthToPeak}{500}
\newcommand{\MethAFirstLawViolationBlendStart}{370}
\newcommand{\MethAFirstLawViolationBlendEnd}{580}
\newcommand{\MethARadialMomentumMedian}{1.8\times10^{-3}}
\newcommand{\MethARadialMomentumMax}{8.9\times10^{-2}}
\newcommand{\MethAConsHamiltonianResidual}{1.0\times10^{-6}}
\newcommand{\MethAResidualEccentricity}{2.1^{+1.1}_{-1.2}\times10^{-4}}
\newcommand{\MethAResidualEccentricityMin}{2.4\times10^{-5}}
\newcommand{\MethAResidualEccentricityMax}{6.0\times10^{-4}}
\newcommand{\MethARadialActionRatio}{2.3^{+3.0}_{-1.8}\times10^{-8}}
\newcommand{\MethARadialActionRatioMax}{1.8\times10^{-7}}
\newcommand{\MethARadialActionBoundFactor}{three}
\newcommand{\MethAModeToleranceCrossingOrder}{6}
\newcommand{\MethAModeCoeffMedianAtPmax}{9\times10^{-7}}
\newcommand{\MethAModeTruncTrainingMedian}{8.7\times10^{-7}}
\newcommand{\MethAModeTruncTrainingFracBelowTol}{94.6}
\newcommand{\MethAActionTruncTraining}{1.9\times10^{-10}}
\newcommand{\MethAClockTruncTraining}{5.2\times10^{-10}}
\newcommand{\MethANoiseFloor}{10^{-6}}
\newcommand{\MethANoiseFloorOrdersBelowTol}{two}

\newcommand{\MethAHpRepresentativeMassRatio}{8}
\newcommand{\MethAHpRepFitError}{2.22^{+11.7}_{-2.21}\times10^{-7}}
\newcommand{\MethAHpRepFitErrorMax}{2.6\times10^{-6}}
\newcommand{\MethAHpRepOrderMin}{1}
\newcommand{\MethAHpRepOrderMax}{8}
\newcommand{\MethAHpRepOrderMedian}{4}
\newcommand{\MethAHpCatalogOrderMin}{4}
\newcommand{\MethAHpCatalogOrderMax}{17}
\newcommand{\MethAHpCatalogOrderMedian}{16}
\newcommand{\MethBTrainMassRatioMin}{1}
\newcommand{\MethBTrainMassRatioMax}{8}

\newcommand{\MethBRatioEarlyDeparturePct}{8}
\newcommand{\MethBRatioLateMin}{0.62}
\newcommand{\MethBRatioLateMax}{1.72}
\newcommand{\MethBSeamStepMedianTwoTwo}{5\times10^{-4}}
\newcommand{\MethBSeamStepMedianHigherMax}{1.5\times10^{-2}}
\newcommand{\MethBActionAnchorOrbits}{four}
\newcommand{\MethBSeamRatioWindowEndOrbits}{five}
\newcommand{\MethBDirectWeightOtherMin}{0.9898}
\newcommand{\MethBDirectWeightTwoTwo}{0.849}
\newcommand{\MethBNModesOtherThanTwoTwo}{six}
\newcommand{\MethBNModes}{seven}
\newcommand{\MethBNTrainSims}{260}
\newcommand{\MethBNearEqualMassSims}{63}
\newcommand{\MethBRingdownSplitTime}{2.06}
\newcommand{\MethBPlungeFitEnd}{12.75}
\newcommand{\MethBMixingCondMax}{1.05}

\newcommand{\MethBMixingOffDiagMax}{0.20}
\newcommand{\MethBDampingAmplificationLow}{10^4}
\newcommand{\MethBDampingAmplificationHigh}{10^5}
\newcommand{\MethBRemnantMassTrendDegree}{four}
\newcommand{\MethBRemnantSpinTrendDegree}{three}
\newcommand{\MethBWaveformTrendDegree}{two}
\newcommand{\MethBRemnantSpinLOORms}{5.3\times10^{-5}}
\newcommand{\MethBRemnantMassLOORms}{1.2\times10^{-4}}

\newcommand{\MethBIMRBlendStart}{-20}
\newcommand{\MethBIMRBlendEnd}{-5}

\newcommand{\MethBRatioRouteAllModeGainPct}{+0.07}
\newcommand{\MethBRatioRouteFactorTwoTwo}{1.0006}
\newcommand{\MethBRatioRouteFactorThreeTwo}{0.9999}
\newcommand{\MethBRatioRouteFactorFourFour}{1.0000}
\newcommand{\MethBRatioRouteNearEqualThreeThree}{1.093}
\newcommand{\MethBRatioRouteNearEqualFiveFive}{0.957}
\newcommand{\MethBRatioRouteNearEqualFiveFiveWorsePct}{5}

\newcommand{\MethBDiagMixingFactorThreeTwo}{1.36}
\newcommand{\MethBDiagMixingFactorFourThree}{1.00}
\newcommand{\MethBDetrendFactorMin}{17}
\newcommand{\MethBDetrendFactorMax}{30}
\newcommand{\MethBDetrendRateTwoOneBefore}{0.44}
\newcommand{\MethBDetrendRateTwoOneAfter}{0.026}
\newcommand{\MethBDetrendRateFiveFiveBefore}{1.35}
\newcommand{\MethBDetrendRateFiveFiveAfter}{0.067}
\newcommand{\ValTrainingSimCount}{260}
\newcommand{\ValOddModeSimCount}{242}
\newcommand{\CampModePairCount}{1748}
\newcommand{\ValNRSurBoxQMax}{4}
\newcommand{\ValNRSurBoxChiMax}{0.8}
\newcommand{\ValNRSurBoxSimCount}{151}
\newcommand{\ValNRSurBoxOutsideCount}{109}
\newcommand{\ValBoxModeCountWord}{six}
\newcommand{\CampBoxQMax}{8}
\newcommand{\CampBoxChiMax}{0.8}
\newcommand{\CampBoxMtotMin}{20}
\newcommand{\CampBoxMtotMax}{120}
\newcommand{\CampAnchoredSimCount}{201}
\newcommand{\ValRhoAAInSample}{620^{+1904}_{-452}}
\newcommand{\ValMedianMMAAInSample}{1.3\times10^{-6}}
\newcommand{\ValRhoAALOO}{421^{+1709}_{-345}}
\newcommand{\ValMedianMMAALOO}{2.8\times10^{-6}}

\newcommand{\ValRhoAdAReleasedLOO}{462^{+1638}_{-372}}
\newcommand{\ValMedianMMAdAReleasedLOO}{2.3\times10^{-6}}
\newcommand{\ValRhoHybSur}{239^{+236}_{-137}}
\newcommand{\ValMedianMMHybSur}{8.7\times10^{-6}}
\newcommand{\ValRhoHybSurN}{248}
\newcommand{\ValRhoBoxAAInSample}{875^{+1978}_{-648}}
\newcommand{\ValMedianMMBoxAAInSample}{6.5\times10^{-7}}
\newcommand{\ValRhoBoxAALOO}{757^{+1754}_{-573}}
\newcommand{\ValMedianMMBoxAALOO}{8.7\times10^{-7}}
\newcommand{\ValRhoBoxAdAInSample}{937^{+1898}_{-702}}
\newcommand{\ValMedianMMBoxAdAInSample}{5.7\times10^{-7}}
\newcommand{\ValRhoBoxHybSur}{301^{+175}_{-162}}
\newcommand{\ValMedianMMBoxHybSur}{5.5\times10^{-6}}
\newcommand{\ValRhoBoxNRSur}{75^{+114}_{-35}}
\newcommand{\ValMedianMMBoxNRSur}{8.9\times10^{-5}}
\newcommand{\ValRhoBoxNRSurNN}{64^{+56}_{-30}}
\newcommand{\ValMedianMMBoxNRSurNN}{1.2\times10^{-4}}
\newcommand{\ValRhoBoxNRSurVTwo}{131^{+268}_{-63}}
\newcommand{\ValMedianMMBoxNRSurVTwo}{2.9\times10^{-5}}

\newcommand{\ValBoxWindowEndAfterPeak}{74}
\newcommand{\ValBoxShortWindowEndAfterPeak}{30}
\newcommand{\ValNRSurStartBeforePeak}{4300}
\newcommand{\ValNRSurEndAfterPeak}{100}
\newcommand{\ValBoxAALOOCloserThanHybSurPct}{88}

\newcommand{\CampTargetCountWord}{five}
\newcommand{\CampDrawsTotalMillion}{2.1}
\newcommand{\CampNRCheckAboveThresholdPoints}{4{,}338}
\newcommand{\CampNRCheckPopulationMillion}{1.66}
\newcommand{\CampNRCheckAboveThresholdPct}{0.26}
\newcommand{\CampTPHMDraws}{460{,}500}
\newcommand{\CampMedianMMTPHM}{4.9\times10^{-4}}
\newcommand{\CampPNinetyMMTPHM}{2.0\times10^{-3}}
\newcommand{\CampMedianMMNRSur}{8.6\times10^{-5}}
\newcommand{\CampPNinetyMMNRSur}{4.2\times10^{-4}}
\newcommand{\CampMedianMMSEOB}{7.6\times10^{-4}}
\newcommand{\CampPNinetyMMSEOB}{3.2\times10^{-3}}
\newcommand{\CampTriangleModeCountWord}{five}
\newcommand{\CampTrianglePoints}{800}
\newcommand{\CampTriangleInclinationCountWord}{nine}
\newcommand{\CampTriangleRows}{7200}
\newcommand{\CampTriangleMedianAAHyb}{2.2\times10^{-5}}
\newcommand{\CampTrianglePNinetyAAHyb}{7.0\times10^{-5}}
\newcommand{\CampTriangleThreshold}{10^{-3}}
\newcommand{\CampTriangleAAHybAboveThresholdRows}{2}
\newcommand{\CampTriangleFaceOnAAHyb}{1.7\times10^{-5}}
\newcommand{\CampTriangleEdgeOnAAHyb}{3.4\times10^{-5}}
\newcommand{\CampTriangleMedianSEOBAA}{3.4\times10^{-4}}
\newcommand{\CampTriangleMedianSEOBHybSur}{3.6\times10^{-4}}
\newcommand{\CampAnchoredMedianNRAA}{8.2\times10^{-6}}
\newcommand{\CampAnchoredMedianNRHybSur}{1.9\times10^{-5}}
\newcommand{\CampAnchoredBothBelowThresholdPct}{99.8}
\newcommand{\CampNRCheckThreshold}{10^{-2}}
\newcommand{\CampNRCheckPoints}{116}
\newcommand{\CampNRCheckMedianAA}{1.95\times10^{-4}}
\newcommand{\CampNRCheckCampaignOverNRMedian}{88}
\newcommand{\CampNRCheckCampaignOverNRPNinety}{675}
\newcommand{\CampNRCheckAACloserThanTargetPct}{93.8}
\newcommand{\CampNRCheckAACloserThanHybSurPct}{94.8}
\newcommand{\CampNRCheckAAOverAdAMedian}{0.955}
\newcommand{\CampTPHMSetPoints}{600}
\newcommand{\CampTPHMSetNearestSimDistance}{0.210}
\newcommand{\CampTPHMSetMMTPHM}{4.3^{+17}_{-3.6}\times10^{-3}}
\newcommand{\CampTPHMSetMMAA}{7.2^{+144}_{-6.8}\times10^{-5}}
\newcommand{\CampTPHMSetMMAdA}{7.7^{+142}_{-7.3}\times10^{-5}}
\newcommand{\CampTPHMSetTPHMWorsePct}{97.7}
\newcommand{\CampTPHMSetEdgeOverFaceTPHM}{2.5}
\newcommand{\CampTPHMSetAAOverAdAMedian}{1.011}
\newcommand{\CampTPHMSetAAOverAdAEffectPct}{1}
\newcommand{\CampModeAAMedianMMTwoTwo}{1.14\times10^{-6}}
\newcommand{\CampModeAdAMedianMMTwoTwo}{1.09\times10^{-6}}
\newcommand{\CampModeRatioTwoTwo}{0.965}
\newcommand{\CampModeAABetterPctTwoTwo}{42}
\newcommand{\CampModeAAMedianMMTwoOne}{2.39\times10^{-5}}
\newcommand{\CampModeAdAMedianMMTwoOne}{2.25\times10^{-5}}
\newcommand{\CampModeRatioTwoOne}{0.998}
\newcommand{\CampModeAABetterPctTwoOne}{45}
\newcommand{\CampModeAAMedianMMThreeThree}{3.30\times10^{-5}}
\newcommand{\CampModeAdAMedianMMThreeThree}{2.67\times10^{-5}}
\newcommand{\CampModeRatioThreeThree}{0.996}
\newcommand{\CampModeAABetterPctThreeThree}{48}
\newcommand{\CampModeAAMedianMMThreeTwo}{1.61\times10^{-5}}
\newcommand{\CampModeAdAMedianMMThreeTwo}{1.66\times10^{-5}}
\newcommand{\CampModeRatioThreeTwo}{0.987}
\newcommand{\CampModeAABetterPctThreeTwo}{45}
\newcommand{\CampModeAAMedianMMFourFour}{4.45\times10^{-5}}
\newcommand{\CampModeAdAMedianMMFourFour}{4.67\times10^{-5}}
\newcommand{\CampModeRatioFourFour}{0.996}
\newcommand{\CampModeAABetterPctFourFour}{46}
\newcommand{\CampModeAAMedianMMFourThree}{4.74\times10^{-5}}
\newcommand{\CampModeAdAMedianMMFourThree}{4.62\times10^{-5}}
\newcommand{\CampModeRatioFourThree}{0.994}
\newcommand{\CampModeAABetterPctFourThree}{47}
\newcommand{\CampModeAAMedianMMFiveFive}{1.56\times10^{-4}}
\newcommand{\CampModeAdAMedianMMFiveFive}{1.51\times10^{-4}}
\newcommand{\CampModeRatioFiveFive}{0.972}
\newcommand{\CampModeAABetterPctFiveFive}{41}
\newcommand{\CampModePooledRatio}{0.991}
\newcommand{\CampModePooledAABetterPct}{45}

\newcommand{\CampModeRatioMin}{0.965}
\newcommand{\CampModeRatioMax}{0.998}

\begin{document}

\title{\NRAAS: a spectral numerical-relativity surrogate family for non-eccentric aligned-spin binary-black-hole waveforms}

\author{Vaishak Prasad}
\affiliation{Institute for Gravitation and the Cosmos, The Pennsylvania State
University, University Park, Pennsylvania 16802, USA}
\affiliation{Department of Physics, The Pennsylvania State University,
University Park, Pennsylvania 16802, USA}
\affiliation{International Centre for Theoretical Sciences, Tata Institute of
Fundamental Research, Bangalore 560089, India}

\date{\today}

\begin{abstract}

We present \NRAAS, a family of numerical-relativity surrogates for
non-eccentric aligned-spin binary-black-hole waveforms inspired by an \emph{action-angle}
parameterization. The model provides both the dynamics: the Hamiltonian values,
azimuthal action $\Jphi(\Phi)$, the orbital clock $\tau(\Phi)$ as functions of
its phasing, and the radiative pieces: the waveforms, their near-analytic
derivatives, and modeling errors associated with each. The surrogated variables
are conditioned into pieces that vary slowly on the radiation-reaction
timescale, represented on shared $h$-adaptive Chebyshev elements, and
regressed across the intrinsic parameters with a Gaussian-process model. A
time-parameterized merger-ringdown set of elements is attached where the
adiabatic description breaks down. Fixed-hyperparameter leave-one-out validation of the central
\NRHJ{}\_\AAf{} model variant against numerical relativity gives a $(2,2)$ unweighted mismatch of
$\AlBlendedLooMismatchTwoTwo$ and an all-mode mismatch of
$\AlBlendedLooMismatchAllModes$ over the full inspiral--merger--ringdown
span.
Sky-averaged and signal-to-noise-ratio-weighted over $\FrontSkyAvgOrientations$ orientations per
simulation, the model's full inspiral--merger--ringdown in-sample mismatch against the SXS numerical-relativity simulations
used for training is $\FrontSkyAvgMismatchLow$ to $\FrontSkyAvgMismatchHigh$ across total
masses $\FrontSkyAvgMassLow$--$\FrontSkyAvgMassHigh\,M_\odot$.
Waveforms are hybridized using post-Newtonian
waveforms. Beyond accuracy, \NRAAS{} provides two capabilities uncommon in numerical-relativity surrogates:
nearly \emph{analytic} derivatives of the strain with respect to the physical
parameters, and a \emph{calibrated} Gaussian-process predictive uncertainty of
the model. A full non-hybridized inspiral--merger--ringdown waveform evaluates in $\FrontSingleCallMs$~ms on
a single CPU call. Batched evaluation takes $\FrontBatchedMsPerWaveform$~ms per waveform
at a batch of $\FrontBatchSize$, measured at a starting frequency of $\FrontBatchedStartFrequencyHz$~Hz.
\end{abstract}

\maketitle

\section{Introduction}
\label{sec:intro}

Gravitational-wave astronomy requires waveform models that are both
faithful to numerical relativity (NR) and fast enough for stochastic inference
over the multi-dimensional intrinsic parameter space. Future gravitational wave data analysis
would be memory- and sometimes compute-limited owing to the long durations of the
signals, prompting the use of gradient-based inference and forecasting (e.g., Fisher, Monte Carlo methods like Hamiltonian, Langevin, flow-based, etc.)
that can exploit derivatives of the strain with respect to the
source parameters~\cite{Iacovelli:2022bbs,Iacovelli:2022mbg,Borhanian:2020ypi,Wong:2023lgb}.
This would require the waveforms to be increasingly differentiable. Fisher
methods have also been a valuable tool to forecast the precision of parameter
estimation for future gravitational wave
detectors~\cite{Finn:1992wt,Cutler:1994ys,Poisson:1995ef,Borhanian:2022czq,Gupta:2023evt,Branchesi:2023mws,Evans:2023euw,Gupta:2023lga,Pandey:2024mlo,ET:2025xjr}.
Differentiable waveform models have been built for gradient-based template
placement and inference~\cite{Coogan:2022qxs,Edwards:2023sak}. At the
signal-to-noise ratios expected in next-generation detectors, the waveform modeling
errors must stay small enough so as to not bias the inferred
parameters of the signal~\cite{Purrer:2019jcp,Kapil:2024zdn}, and it is useful to carry an estimate of that error into the
inference procedure~\cite{Moore:2014pda,Doctor:2017csx,Williams:2019vub}.
Signals that stay in the detector sensitivity band for many orbits also need waveforms that start well
before the finite-length NR data available in catalogs. For models trained on finite-length NR
waveforms, hybridization with post-Newtonian (PN)
waveforms~\cite{MacDonald:2011ne} becomes necessary. These requirements have so far been studied
separately in various capacities. \NRHJ{} is built to meet all four requirements: accuracy against numerical
relativity, hybridization to low frequency, analytic parameter derivatives,
and a calibrated estimate of its own error.

NR~\cite{Boyle:2019kee,Scheel:2025jct,Healy:2022wdn,Ferguson:2023vta} is considered the gold standard for modeling
gravitational waveforms from systems such as binary mergers. However, it is far
too expensive to integrate the full Einstein's field equations and evaluate inside a stochastic sampler used to infer
the Bayesian posteriors of the source parameters. The waveform models used routinely
are therefore fast, semi-analytic, post-Newtonian-informed models, calibrated to NR. The effective-one-body
family~\cite{Buonanno:1998gg,Buonanno:2000ef}, including the
\textsc{SEOBNRv5}~\cite{Pompili:2023tna,Ramos-Buades:2023ehm,Mihaylov:2023bkc} and
  \textsc{TEOBResumS}~\cite{Nagar:2018zoe,Nagar:2024dzj}, and
  the phenomenological \textsc{IMRPhenom} family in frequency and time domain
~\cite{Husa:2015iqa,Khan:2015jqa,Pratten:2020fqn,Garcia-Quiros:2020qpx,Pratten:2020ceb,Estelles:2021gvs,Thompson:2023ase}
  are some examples of modes that provide sufficiently accurate waveforms at a fraction of the cost of full NR simulations,
  making it possible to carry out parameter estimation of signals observed in LIGO-Virgo-KAGRA network of detectors.
  Surrogate models bridge the gap differently, by interpolating a training set
  of NR simulations to arbitrary points of the parameter space at negligible
  cost~\cite{Field:2013cfa,Tiglio:2021ysj}, aiming to produce an accurate estimate of the
  waveforms produced by Numerical Simulations themselves, without having to run
  NR simulations that solve for the full spacetime geometry.
  Established surrogates such as \textsc{NRSur7dq4}~\cite{Varma:2019csw} and
  \textsc{NRHybSur3dq8}~\cite{Varma:2018mmi} achieve close-to-NR-level
  accuracy through empirical interpolation bases fit in selected time nodes
  to a co-orbital or
  co-precessing decomposition of the waveform. Surrogate modeling now spans
  many kinds of models: the first NR surrogates and their precessing
  successors~\cite{Blackman:2015pia,Blackman:2017dfb,Blackman:2017pcm},
  reduced-order models of analytic approximants~\cite{Purrer:2014fza,Purrer:2015tud,Cotesta:2020qhw,Gadre:2022sed},
  extensions to large mass ratios~\cite{Rifat:2019ltp,Islam:2022laz,Yoo:2022erv,Rink:2024swg}
  and to near-extremal spins~\cite{Walker:2022zob}, surrogates carrying memory and
  Cauchy-characteristic evolution~\cite{Yoo:2023spi}, to ringdown~\cite{MaganaZertuche:2024ajz}, and eccentric
  binaries~\cite{Islam:2021mha,Nee:2025nmh,Maurya:2025shc,Islam:2026blk,Islam:2025llx}. The cross-parameter
  regression across the compact binary parameter space itself has been carried out with polynomial fits, Gaussian
  processes~\cite{Doctor:2017csx,Williams:2019vub}, neural
networks~\cite{Chua:2018woh,Khan:2020fso,Tissino:2022thn,GramaxoFreitas:2024bpk,GramaxoFreitas:2026jgl,Theodoropoulos:2026wkj,Whittall:2026yvp},
and other machine-learning regressors~\cite{Schmidt:2020yuu,Barsotti:2021wks}.

This paper introduces \NRAAS, a \emph{family} of NR surrogates for aligned-spin
binary black holes inspired by an \emph{action-angle} parameterization. The
inspiral of a quasi-circular binary in PN theory is an adiabatic
process with two well-separated timescales: a fast orbital timescale and a slow
radiation-reaction timescale on which the orbit steadily shrinks. We exploit
this directly, parameterizing each waveform by the fast orbital \emph{angle},
i.e., the orbital phase $\Phi$ obtained from the $(2,2)$ mode, and reducing every
spherical-harmonic mode to a co-orbital variable that varies slowly in $\Phi$.

The construction is inspired by Hamilton--Jacobi theory, where the equations of
motion of the system simplify when there exist $N$ independent constants of
motion in involution with each other. A canonical transformation to the action
variables then implies that the conjugate angle coordinates advance uniformly,
and, for a multiply periodic system, restricting the motion of the system to
torus-like regions of the phase space.
The action variables also serve as a natural set of variables to label each
phase trajectory. For a Newtonian circular orbit, the angular momentum is
a constant of motion, and is one of the action variables. The orbital phase is the
conjugate angle. Demodulating the dynamics with the fast angle therefore leaves
quantities that depend on the actions alone. Actions are, moreover, adiabatic
invariants: their response to slow external forcing of the system averages away over the fast angle,
rendering them the most slowly varying, and hence the most compactly
representable, quantities in the problem. A surrogate built in this frame has
to represent only what survives the removal of the fast phase, and the expectation is that
it is the smoothest description of the waveform available. It is with this motivation that we
attempt to build a surrogate model. In Section~\ref{sec:hjmotivation}
we describe this application to the construction of the surrogate.

The name of the surrogate model family is chosen accordingly, ``HJ'' standing
for the Hamilton--Jacobi inspired construction. Among the flavors, ``AA'' stands for
action-angle and ``AdA'' for adiabatic-angle. The variants \AAint{} and \AAintPN{}
are not discussed here. These models, which borrow more elements from the HJ theory, though
less accurate,  helped inform the development of the accurate variants presented here,
and will be described in a
companion paper. ``3dq8'' denotes the
three-dimensional aligned-spin parameter space $(\log q,\chi_{1z},\chi_{2z})$
validated to mass ratio $q\le \FrontTrainingMaxMassRatio$, and is borrowed from the existing established
surrogate family nomenclature for consistency.

The difference between the model variants is in which variables are surrogated,
and which of those are derived from the surrogated quantities.
\AAf{} fits the actions, the clock $\tau(\Phi)$, which gives the time as a function of the orbital phase,
and the PN-normalized mode ratios, while \AdA{} fits the clock and the co-orbital modes directly
(Sec.~\ref{sec:split}). \AAf{} and \AdA{} have nearly the same accuracy in fixed-hyperparameter leave-one-out (LOO) validation, with \AdA{} marginally more
accurate at the median. \AdA{} is faster, but it is yet to be hybridized at the time of writing this paper.
Every figure and number quoted in this paper is for the \AAf{} variant unless another model is
explicitly named.

\NRAAS{} contributes an action-angle conditioning whose slow variables are
the surrogated content. These slow variables are represented on
spectral elements for each parameter / numerical datapoint. A calibrated Gaussian-process
regression over their spectral coefficients provides cross-parameter interpolation,
and analytic parameter derivatives of the physical
strain. The remainder of this section compares the model against existing works.
The regression and modeling details are described in Secs.~\ref{sec:conditioning}--\ref{sec:gradients}.

\NRHJ{} differs from the existing differentiable waveform models in certain ways.
Sec.~\ref{sec:related} and Table~\ref{tab:landscape} state the
comparison.

The organization of the paper is as follows. Section~\ref{sec:trainingset}
describes the training set. Section~\ref{sec:conditioning} describes the
action-angle motivation and conditioning, the architecture of the surrogate, and
the physically motivated inspiral/merger--ringdown boundary.
Section~\ref{sec:construction} details the surrogate build and the conservative
dynamics action-variable content of the surrogate.
Section~\ref{sec:hybridization} presents the action-space native hybridization to
arbitrarily low starting frequency. Section~\ref{sec:aligned} reports
in-sample and LOO validation against NR simulations, the latter held out of the training set,
and comparisons with existing
surrogates, including the distinguishability of the models from NR
(Sec.~\ref{sec:distinguishability}), and Sec.~\ref{sec:campaign} reports a
cross-model campaign of model-versus-model mismatches,
with the large disagreements resolved by comparison
against NR. Section~\ref{sec:gradients} presents the analytic gradients,
calibrated uncertainty, and a differentiable parameter-estimation
demonstration. Section~\ref{sec:performance} reports computational performance,
and Section~\ref{sec:software} describes the software, its public interface and
its availability. Section~\ref{sec:discussion} discusses limitations and
extensions, and Appendix~\ref{sec:remnant} discusses the remnant map.

Throughout this paper, a statistic drawn from a distribution is quoted as
its median, with the distances to the \FrontSpanLowerPercentile{} and \FrontSpanUpperPercentile{} percentiles of that
distribution as sub- and superscripts. \section{The surrogate modeling landscape}
\label{sec:related}

Three properties can be used to distinguish the currently available differentiable waveform models:
i) Whether a model is differentiable, ii) whether it runs natively on graphics processing units (GPUs), and
iii) whether it carries a per-evaluation error estimate.
\NRHJ{} provides all three (Table~\ref{tab:landscape}). Its design is guided
by the requirements anticipated of next-generation gravitational-wave data analysis,
providing a foundation for continued development, despite its restriction currently to aligned-spin, non-eccentric parameter space.
It differs from currently available models in the following ways:
\begin{enumerate}
    \item Currently available NRSur models are time-domain surrogates, that
    choose an optimum set of time nodes using an empirical basis to interpolate
    across waveforms from different systems. \NRHJ{} instead interpolates over
    the binary orbital phase in the
    co-precessing frame (here, same as the inertial frame, as the training set is non-precessing),
    naturally sampling at a lower rate in time in the early
    inspiral, and at a higher rate closer to merger, in a manner that is
    tailored to each system's rate of change of adiabaticity $\epsilon$:
    \begin{equation}
        \epsilon=\dot\omega/\omega^2,
    \end{equation}

    with $\omega$ the orbital frequency of the binary.
    The angle parameterization also ensures that the surrogate continues to retain
    the accuracy of the underlying NR simulations as the lengths of the simulations
    increase in the catalogs. Waveforms currently in use by data analysis work in the frequency or the time domain.
    Therefore, a light root-finding step is implemented to
    represent the waveforms in the time domain at evaluation time.

    \item Mainstream differentiable-waveform codes that currently exist, like ripple, gwfast, jim,
    etc., ~\cite{Edwards:2023sak,Iacovelli:2022mbg,Wong:2023lgb} are mostly JAX re-implementations of
    closed-form frequency-domain approximants, with no dynamics to integrate.
    The publicly available differentiable NR surrogates are JaxNRSur~\cite{JaxNRSur} and the more recent neural network surrogate model~\cite{Purrer:2026wgp}.
    JaxNRSur is a JAX/Equinox reimplementation of
    \textsc{NRHybSur3dq8} and \textsc{NRSur7dq4} that obtains its gradients by
    auto-differentiation. Auto-differentiation, although very convenient, can fail out of the box when
    root finding is involved (as in our case, inverting the phase-to-time map),
    and is not guaranteed to yield gradients that are accurate, well-behaved,
    and physical at all points of the parameter space. Neural-network emulators of NR
    waveforms~\cite{GramaxoFreitas:2024bpk,Purrer:2026wgp,Modrekiladze:2026mcr}, some of
    them differentiable by construction~\cite{Purrer:2026wgp,Modrekiladze:2026mcr},
    form a separate branch of this effort, which do not yet expose an
    interpolation or modeling uncertainty yet.
    \NRHJ{} takes a different approach. Instead of
    using auto-differentiation, it differentiates the analytic Chebyshev trend functions and
    the Gaussian-process-regression (GPR) mean in closed form, and carries the
    implicit-function-theorem transport for the phase-to-time inversion. The accuracy is
    guaranteed by the underlying model due to the factoring out of known post-analytic structure,
    careful handling of domain edges and coordinates, and the gradients are hand-derived and checked to be well-behaved.

    \item \NRHJ{} is \emph{fused into a device-native \textsc{Kokkos} kernel},
    providing performance portability. The core codebase is single-language,
    with no parallel JAX implementation or XLA runtime to maintain separately.

    \item The JAX implementations of surrogate models do not expose
    their interpolation uncertainty. In \NRHJ, the GPR predictive variance is
    calibrated, propagated, and exposed to the user application programming interface (API).
\end{enumerate}

 \NRHJ{} therefore needs no XLA runtime, but this comes at the cost of deriving
 the spectral derivative rules that automatic differentiation
 would otherwise generate automatically without human intervention. In exchange for this effort, the \NRHJ{}
 evaluator buys better control over the code path and optimization, deploys natively in a single
 language, and exposes its error envelope. Among
 surrogates, to the best of our knowledge, only the remnant surrogate model SurfinBH~\cite{Varma:2018aht} emits
 a per-evaluation uncertainty.

\begin{table*}[t]
  \centering
  \caption{Comparison of \NRHJ{} with other differentiable and GPU-capable waveform models.
  The column ``Differentiable'' depicts how the derivatives are obtained: analytic rules or
  automatic differentiation. ``GPU'' depicts the GPU backend capability.
  FD denotes frequency domain, and IFT the implicit-function theorem.
  ``Error-aware'' states whether the model returns an uncertainty with each evaluation.
  \NRHJ{} is the only model in the table that combines an NR surrogate, analytic derivatives
  computed on the GPU, PN hybridization, and a propagated predictive uncertainty.}
  \label{tab:landscape}
  \begin{tabular}{lllll}
    \toprule
    Code & Model class & Differentiable & GPU & Error-aware \\
    \midrule
    \textbf{\NRHJ} (this work)
      & NR Sur ($h$-adaptive spectral, adiabatic-ang)
      & analytic (Cheb./GPR $+$ IFT) & \textsc{Kokkos} & yes (GPR envelope) \\
    JaxNRSur~\cite{JaxNRSur}
      & NR Sur (NRHybSur3dq8/7dq4)
      & automatic (JAX) & XLA & no \\
    gwsurrogate / NRSur~\cite{Field:2013cfa,Varma:2019csw,Field:2025isp}
      & NR Sur
      & no & no & offline only \\
    ripple / gwfast~\cite{Edwards:2023sak,Iacovelli:2022mbg}
      & closed-form FD (Phenom, TaylorF2)
      & automatic (JAX) & XLA & no \\
    surfinBH~\cite{Varma:2018aht}
      & remnant surr. (GPR)
      & no & no & $1\sigma$, not diff. \\
    BBHx / \textsc{few}~\cite{Katz:2021uax,Katz:2021yft,Chapman-Bird:2025xtd}
      & closed-form FD / interpolation
      & no & CUDA & no \\
    \bottomrule
  \end{tabular}
\end{table*}
 \section{The Training set}
\label{sec:trainingset}

The training set~\cite{NRHJSurTrainingSet} consists of $\MethATrainingSimCount$ simulations drawn from the aligned-spin, non-eccentric, quasi-circular simulations of the SXS
catalog~\cite{Mroue:2013xna,Boyle:2019kee,Scheel:2025jct,SXSCatalogData:v3.0.0}, one per parameter point (the selection is described below), including simulations first presented in Refs.~\cite{Blackman:2015pia,Kumar:2015tha,Chu:2015kft,Blackman:2017dfb,LIGOScientific:2016ebw,Varma:2018sqd,Varma:2018mmi,Varma:2019csw}. The simulations the produced the catalog were run with the
Spectral Einstein Code~\cite{SpECwebsite} (methods in
Refs.~\cite{Lindblom:2005qh,Szilagyi:2009qz,Scheel:2006gg,Hemberger:2012jz,Pfeiffer:2002wt,Lovelace:2008tw,Pfeiffer:2007yz,Mroue:2010re,Scheel:2014ina}),
and their waveforms are extrapolated to future null
infinity~\cite{Boyle:2009vi}, and obtained with the \texttt{sxs} package~\cite{sxspackage}. We write the component masses as $m_1\ge m_2$, with the total mass $M$, mass ratio $q$,
symmetric mass ratio $\nu$ and mass difference $\delta$ defined by
\begin{align}
  M &= m_1+m_2, \qquad q = \frac{m_1}{m_2}, \nonumber\\
  \nu &= \frac{m_1 m_2}{M^2} = \frac{q}{(1+q)^2}, \qquad
  \delta = \frac{m_1-m_2}{M} = \frac{q-1}{q+1}.
  \label{eq:massparams}
\end{align} We choose non-deprecated binary-black-hole simulations with
in-plane spins below $\MethAInPlaneSpinTol$, reference eccentricity below $\MethAEccentricityTol$,
mass-ratio $q\le\MethAMassRatioCut$, and at least $\MethAMinOrbitsCut$ orbits after the reference time, which leaves
$\MethACandidateSimCount$ simulations at $\MethADistinctPointCount$ distinct parameter points. One simulation per
point is kept, the longest one, unless its NR resolution error (the mismatch between its two highest
resolution levels) is large compared with the model's LOO error at that point, in which case the next-longest is
taken. Each such replacement lowered the LOO error at its point. \MethAReplacementCount{} such replacements were
chosen at $q=\MethAReplacementMassRatioHigh$ and at $q=\MethAReplacementMassRatioLow$. These simulations span $q=\MethAMassRatioMin$ to $\MethAMassRatioMax$,
spins $\MethASpinMin$ to $\MethASpinMax$, and $\MethAOrbitsMin$ to $\MethAOrbitsMax$ orbits
(median $\MethAOrbitsMedian$), counted from the reference time onward with the relaxed masses
and spins as coordinates. The odd-$m$ PN-ratio radiative variables of Sec.~\ref{sec:radiative}
are trained on $\MethAOddRatioTrainingCount$ of them. The $\MethAExchangeSymmetricCount$ equal-mass, equal-spin simulations are excluded because the
odd-$m$ modes and their PN normalizers vanish together there. As those carry
the catalog's highest spins, the high-spin edge of the odd-$m$ ratio domain ends at $|\chi|=\MethAOddRatioSpinEdge$
while the even-$m$ edge is at $\MethAEvenRatioSpinEdge$.

\section{The adiabatic-angle parameterization}
\label{sec:conditioning}

\subsection{Hamilton's equation, adiabatic invariance, and the
first law}
\label{sec:action}
\label{sec:hjmotivation}

This subsection describes the conservative dynamics architecture of the numerical
relativity surrogate family. The azimuthal action $\Jphi(\Phi)$,
determined empirically from each NR simulation, is one of the quantities
the surrogate models and uses.

\paragraph{Action-angle variables.}
For a system with $N$ degrees of freedom possessing $N$ independent constants
of motion in involution, the Hamilton--Jacobi equation separates and
Hamilton's characteristic function generates a canonical transformation to momenta that are constants of motion.
For multiply periodic motion, the canonical choice is the
set of \emph{action} variables
\begin{equation}
  J_i = \frac{1}{2\pi}\oint p_i\,\mathrm{d}q_i ,
  \label{eq:actiondef}
\end{equation}
with $H=H(J_i)$ and the conjugate \emph{angles} advancing linearly,
\begin{equation}
  \dot\theta_i = \omega_i(J) = \frac{\partial H}{\partial J_i} .
  \label{eq:aafreq}
\end{equation}
The two-body problem is exactly integrable in the test-mass limit, where the
Kerr actions and frequencies are the standard basis for extreme-mass-ratio
waveforms~\cite{Carter:1968rr,Schmidt:2002qk}. At comparable masses, the
conservative PN dynamics is integrable order by order, with the
action-angle reduction explicit at 2PN~\cite{Damour:1988mr},
3PN~\cite{Damour:1999cr} and 4PN~\cite{Damour:2014jta}. The resulting
Delaunay Hamiltonian $\widehat H(J_r,\Jphi)$ is the analytic counterpart of
the function we model, restricted to circular ($J_r=0$) orbits. The actions are
the adiabatic invariants of the dynamics, and we build the surrogate model on these
features (called the "skeleton" from here on).

\paragraph{The fast/slow decomposition as the action-angle decomposition.}
The conditioning of Eq.~\eqref{eq:reconstruct}
(Sec.~\ref{sec:conditioning}) is the effect of Eq.~\eqref{eq:aafreq} applied to the
radiation field. The field of a multiply periodic orbit decomposes into
harmonics of the angles with coefficients that are functions of the actions
alone,
\begin{equation}
  h=\sum_{n}A_{n}(J)\,e^{-\mathrm{i}n\theta},
  \label{eq:angleharmonics}
\end{equation}
and for a circular orbit
$\Phi$ is the angle conjugate to $\Jphi$ up to an additive constant.
Factoring out the fast phase therefore leaves functions of the actions, which
vary only on the radiation-reaction timescale. The co-orbital variables, the
clock $\tau(\Phi)$ and the action can all be carried on one shared spectral element
set expanded in one variable $\Phi$.

\paragraph{Hamilton's equation for the angle and the first law.}
Hamilton's equation for the orbital angle $\phi$ is
\begin{equation}
  \omega = \frac{\partial H}{\partial \Jphi},
  \label{eq:hamiltonangle}
\end{equation}
and the first law of binary mechanics follows,
\begin{gather}
  \label{eq:firstlaw}
  \mathrm{d}M=\omega\,\mathrm{d}\Jphi
    +\Omega_1\,\mathrm{d}S_1+\Omega_2\,\mathrm{d}S_2,\\
  \mathrm{d}E=\omega\,\mathrm{d}\Jphi,
\end{gather}
where $\Omega_a$ and $S_a$ are the horizon angular velocities and spins of the two black holes.
At fixed masses and aligned spins, there is one relevant
action $\Jphi$. Once the on-shell Hamiltonian is identified with the mass term of the first law of
thermodynamics, Hamilton's equation for its angle, Eq.~\eqref{eq:hamiltonangle},
becomes the first law of binary mechanics Eq.~\eqref{eq:firstlaw},~\cite{LeTiec:2013iux,LeTiec:2015kgg,Fujita:2016igj}.
Here we assume that the mass-variation terms $\sum_a z_a\,\delta m_a$ (redshifts $z_a$ and mass variations $\delta m_a$ of the two bodies)
of the generalized first law of binary dynamics~\cite{Iyer:1994ys,Friedman:2001pf,LeTiec:2011ab} vanish approximately.

\subsection{Fast angle and co-orbital slow surrogated variables}

Let $h_{\ell m}(t)$ be the spin-weighted spherical-harmonic modes of an NR
waveform, devoid of junk radiation by starting at the metadata reference time $t_0$.
We take the orbital phase from the dominant mode,
\begin{align}
  \phi_{22}(t) &= \operatorname{unwrap}\arg h_{22}(t), \\
  \Phi(t) &= \tfrac{1}{2}\,\bigl|\phi_{22}(t) - \phi_{22}(t_0)\bigr| ,
  \label{eq:orbphase}
\end{align}
$\Phi$ is the (monotonically increasing) orbital \emph{angle}, the fast
coordinate conjugate to the azimuthal angular momentum of the system \Jphi (the
action coordinate of the adiabatic inspiral). Each mode is then demodulated into
a co-orbital surrogated variable,
\begin{equation}
  c_{\ell m}(\Phi) = h_{\ell m}\,e^{-\mathrm{i}\,m\,s\,\Phi},
  \label{eq:coorbital}
\end{equation}
which removes the fast orbital oscillation and leaves a complex function that
varies only on the slow radiation-reaction timescale. The sign $s$ of the $(2,2)$
phase evolution is recorded once. We find it to be $\MethAPhaseSign$ for every simulation
of the training set, consistent with the conventions of
Refs.~\cite{Boyle:2019kee,Scheel:2025jct}. The co-orbital decomposition is also used by the NRSur models~\cite{Blackman:2017dfb,Varma:2019csw}.
Additionally, we divide the radiative modes by their PN amplitudes, at 3.5PN order and at 4PN for the $(2,2)$
(the PN normalizers, Sec.~\ref{sec:radiative}), and model only the ratio. This makes the surrogated variables smoother.

Alongside the mode variables, we
carry four slow functions of $\Phi$. The first is the "clock", the time-of-phase
relation $\tau(\Phi)$, with time measured from the amplitude peak in units of
the total mass $M$. The second is the flux-integrated azimuthal action
$\Jphi(\Phi)$, whose profile is derived from the Bondi flux integral using the NR data,
with only its integration constant fixed by 4PN expressions. The 4PN reference is chosen
over the Arnowitt--Deser--Misner values to ensure that the same variable
serves the hybridization join (Sec.~\ref{sec:hybridization}). It is also the main
fitted quantity of the dynamics (Sec.~\ref{sec:action}). The third is the PN
frequency parameter $x(\Phi)=(M\omega)^{2/3}$. The clock supplies the
physical time axis at reconstruction. Being a phase derivative and therefore
noise-amplified, $x$ is not used as an abscissa for the surrogate. Instead, its use at evaluation time is
limited to the computation of velocity $v=\sqrt{x}$ at which the PN normalizers of the
PN ratio method used to construct the radiative modes (Sec.~\ref{sec:radiative}) are evaluated.
The fourth is the value of the conservative numerical Hamiltonian $E(\Phi)$ (normalized so that $E(\Phi_0)$ is the PN binding energy),
obtained in the same way as the action, from the radiated flux, and it enables the computation of Hamilton's equations numerically,
Its time derivative $\dot E$ is used in the first-law residual of Eq.~\eqref{eq:firstlawresidual}.
Alongside these, the model carries one further fitted quantity that is a single number for each binary system, and was critical for the successful spectral representation:
the coefficient of the leading non-adiabatic correction at the plunge. This is a parameter that we empirically discover in modeling the post-adiabatic dynamics,
which is also surrogated. Together, the co-orbital modes and these functions form the set of surrogated variables of the model.

\subsection{Peak alignment}
\label{sec:peak}

Waveforms are aligned at the peak of the root-sum-square of the modeled
positive-$m$ mode amplitudes. The negative-$m$ amplitudes are modeled using the
reflection symmetry, as these systems are non-precessing. The phase coordinate is transformed to
$\Phi\to\Phi-\Phi_{\rm peak}$ (zero at the peak, negative through inspiral), the
time becomes $\tau=t-t_{\rm peak}$, and a constant frame rotation is applied to the waveform, so
that $c_{22}$ is real and positive at the common end of the inspiral--merger
portion of the training domain. The residual $\pi$ ambiguity of this rotation symmetry, which
would flip the sign of odd-$m$ modes, is fixed geometrically, by comparing the
$(2,2)$ phase with the angle of the separation vector between the two horizons
in the publicly released catalog's trajectory data. Fixing it in this way makes the odd sector
antisymmetric under body exchange. All $\MethATrainingSimCount$ training
simulations were conditioned to these conventions. Where the horizon data are
missing or fail their quality check, the code falls back to fixing the sign
of the $(2,1)$ mode to the sign of its leading PN behavior.
With this alignment, the slow variables are smooth functions of
the intrinsic parameters at fixed $\Phi$, ensuring good conditioning
of the variables for the regressing across the BBH parameter space (cross-parameter regression)
of Sec.~\ref{sec:construction}.

\paragraph{Frame.}
The energy $E$ and the action $\Jphi$ are Bondi-like quantities, defined only relative to a
Bondi--Metzner--Sachs (BMS) frame~\cite{Bondi:1962px,Sachs:1962wk,Flanagan:2015pxa}.
On the restriction to circular orbits (where the radial action $J_r = 0$), they are free of the PN coordinate
choice~\cite{Damour:1999cr}. The radiative content of the waveform model is specified in the SXS catalog's frame
(post-junk, relaxed masses and spins, linear center-of-mass drift
removed~\cite{Woodford:2019tlo}). The correction removes a constant motion of the center of mass (CoM) of the system.
CoM motion beyond that constant drift,
e.g., the accumulating recoil from radiated linear momentum, residual
wobble of the coordinate center of mass due to boundary reflections,
and junk-induced supertranslation content, can survive the constant-drift correction.  The effects of such drifts have been understood in the literature~\cite{Woodford:2019tlo,Boyle:2015nqa,Mitman:2022kwt,Khairnar:2026mtm,Sun:2025quw,Ficarra:2026cej}.
A displaced origin mixes each mode with its $m\pm1$ neighbors. This would, for instance, leak the $(2,2)$ frequency into the smaller odd-$m$ modes.
In the co-orbital variable of the neighboring mode, say, $c_{21}=h_{21}e^{-\mathrm{i}s\Phi}$,
the leaked component therefore appears at exactly one relative cycle per orbit (as $m=2 -1 = 1$). This identifies the component as residual leakage: genuine co-orbital
content is secular, while leaked content lies at the integer harmonics of the orbital frequency.
BMS-frame-fixed waveforms~\cite{Mitman:2021xkq} are expected to remove the residual leakage.
The masses and spins, derived from the quasi-local horizon quantities, are assumed to be early enough in the inspiral
to be invariant under a choice of time coordinate, and thus approximately gauge invariant in the adapted tetrad frame~\cite{Dreyer:2002mx,Ashtekar:2004cn,Ashtekar:2021kqj}.

\subsection{The inspiral/merger--ringdown boundary}
\label{sec:boundary}

The action-angle description is only valid while the inspiral is adiabatic. We
make this boundary quantitative using two independent diagnostics that concur. The first is
the adiabaticity ratio $\epsilon=\dot\omega/\omega^2$, which is the dimensionless ratio
of the orbital-frequency evolution rate to the square of the orbital frequency.
It is a measure of the extent of variation of the frequency drifts within one orbit. Through the
inspiral, $\epsilon\ll1$, suggesting that the fast orbital angle is a good abscissa for surrogating
the variables, and the co-orbital variables vary only on the slow radiation-reaction timescale.
Approaching the merger, the frequency changes by order unity within a single cycle,
and the timescale separation collapses ($\epsilon=\dot\omega/\omega^2\to 1$). This is the epoch where the radiation reaction
timescale becomes comparable to the orbital timescale, and the action-angle representation breaks down.

This leads naturally to the second diagnostic, the action-variable diagnostic, derived from Hamilton's equations
for the conservative sector, i.e., the first law of binary mechanics discussed above, using which
a dimensionless first-law residual can be constructed:
\begin{equation}
  \Bigl| 1 - \omega\,\dot J / \dot E \Bigr| .
  \label{eq:firstlawresidual}
\end{equation}
This residual is found to be insensitive to where in the mode space ($\ell, m$) the mode sum is truncated.
In PN theory, there exists an approximate helical Killing vector~\cite{Detweiler:1993eq,Friedman:2001pf,LeTiec:2011ab,Rincon-Ramirez:2026tbo}.
The existence of such an approximate helical symmetry implies that each mode radiates energy and angular momentum in the same ratio $\omega$:
\begin{equation}
    \dfrac{\dot{E}}{\dot{L}} = \dfrac{\dot{E}_{\ell m}}{\dot{L}_{\ell m}} \approx \omega
\end{equation}
Therefore, a mode left out of the sum does not contribute to both the numerator and denominator of the expression above.
The residual of the first law shown above in Eq.~\eqref{eq:firstlawresidual} measures the rate at which radiation sources the unmodeled radial degree of
freedom, excited by the external radiation reaction forces in the post-adiabatic
inspiral, and thus the deviation from the approximate helical symmetry of the spacetime.
Through the inspiral, the median of the first-law residual over the training simulations is estimated to be $\sim\!\MethAFirstLawResidualInspiral$. From
this inspiral level to the amplitude peak, it grows by a factor of about $\MethAFirstLawGrowthToPeak$,
tracking the breakdown of adiabaticity
(measured and discussed in detail in a companion paper). We use the first law here as a critical diagnostic used to inform the surrogate
architecture, by demarcating the inspiral/merger--ringdown boundary, where the model switches its fitting
variable from the action-angle inspired orbital phase to the geometric time. The
angle-parameterized inspiral is then truncated near
the adiabatic edge. The merger--ringdown, where the fast angle is no longer a good abscissa,
is modeled separately in time (Sec.~\ref{sec:mr}). In the next para, we discuss more details on how the first law
along with the quantification of the radial action helps justify using the choice of the inspiral/merger--ringdown boundary.

\paragraph{Validity of a single action representation}
A generic bound aligned-spin orbit has two actions, $(J_r,\Jphi)$. The
training set from the SXS catalog is quasi-circular. As a result, the radial momentum, which vanishes on circular orbits,
grows toward merger ($|p_r|/\mu$ median $\MethARadialMomentumMedian$, max $\MethARadialMomentumMax$), with a
monotonic inward drift driven by radiation reaction instead of a periodic
radial oscillation as in the case of eccentric dynamics. The radial action integral accumulates only over closed
radial cycles, which is exclusively sourced by the residual eccentricity.

The effect of the residual eccentricity on the radial action can itself be
quantified, using the SXS metadata. For a nearly
circular orbit of orbital eccentricity $e$, the Delaunay actions give~\cite{Damour:1988mr,Damour:1999cr}
\begin{equation}
  \frac{J_r}{\Jphi}=\frac{1}{\sqrt{1-e^{2}}}-1\simeq \frac{e^{2}}{2},
  \label{eq:jrdelaunay}
\end{equation}
An eccentricity-selection tolerance of $e\le\MethAEccentricityTol$ therefore bounds the radial
action at $J_r/\Jphi<\MethARadialActionBound$.

The residual eccentricities measured
from the strain across the $\MethATrainingSimCount$ training simulations are
$\MethAResidualEccentricity$ (full range $\MethAResidualEccentricityMin$ to
$\MethAResidualEccentricityMax$), translating to a measured radial
action $J_r/\Jphi=\MethARadialActionRatio$, at most
$\MethARadialActionRatioMax$, a factor of \MethARadialActionBoundFactor{} under the bound.
This helps us validate the data quality of the chosen training set.

Toward merger, the circular orbit first law, Eq.~\eqref{eq:firstlaw}, fails as the
measured radial momenta depart from it, and the departure is the radial-work term
$W$ (see Table~\ref{tab:skeleton}):
\begin{equation}
  \frac{\mathrm{d}E}{\mathrm{d}\Phi}
    = \omega\,\frac{\mathrm{d}\Jphi}{\mathrm{d}\Phi} + W ,
  \label{eq:firstlawW}
\end{equation}
a term added to the circular first law that accounts for the radiative work done into the residual radial motion by radiation reaction.
The work term's $\epsilon^{2}$ power is fixed by the adiabatic scaling, and only its
scalar coefficient $k$ is fitted,
\begin{align}
  W =& k\,\epsilon^{2}\,\frac{\mathrm{d}E}{\mathrm{d}\Phi} ,
  \qquad\text{equivalently}, \\
  \omega\,\frac{\mathrm{d}\Jphi}{\mathrm{d}\Phi} &= E'\,(1-k\,\epsilon^{2}) ,
  \label{eq:plungework}
\end{align}
with a prime denoting $\mathrm{d}/\mathrm{d}\Phi$ ($E'=\mathrm{d}E/\mathrm{d}\Phi$) and $\epsilon=-\tau''/\tau'$ from the
spectral fit (developed in the companion paper). Integrating the circular
first law alone from the PN reference value $E(\Phi_0)$,
\begin{align}
  H_{\rm cons}(\Phi)&=E(\Phi_0)
    +\int_{\Phi_0}^{\Phi}\omega\,\frac{\mathrm{d}\Jphi}{\mathrm{d}\Phi'}\,
      \mathrm{d}\Phi' \\
    &= E(\Phi_0)+\int_{\Phi_0}^{\Phi}E'\,(1-k\,\epsilon^{2})\,\mathrm{d}\Phi' ,
  \label{eq:hcons}
\end{align}
and comparing against the measured $E(\Phi)$ gives median
$|H_{\rm cons}/E-1|=\MethAConsHamiltonianResidual$. This difference can be attributed to the integrated work along the radial degree of freedom.

This augmented first law is chosen to define the spectral representation boundary,
and the added term extends the validity of the first-law description deeper into the merger.
Without $W$, the first-law residual at the inspiral/merger--ringdown boundary $[\MethABlendStart,\MethABlendEnd]\,M$ is $\MethAFirstLawViolationBlendStart$--$\MethAFirstLawViolationBlendEnd$
times its inspiral level (catalog median).
On including it, the energetics remain consistent up to the
attachment of the time-parameterized merger--ringdown element. This is one of
the critical aspects of the model that allows the late inspiral to be
represented accurately, without which this hybrid design would not have been feasible.
The role of $W$ parallels two existing approaches in the literature. We note that it is close in spirit
to the next-to-quasi-circular corrections of
the effective-one-body models~\cite{Damour:2007xr,Damour:2009kr}, which
repair the quasi-circular \emph{waveform} near merger with NR-calibrated
coefficients, and to the post-adiabatic expansions of the two-timescale
treatment of inspirals~\cite{Hinderer:2008dm}, which systematically add corrections to
the adiabatic evolution in powers of the slow-evolution parameter. Here the
correction term acts only on the energetics of the first law, and helps determine
the boundary of the phase representation, which is critical to waveform modeling. The
waveform through merger is then continued by the separately fitted
set of merger--ringdown elements, represented in time.

\section{Surrogate numerical construction}
\label{sec:construction}

\subsection{The dynamics}
\label{sec:split}

The surrogate model makes use of post-Newtonian, and action-angle structure on various occasions,
but nevertheless, incurs no loss in generality by modeling the evolution provided by numerical relativity
left over after factoring out the known physics.

The division between fitted content and fixed structure, used throughout
the paper, is summarized in Table~\ref{tab:skeleton}.

\begin{table}[t]
  \centering
  \caption{The surrogated dynamics. Every function of the NR data is a
  surrogate variable (spectral expansion over shared elements, Gaussian-process
  regression over $(q,\chi)$).}
  \label{tab:skeleton}
  \begin{tabular}{p{0.30\linewidth}p{0.62\linewidth}}
    \toprule
    Role & Objects \\
    \midrule
    \textbf{Skeleton} (structure, never fitted) &
      The action-angle relation $\dot\Phi=\omega=\partial H/\partial\Jphi$,
      equivalently, at fixed masses, the first law
      $\mathrm{d}E=\omega\,\mathrm{d}\Jphi$ (one relation, not
      two~\cite{LeTiec:2015kgg,Fujita:2016igj}; Sec.~\ref{sec:hjmotivation}),
      with
      $\dot\Jphi=-\mathcal{F}$ supplied by the measured flux; the radial
      term $W=k\,\epsilon^2E'$, whose $\epsilon^2$ power is fixed by the
      adiabatic scaling and whose coefficient alone is fitted; PN mode
      structure (the normalizers of the radiative ratios). \\
    \textbf{Content} (fitted from NR) &
      $\Jphi(\Phi)$; the clock $\tau(\Phi)$; $E(\Phi)$, measured per
      simulation from the fluxes; the coefficient $k$; the complex
      radiative ratios $R_{\ell m}=c_{\ell m}^{\rm NR}/h_{\ell m}^{\rm PN}$. \\
    \bottomrule
  \end{tabular}
\end{table}

More details of the conservative dynamics will be presented in another paper.
In this paper, the \AAf{} model fits $\Jphi(\Phi)$, $E(\Phi)$, $\tau(\Phi)$
and the ratios of the NR modes to their PN normalizers, and exposes the dynamics $(J,E,\omega)$ alongside the strain.
In contrast, the \AdA{} model fits $\tau(\Phi)$ and the co-orbital modes directly.

\subsection{Shared spectral elements}

The slow variables that are smooth in $\Phi$ are represented on several ($h$)
spectral element decompositions shared across training waveforms. Domain
decomposition has been used before in surrogate construction, in parameter
space~\cite{Cerino:2022dhr} and more recently in time~\cite{Ravishankar:2026akf}.
Ref.~\cite{Cerino:2022dhr} divides the parameter space and builds a local reduced basis
in each part, whereas \textsc{NRSur7dq4v2} models the inspiral and the merger--ringdown on two
overlapping time intervals, improving the accuracy in both the domains. Here the phase axis is
divided into many elements shared by all training simulations, and the parameter dependence is a single
regression over the entire aligned-spin BBH parameter space. Within each
spectral element, the slow variables are represented by a $p$-order spectral
expansion in Chebyshev polynomials. In order to do this, a common global phase domain is first
identified across the available aligned-spin numerical simulations. This
amounts to discarding excess cycles from longer simulations that
satisfy the selection criteria, and is necessary for uniformity across the
parameter space. The other possibility is to hybridize the individual numerical
simulations to equalize the lengths, inheriting the full lengths of the simulations,
and will be explored in a future release.

Starting from the common phase domain across simulations, a spectral element is
accepted when, on that element, a Chebyshev expansion of order $p$ is sufficient to
simultaneously approximate multiple dynamical variables: the frequency variable $x$, the clock $\tau(\Phi)$,
and the co-orbital modes $c_{\ell m}$ of every training simulation. On an element $e$, each of these
variables $y\in\{x,\tau,c_{\ell m}\}$ of each simulation is expanded as
\begin{equation}
  y(\Phi) \approx \sum_{k=0}^{p} c^{(y,e)}_{k}\,T_k\bigl(\xi_e(\Phi)\bigr),
  \label{eq:chebexpand}
\end{equation}
where $T_k$ are the Chebyshev polynomials and $\xi_e$ maps the element onto $[-1,1]$. The
reconstruction residual of the variable on the element is the largest difference between the
expansion and the NR data, normalized by a global (not per element) scale $S_y$ of the variable,
\begin{align}
  r_{y,e} &= \frac{1}{S_y}\,\max_{\Phi\in e}\Bigl|\sum_{k=0}^{p} c^{(y,e)}_{k}\,T_k\bigl(\xi_e(\Phi)\bigr) - y_{\rm NR}(\Phi)\Bigr|,
  \label{eq:residual}\\
  S_{c_{\ell m}} &= \max_{\ell' m'}\max_{\Phi}|c_{\ell' m'}|,\quad
  S_x = \max_{\Phi}|x|,\quad
  S_\tau = \max_{\Phi}|\tau|,
  \nonumber
\end{align}
with the maxima in $S_y$ taken over the simulation's inspiral domain. The element is accepted when
\begin{align}
  \max_{\text{simulations}}\;\max_{y}\;\frac{r_{y,e}}{\varepsilon_y} &\le 1,
  \label{eq:accept}\\
  \varepsilon_{c_{\ell m}} = \varepsilon_x = \MethAModeWorkingTolerance,\qquad
  \varepsilon_\tau &= \MethAClockWorkingTolerance,
  \nonumber
\end{align}
where $\varepsilon_y$ is the \emph{working tolerance} of the variable. The coefficient magnitudes are
monitored as $p$ grows to $p_{\max}$ using a spectral decay indicator~\cite{Mavriplis:1994}.
The scale of every mode in Eq.~\eqref{eq:residual} is the maximum of the largest co-orbital mode,
in practice the $(2,2)$ mode. This means that a weak mode is required to match its residual to only a fixed
fraction of the $(2,2)$ mode and not of its own amplitude. This weights each mode by its amplitude relative to the dominant mode, i.e., by its
contribution to the observed strain.
This normalization choice ensures that subdominant modes, such as the odd-$m$ modes at equal mass, which
vanish, are not resolved beyond the observational requirements. The working tolerance of the mode variables,
$\MethAModeWorkingTolerance$, is set at the estimate of the relative noise floor
of the NR data from residual eccentricity and waveform-extraction noise; that of the clock
$\tau(\Phi)$ is $\MethAClockWorkingTolerance$, for the reason given below.

\begin{figure*}[t]
  \centering
  \includegraphics[width=0.98\linewidth]{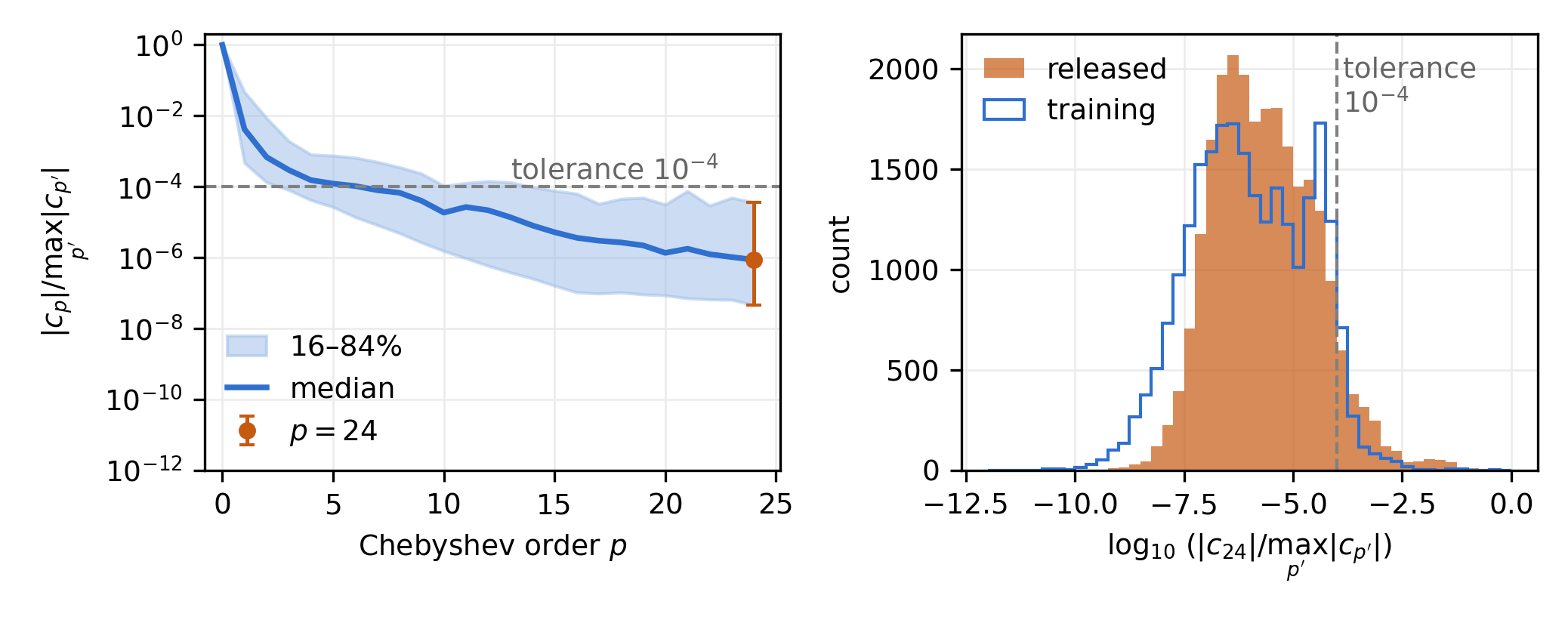}
  \caption{Spectral convergence of the shared Chebyshev
  representation of the \AAf{} model, restricted to the radiative sector (the co-orbital ratio
  modes $R_{\ell m}$ of the \AAf{} surrogated-variable set).
  The working tolerance of the mode variables, $\MethAModeWorkingTolerance$, is marked. \emph{Left:} Expansion
  coefficient magnitude versus Chebyshev order (median and $16$--$84\%$ band)
  over all $\MethASpectralRowCount$ (mode variable, element, waveform) coefficients of
  the conditioned training set, each normalized as in Eq.~\eqref{eq:truncation}. They cross the working tolerance near
  polynomial order $p\approx\MethAModeToleranceCrossingOrder$, and fall to $\sim\!\MethAModeCoeffMedianAtPmax$ by $p=\MethAInspiralOrder$, as
  they approach the NR noise floor.  \emph{Right:} the
  truncation error, $|c_{\MethAInspiralOrder}|/\max_{p'}|c_{p'}|$, of the \AAf{}
  model, with median
  $\MethAModeTruncReleasedMedian$. $\MethAModeTruncReleasedFracBelowTol\%$ of elements are below the working
  tolerance. The two panels are different
  populations, and their medians are not comparable: on the training side the
  same statistic has median $\MethAModeTruncTrainingMedian$ with $\MethAModeTruncTrainingFracBelowTol\%$ below the working
  tolerance.
  The gap between them is the cross-parameter regression of
  Sec.~\ref{sec:regression} (and not the truncation error).}
  \label{fig:spectral}
\end{figure*}

The \emph{truncation error} estimates the error of terminating the expansion
at the order $p$ an element retains. It is the magnitude of the last coefficient of
Eq.~\eqref{eq:chebexpand}, normalized to the largest coefficient of the same element,
\begin{equation}
  T_{y,e} = \frac{|c^{(y,e)}_{p}|}{\max_{k}|c^{(y,e)}_{k}|},
  \label{eq:truncation}
\end{equation}
which is the normalization chosen in Fig.~\ref{fig:spectral}. The \emph{NR noise floor} is the level below which the NR modes carry no
information, where the spectral decay flattens in the decay indicator. We measure it as the median of
$T_{y,e}$, taken over all modes, elements, and training simulations, at $p=\MethAInspiralOrder$. It lies
near $\MethANoiseFloor$, \MethANoiseFloorOrdersBelowTol{} orders below the working tolerance. Therefore, the tolerance is
conservative with respect to it.

When the numerical variables to be surrogated are expanded on the shared spectral elements, the
spectral decay indicator was found not to decay with the same slope to all 24 polynomial orders.
Thus, we arrive at an $h$ refinement rule for identifying the element boundaries.
If an element cannot meet its working tolerance within the initial fast-converging
regime, it is bisected, and the procedure repeats on each bisected element until every
element meets the working tolerance within that regime, or is narrower than
$\MethAMinElementWidth$ in the phase domain $\Phi$. The latter condition ensures a logical
termination point for spectral refinement and avoids fitting to the numerical noise floor.

The inspiral models use h-refinement with every element kept at
order $p=\MethAInspiralOrder$, which gives $\MethAInspiralElementCount$ shared elements for \AAf{}.
An adaptive shared $hp$ refinement technique was attempted, but we found that most elements need
up to 24 polynomials to achieve sufficient representation accuracy demanded by next-generation detectors.
After the fact, we therefore freeze the design to a fixed-order adaptive h-refined domain\footnote{As described later, it turns out that 24 p orders and h refinement is much more than what GPRs can achieve for this problem}.
The merger--ringdown
elements of Sec.~\ref{sec:mr} are placed with the same decay indicator, and are also
kept at $p=24$. Although the element structure is the same, each
surrogated variable demands its own working tolerance.
In particular, that of $\tau(\Phi)$ is $\MethAClockToleranceFactor$ times tighter than the mode variables',
because a time error $\delta\tau$ becomes a strain phase error $\sim m\,\omega\,\delta\tau/2$
on reconstruction, and relative-to-peak accuracy is not sufficient. Each waveform
then reduces to a set of Chebyshev coefficients on the shared elements.
The spectral convergence of this representation is shown in Fig.~\ref{fig:spectral},
measured on the mode variables, whose working tolerance is $\MethAModeWorkingTolerance$.
Aggregated over their (mode variable, element) conditioned coefficient vectors, each normalized
as in Eq.~\eqref{eq:truncation}, the median coefficient magnitude falls below the $\MethAModeWorkingTolerance$
working tolerance near order $p\approx\MethAModeToleranceCrossingOrder$ and falls to $\sim\!\MethAModeCoeffMedianAtPmax$ by $p=\MethAInspiralOrder$.
Its decay slows as it approaches the NR noise floor.

\begin{figure*}[tp]
  \centering
  \includegraphics[width=\linewidth]{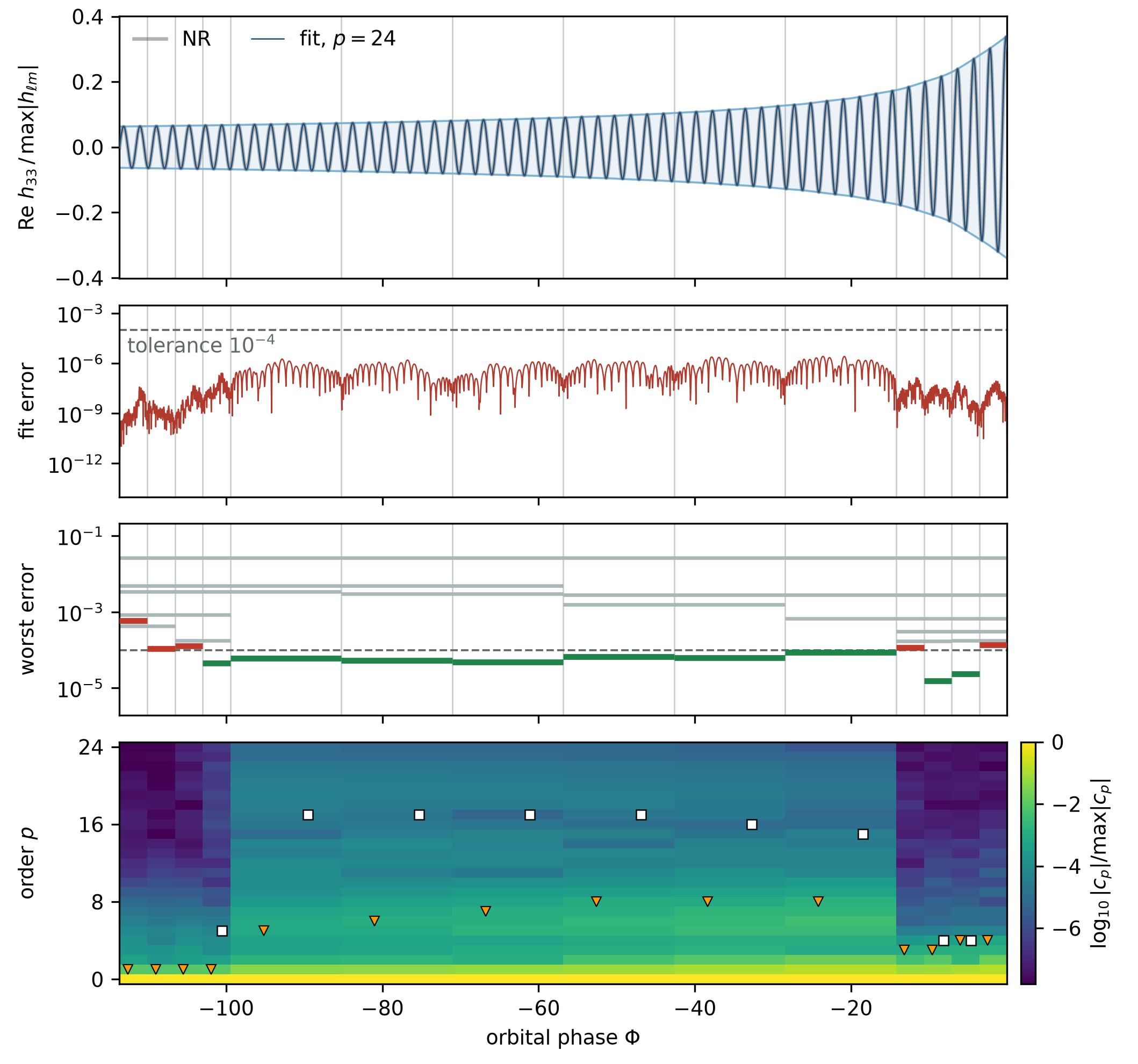}
  \caption{Converged mesh of the \AAf{} model for a representative $q=\MethAHpRepresentativeMassRatio$ system
  (SXS:BBH:2700), shown on the inspiral-merger portion of the $(3,3)$ mode. \emph{Top:} the real part of the mode plotted against the orbital
  phase $\Phi$ with the $\MethAInspiralElementCount$ element boundaries (vertical lines).
  \emph{Second:} the pointwise error of the order-$\MethAInspiralOrder$ fit, everywhere below the $\MethAModeWorkingTolerance$
  working tolerance. It is
  $\MethAHpRepFitError$, the median over the sampled phase with
  the 5th and 95th percentiles as the interval. Its largest single excursion is $\MethAHpRepFitErrorMax$.
  \emph{Third:} every element examined during the refinement, drawn as a horizontal segment over
  its phase interval at the height of its largest normalized residual, Eq.~\eqref{eq:residual}, over
  all variables and training simulations; the dashed line is the working tolerance. Grey elements
  were bisected, green ones met the working tolerance, and red ones stopped at the minimum width
  of $\MethAMinElementWidth$. \emph{Bottom:} the log coefficient magnitude versus order for
  each element, with the order at which this mode meets the working tolerance (triangles, $\MethAHpRepOrderMin$ to $\MethAHpRepOrderMax$
  with a median of $\MethAHpRepOrderMedian$) and the order from which the whole training set stays below it (squares,
  $\MethAHpCatalogOrderMin$ to $\MethAHpCatalogOrderMax$ with a median of $\MethAHpCatalogOrderMedian$). Every element keeps all $\MethAInspiralCoeffsPerElement$ coefficients.}
  \label{fig:hprefine}
\end{figure*}

On the training data side, the truncation error has a median of $\MethAModeTruncTrainingMedian$, with
$\MethAModeTruncTrainingFracBelowTol\%$ of elements below the working tolerance. For the coefficients returned by
the \AAf{} model at the same training points, the same statistic has a median of
$\MethAModeTruncReleasedMedian$, with $\MethAModeTruncReleasedFracBelowTol\%$ of elements below the working tolerance. The coefficients the model evaluates are
the regressed ones, not the ones the indicator measured, and the cross-parameter
regression of Sec.~\ref{sec:regression} adds error of its own on top of
the truncation error. Both the training and the GPR-predicted populations
are shown in Fig.~\ref{fig:spectral}.
Figure~\ref{fig:hprefine} shows, for the \AAf{} model, the converged mesh and the
spectral representation fit error for a representative $q=\MethAHpRepresentativeMassRatio$ system.
Both figures were measured on the \AAf{} model. The refinement,
the tolerance settings, and the decay indicator are shared by all the flavors of
the family.

\paragraph{Actions are the most compressible slow variables.}
An action's first-order response to slow variation averages to zero over the
fast angle. Therefore, actions should need the fewest degrees of freedom of any surrogated variable
carried through the inspiral. In concordance with this expectation, on all \MethATrainingSimCount{} simulations and the $\MethAInspiralElementCount$ shared elements
of the \AAf{} model, $\Jphi$ is found to have the smallest truncation error of the four
candidate surrogated variables: $\MethAActionTruncTraining$, against $\MethAClockTruncTraining$
for $\tau$, $1.1\times10^{-9}$ for $E$ and $5.9\times10^{-8}$ for $x$.

\subsection{Cross-parameter regression}
\label{sec:regression}

Every shared-element coefficient is regressed over the intrinsic parameters of the aligned-spin parameter space. For
the three-dimensional aligned-spin model, the training set is irregularly scattered in $(\log
q,\chi_{1z},\chi_{2z})$. Before the cross-parameter fit,
the parameters are mapped to the coordinates on which the GPR trend and the kernel act.
The coordinate chart for mass ratio is chosen to be
$\log q$. This choice makes the training density more uniform over
$q=\MethBTrainMassRatioMin$--$\MethBTrainMassRatioMax$, and body exchange ($q\to1/q$) becomes a sign flip of the
coordinate, making it easy to enforce expected physical symmetries in the parameter space (modeled by the respective spectral basis functions).
The charts for the spins are based on their symmetric and antisymmetric combinations,
\begin{equation}
  \chi_s=\tfrac12\left(\chi_{1z}+\chi_{2z}\right), \qquad
  \chi_a=\tfrac12\left(\chi_{1z}-\chi_{2z}\right).
  \label{eq:spincombos}
\end{equation}
Each coordinate is then brought to order unity
by a linear map. The coordinate $\chi_s$, which is unchanged under exchange, is
mapped onto $[-1,1]$ over its training range, while $\log q$ and $\chi_a$
are divided by their largest training magnitudes with no offset. This ensures that
the equal mass symmetry violation stays exactly at zero, and body exchange remains an exact
reflection of the coordinates, the property the exchange-symmetric
construction at the end of this subsection exercises.
A global, total-degree polynomial covering the entire parameter space, formed
from products of Chebyshev polynomials in each dimension, is
feature-limited~\cite{Boyd:2001}. More importantly, it \emph{oscillates where
the catalog is sparse at high degree}~\cite{Cohen:2013}.

We therefore regress with a Gaussian process~\cite{Rasmussen:2006}. A low-degree product-Chebyshev
trend captures the bulk parameter dependence, and a squared-exponential kernel
regressor fits the residual,
\begin{equation}
  C_j(\theta) = \underbrace{\textstyle\sum_a
    T_a(\theta)\,\beta_{ja}}_{\text{trend}}
  \; + \;
  \underbrace{\textstyle\sum_i k(\theta,\theta_i)\,\alpha_{ji}}_{\text{GPR
    residual}},
  \label{eq:gpr}
\end{equation}
Here $i$ labels the training simulation, and $j$ the
global coefficient index (across surrogated variables and across different elements
and polynomial orders). For each coefficient $j$ the kernel weights $\alpha_j$
are found by solving
the linear system $(K+\sigma^{2}I)\,\alpha_j = C_j - T\beta_j$, where $C_j$
is the vector of that coefficient's values at the training points (length equal
 to the training set), $T\beta_j$
is the fitted trend evaluated at the same points (its coefficients $\beta_j$
are fitted first, by least squares to $C_j$), $K_{ii'}=k(\theta_i,\theta_{i'})$
is the kernel matrix over pairs of training points, and $\sigma^{2}$ is the
nugget. The kernel matrix over the training points depends only on the hyperparameters,
and not on which coefficient is being fitted. It is factored once per hyperparameter choice
and shared by every coefficient using that choice. The odd-$m$ ratio fits exclude the
equal-mass, equal-spin configurations, where the odd-$m$ modes carry no
power. The merger--ringdown fits keep these configurations, because they fit the odd-$m$ modes
directly rather than as ratios. The exchange-antisymmetric kernel makes the fitted odd-$m$ modes
vanish at equal mass and equal spin, as they must.

Kernel hyperparameters are selected per surrogated variable by minimizing the closed-form LOO
prediction error of its coefficients over a grid of anisotropic length scales and nuggets. For a Gaussian-process
mean, the LOO errors of all training points follow from one factorization of the kernel
matrix, with no refitting~\cite{Rasmussen:2006,Sundararajan:2001}. Smooth variables such as $\tau$ and the $(2,2)$ mode
select long length scales and small nuggets, while the subdominant modes
select shorter scales and larger nuggets. This removes the oscillation of the high-degree polynomial fit at sparsely sampled
corners of the parameter space, such as $q=8$, $\chi=(-0.8,-0.8)$. As the Gaussian process carries a
posterior variance, it can also be used to obtain the calibrated error envelope of
Sec.~\ref{sec:gradients}.

\paragraph{Leave-one-out tests.}
\label{sec:loo}
Five kinds of LOO tests are used in this paper.
\begin{itemize}
\item The \emph{closed-form LOO} of one surrogated variable gives the prediction errors of its
  coefficients at all training points from one factorization of the kernel matrix, and selects the
  hyperparameters (above).
\item In a \emph{fixed-hyperparameter LOO} test, the inspiral and merger--ringdown regressions are refit on
  the remaining $\AlLooTrainingSetSize$ simulations with the kernel hyperparameters, the element mesh and the
  remnant map held at the model's values, and the held-out simulation is scored against the refit
  model.
\item In a \emph{full-retraining LOO} test, the inspiral hyperparameters are re-selected on the
  $\AlLooTrainingSetSize$ simulations as well, and the merger--ringdown model is attached unchanged.
\item The \emph{LOO error of a single fit} is the held-out prediction of one surrogated variable minus its value
  from the NR simulation, with that fit refit on the remaining simulations and its hyperparameters fixed. It sets
  the blend calibration scalars (Sec.~\ref{sec:radiative}) and tests the error forecast
  (Sec.~\ref{sec:gradients}).
\item The \emph{LOO of the remnant map} refits the map on the remaining simulations, with its hyperparameters
  fixed, and compares its prediction with the remnant mass and spin of the SXS metadata.
\end{itemize}
The hybrid dephasing test (Fig.~\ref{fig:hybdephasing}) and the mode-completeness test
(Fig.~\ref{fig:modecomplete}) use full-retraining LOO. Every other LOO mismatch in this paper is
fixed-hyperparameter LOO.

\paragraph{Exchange symmetry.}
\label{sec:symmetry}
Exchanging the two black holes is a physical symmetry of a nonprecessing
binary. Under it, $\log q$ and $\chi_a$ change sign, $\chi_s$ is unchanged,
and each mode acquires a factor $(-1)^m$. The even-$m$ modes are therefore
symmetric under exchange and the odd-$m$ modes, in particular $(2,1)$
and $(3,3)$, are antisymmetric. The odd-$m$ modes vanish only at
equal mass \emph{and} equal spin. At equal mass with unequal spins they stay
finite, driven by the spin asymmetry $\chi_a$. The fit enforces this
exactly. In the scaled coordinates the exchange $P$ is a reflection. In the
fit of an exchange-odd variable, the kernel is replaced by its antisymmetric
combination, and in the fit of an exchange-even variable by its symmetric
combination,
\begin{align}
  k_{\rm odd}(x,x')&=\tfrac{1}{2}\left[k(x,x')-k(x,Px')\right],\\
  k_{\rm even}(x,x')&=\tfrac{1}{2}\left[k(x,x')+k(x,Px')\right].
  \label{eq:paritykernels}
\end{align}
The trend is restricted to the
Chebyshev terms of the same exchange parity. For an odd-$m$ mode, the
reconstructed mode then carries the
$(-1)^m$ sign flip between mirror configurations exactly, vanishes exactly on
the equal-mass, equal-spin line at any spin magnitude, and its parameter
derivatives inherit the same parity, which matters for gradient-based
sampling. Both models \AAf{} and \AdA{} carry this parity-kernel construction in
every group of fitted variables.

\subsection{Radiative sector}
\label{sec:radiative}

In the \AAf{} flavor, each co-orbital mode is fitted twice: once directly as
$c_{\ell m}$, and once as its ratio to the PN mode amplitude. The two
fits are then combined in a \emph{blending} procedure. In the ratio, the known PN amplitude acts as a
normalizer. Therefore, only the departure from PN theory is fitted, which is slowly varying
and well conditioned.
We fit the complex NR-to-PN \emph{ratio}
\begin{equation}
  R_{\ell m}(\Phi;\theta) = \frac{c_{\ell m}^{\rm NR}}{h_{\ell m}^{\rm PN}}.
  \label{eq:ratio}
\end{equation}
The surrogated model thereby
inherits the symmetries and the boundary behavior of the PN radiative
amplitudes, which is expected to improve the ability of the surrogate model to predict data
at locations where the NR data is sparse and noisy. Large, sudden departures
from the PN amplitudes signify NR data quality issues, and were useful in excluding
some simulations from the training set. In each PN mode amplitude, the leading-order dependence on $v$, $\nu$ and $\delta$
factors out of the PN series~\cite{Borhanian:2019kxt,Caceres-Barbosa:2025lhr}. These factors guided the choice of coordinates
and the fitting routine of the \AdA{} model. The fitted ratio is close to unity early in the
inspiral (median departure below $\MethBRatioEarlyDeparturePct\%$ in the first orbit) and departs from it
smoothly as $v$ grows, spanning $\MethBRatioLateMin$--$\MethBRatioLateMax$ five orbits before the end of the
inspiral. The drift away from $R_{\ell m}=1$ is the accumulated systematic error
of the PN amplitude, which the fit models. In practice, we find that neither the PN-normalized ratio
fit nor the direct fit to the modes suffices. The ratio fit amplifies NR noise
where the PN amplitude is small, while the direct fit fails for the $(3,3)$ mode
at near-equal masses and high spins, where the PN structure of the normalizer
carries information the direct interpolation cannot recover. Therefore, we take a
hybrid approach in the \AAf{} model, in which the two fits are blended smoothly
with weights set by each fit's calibrated predictive GPR variance, so that each
mode, in each region of the parameter space, takes the more accurate fit. The
\AdA{} flavor uses only the direct fit.

For the $(5,5)$ mode,
the normalizer has its mass-asymmetry prefactor divided out before the ratio is
fitted. The ratio also tends to $1$ at low frequencies, which makes the
match to the PN extension (Sec.~\ref{sec:hybridization}) well conditioned
in the \AAf{} model. The extension
is matched to the blended mode content through the median of $R_{\ell m}$ over the first \MethBSeamRatioWindowEndOrbits{}
orbits inside the domain, which leaves the modes slightly discontinuous at the hybridization
boundary, by a median of $\MethBSeamStepMedianTwoTwo$ on $(2,2)$ and at most $\MethBSeamStepMedianHigherMax$ on the higher modes.
The action $\Jphi$ of the PN extension is set equal to the model's action \MethBActionAnchorOrbits{} orbits inside the domain. The ratio is fitted in $(\operatorname{Re},
\operatorname{Im})$ instead of in (amplitude, phase), to avoid coordinate artifacts at the zero
crossings of the amplitude. This is also mentioned and used in Ref.~\cite{Varma:2018mmi}.
An odd-$m$ amplitude has two competing sources: the mass asymmetry
$\delta$ and the antisymmetric spin combination $\Sigma=(m_2\chi_{2z}-m_1\chi_{1z})/M$.
In configurations where they oppose, the $(2,1)$ amplitude passes close to zero at
a finite frequency. At such a crossing $|R|$ has a kink and $\arg R$ jumps by
$\pi$, while $\operatorname{Re}\,R$ and $\operatorname{Im}\,R$ pass through zero smoothly.

Fitting the real and imaginary parts also preserves the exchange treatment of
Sec.~\ref{sec:symmetry}. Under body exchange, the odd-$m$ mode and its PN
normalizer both change sign, and their ratio is unchanged. The reconstructed mode then takes
both its $(-1)^m$ sign change and its zero at equal mass and equal spin from the PN prefactor,
and the fit does not need to learn either property. At equal mass and equal spin, the mode and
its normalizer both vanish and the ratio is undefined. Therefore, these configurations are
excluded from the odd-$m$ training set (Sec.~\ref{sec:regression}). The normalizers are the aligned-spin PN mode
amplitudes at a uniform $3.5$PN order in the non-spinning and the spinning
sectors alike~\cite{Henry:2022ccf}, with the spin sector cross-checked
against an independent derivation~\cite{Henry:2022dzx}.
One term is kept beyond the uniform 3.5 PN order normalizers. The $(2,2)$
amplitude is normalized with $4$PN~\cite{Blanchet:2023sbv},
which is the only $4$PN amplitude result available and is non-spinning by construction.
 The binding energy and the fluxes are transcribed from PostNewtonian.jl~\cite{PostNewtonianjl}, which collects the PN
results of Refs.~\cite{Blanchet:2013haa,Jaranowski:2013lca,Bini:2013zaa,Bohe:2012mr,Bohe:2015ana,Marsat:2013caa,Marsat:2014xea,Blanchet:2023sbv},
and agree with that package to a worst relative difference of $4.4\times10^{-16}$. The angular momentum is derived
from the energy through the first law, and reproduces the closed-form expression of Ref.~\cite{Blanchet:2024mnz} to
$4\times10^{-16}$, an external check.

\paragraph{The odd-even mode treatment}
The main difference between the \AdA{} and \AAf{} models is that \AdA{} fits
directly to the radiative variables without the PN normalizers. Fitting
the PN-normalized ratio  $R_{\ell m}$ has the distinctive advantage of being
able to handle the boundaries and corners of the domains where the amplitudes
of certain modes vanish, and to preserve their symmetries. The PN theory was decisive at various stages of the surrogate modeling
process, both in fitting the spectral coefficients using GPR, and in choosing the
coordinate system in which the surrogate is modeled. For the direct fits to the odd-$m$
modes (without the PN normalizers), and for all fits of \AdA{} and of the
merger--ringdown, the coordinate $\chi_s$ is replaced by the
PN reduced spin~\cite{Ajith:2011ec}
\begin{equation}
  \hat{\chi} = \chi_s + \frac{\delta\,\chi_a}{1 - \tfrac{76}{113}\nu},
  \label{eq:reduced_spin}
\end{equation}
which is the spin combination the PN phasing depends on at
leading order in the spin--orbit coupling, and this choice led to generally
better GPR fits. The other coordinates remain the same, i.e., these fits use
$(\log q,\hat{\chi},\chi_a)$. Like $\chi_s$, $\hat{\chi}$ is even under body
exchange, because $\delta$ and $\chi_a$ change sign together. The exchange
operator, the reflection property of the scaled coordinates the GPR is modeled
in, and the parity-kernel construction of Sec.~\ref{sec:symmetry} therefore all
carry over unchanged. In \AAf{}, the direct fit to the odd-$m$ modes uses $\hat{\chi}$, while the
odd-$m$ PN-normalized (ratio) fit and all even-$m$ fits use $\chi_s$ as described
in Sec.~\ref{sec:regression}.

\paragraph{The combination.}
For each mode, the co-orbital coefficient returned by the model is the inverse-variance
combination of the two fits,
\begin{equation}
  c_{\ell m} = \frac{c^{\rm D}_{\ell m}/v^{\rm D}_{\ell m}
                   + c^{\rm R}_{\ell m}/v^{\rm R}_{\ell m}}
                  {1/v^{\rm D}_{\ell m} + 1/v^{\rm R}_{\ell m}},
  \label{eq:blend}
\end{equation}
where $c^{\rm D}$ is the directly fitted coefficient, $c^{\rm R}$ is the
fitted ratio multiplied back by its PN normalizer, and
$v^{\rm D}$ and $v^{\rm R}$ are the two predictive variances brought to common
units. Each variance carries one calibration scalar per fit per mode, the median over the training
simulations of the ratio of that fit's LOO error to its predicted variance (value at least one), stored in the
model file. This is a standard recalibration of regression uncertainties~\cite{Kuleshov:2018,Levi:2022}, and it
makes the two regressions' error estimates comparable before they are combined.
Reconstruction is otherwise unchanged: the combined coefficient enters
Eq.~\eqref{eq:reconstruct} exactly as a single fit would. On the exchange-symmetric locus ($q=1$, $\chi_{1z}=\chi_{2z}$), the odd-$m$ modes and their PN normalizers vanish, and
Eq.~\eqref{eq:blend} becomes $0/0$. The model returns the exact limit there,
using the ratio-fit term $c^{\rm R}_{\ell m}=0$.

\paragraph{Contribution of the PN normalized ratios.}
The weights are strongly unequal, and in favor of the direct fit. At $q=4$,
$\chi=(0.3,-0.2)$, the direct fit carries $\MethBDirectWeightOtherMin$ or more of the weight on
\MethBNModesOtherThanTwoTwo{} of the \MethBNModes{} modes, and $\MethBDirectWeightTwoTwo$ on the $(2,2)$. Measured in-sample, on the model
waveform against NR over the $\MethBNTrainSims$ training simulations, over
the full inspiral--merger--ringdown (IMR) and with the same mismatch definition throughout,
removing the ratio fit entirely changes the median per-simulation all-mode
mismatch ratio by $\MethBRatioRouteAllModeGainPct\%$,
and the even modes do not change. The $(2,2)$ moves by a factor $\MethBRatioRouteFactorTwoTwo$, the
$(3,2)$ by $\MethBRatioRouteFactorThreeTwo$ and the $(4,4)$ by $\MethBRatioRouteFactorFourFour$. The one configuration class
where the ratio fit contributes is the $(3,3)$ at mass ratios below
$1.02$, where it improves that mode by a factor $\MethBRatioRouteNearEqualThreeThree$ over $\MethBNearEqualMassSims$ simulations,
while the $(5,5)$ mismatch becomes about $\MethBRatioRouteNearEqualFiveFiveWorsePct\%$ larger (a factor $\MethBRatioRouteNearEqualFiveFive$). The improvement in
accuracy from the ratio fit is therefore limited to near-equal masses, where the mass asymmetry $\delta$ approaches zero.

The PN-normalized ratio method makes the radiative sector inherit the
PN symmetries and boundary behavior exactly, and was used to diagnose
and improve the direct fit. It also provides a natural way to join the NR-informed
modes to the PN extension at the hybridization boundary
(Sec.~\ref{sec:hybridization}). In the bulk of the parameter space its
contribution to the accuracy of the \AAf{} model is small. It helps near equal
masses, where it improves the $(3,3)$ mode, and the faithfulness to NR in the
radiative sector comes mostly from the direct fit method. The blended
evaluation is more than twice as slow as the direct fit alone, and in fixed-hyperparameter LOO it is more
accurate than the direct fit on only $\AlBlendedBetterThanDirectLooPctTwoTwo\%$ of the training simulations for the $(2,2)$
mode and $\AlBlendedBetterThanDirectLooPctAllModes\%$ for all modes (Table~\ref{tab:blendloo}). We nevertheless blend
the two fits, for the robustness this gives and for the value of the ratio fit in
diagnostics and in predicting failures.

\subsection{Merger--ringdown elements}
\label{sec:mr}

Past the adiabatic edge, the waveform is modeled directly in time on a uniform
grid $\tau$ relative to the amplitude peak of Sec.~\ref{sec:peak}. This segment is split
once more, at $\tau_s=\MethBRingdownSplitTime\,M$, the first element boundary after the peak.
In the elements before $\tau_s$, the binary is still plunging. The mode frequencies
are still chirping, and they depend more on the progenitor
parameters through the ongoing
dynamics rather than on the parameters of the remnant.
Given the absence of PN modeling in the merger regime,
the aligned strain modes are used directly as the fitting variables
in this segment. After
$\tau_s$, the radiation is treated as ringdown, and the surrogate can again
be informed by knowledge from perturbation theory. We apply two operations to make the
fitting variables slowly varying.

First, the strain is transformed from the spin-weighted spherical basis into the
spin-weighted \emph{spheroidal} (SWSpH) basis, the angular eigenfunctions of the
perturbed Kerr remnant~\cite{Teukolsky:1973ha}. The spheroidal harmonic
${}_{-2}S_{\ell m}$ at spheroidicity $a\omega_{\ell m 0}$ expands on spherical
harmonics as
\begin{equation}
  {}_{-2}S_{\ell m} = \sum_{\ell'} C^{(\ell m)}_{\ell'}\,{}_{-2}Y_{\ell' m}.
  \label{eq:swsph_expand}
\end{equation}
The expansion couples only those spherical harmonics that share the azimuthal
index $m$. All modes are therefore transformed to the spheroidal basis, one block
of equal $m$ across the available $\ell$s at a time, with each block restricted to
the modeled modes. For $m=2$,
\begin{equation}
  \begin{pmatrix} h_{22} \\ h_{32} \end{pmatrix}
  =
  \begin{pmatrix}
    C^{(22)}_{2} & C^{(32)}_{2} \\
    C^{(22)}_{3} & C^{(32)}_{3}
  \end{pmatrix}
  \begin{pmatrix} A_{22} \\ A_{32} \end{pmatrix},
  \label{eq:swsph_m2}
\end{equation}
and the $m=3$ block has the same form with $\ell\in\{3,4\}$. The blocks with a
single modeled mode ($m=1,4,5$) reduce to a rescaling. The off-diagonal
coefficients have median magnitudes near $0.1$ over the training remnant spins and
rise to $\MethBMixingOffDiagMax$ for the most rapidly spinning remnants, and the blocks are well
conditioned (condition number at most $\MethBMixingCondMax$). The mode mixing due to the
spin-weighted spheroidal transform matters for the subdominant mode of each block.
Repeating the fixed-hyperparameter LOO study with the off-diagonal coefficients set to zero,
and every other setting held fixed, leaves the LOO mismatch statistics of
the dominant modes $(2,2)$ and $(3,3)$ unchanged, and degrades the LOO
mismatch of the $(3,2)$ by a median factor of $\MethBDiagMixingFactorThreeTwo$, while that of the $(4,3)$ is
unchanged ($\MethBDiagMixingFactorFourThree$).

Second, each spheroidal amplitude is detrended by the oscillatory part of the
fundamental quasinormal frequency of its own mode,
\begin{equation}
  \tilde A_{\ell m}(\tau) = A_{\ell m}(\tau)\,
    \exp\!\left(-\mathrm{i}\,s\,
      \bigl|\operatorname{Re}\omega_{\ell m 0}(\chi_f)\bigr|\,
      \tau/M_f\right),
  \label{eq:qnm_detrend}
\end{equation}
with $s$ the phase sign of Eq.~\eqref{eq:coorbital}. Only the
real part of the frequency is used. The ringdown mode phases are
detrended while the damping is left untouched, and so are the amplitudes. The
reason for this choice is that dividing out the exponential decay of every mode by
that of the fundamental mode would multiply the late-time data, where the raw
numerical modes are the weakest and noisy, by a factor of order $\MethBDampingAmplificationLow$--$\MethBDampingAmplificationHigh$,
amplifying the NR noise floor
instead of the signal. We use the magnitude of the real part of the
quasinormal frequency of the fundamental quasinormal mode to detrend the late ringdown
waveform, because with the retrograde continuation $\omega_{\ell m n}(-a) = -\omega^{*}_{\ell,-m,n}(a)$
the real part jumps from $+|\operatorname{Re}\omega_{\ell m 0}|$ to $-|\operatorname{Re}\omega_{\ell m 0}|$ when the
remnant spin $\chi_f$ changes sign, while its magnitude is continuous.
The residual $\tilde A_{\ell m}$ is the variable represented on the spectral elements, and
it is kept complex, as in the inspiral (Sec.~\ref{sec:radiative}). Before the detrend, the
phase of $A_{\ell m}$ advances at the quasinormal frequency, and its real and imaginary
parts oscillate through many cycles over the segment, which a smooth basis represents
poorly. After the detrend, they vary slowly and can be fitted directly. Over
the \MethBNModes{} modeled modes, the detrend reduces the median rate of phase change by
a factor of $\MethBDetrendFactorMin$ to $\MethBDetrendFactorMax$, where the median is taken over the $\MethBNTrainSims$ training
simulations for each mode. For the slowest mode, $(2,1)$, the median rate is
$\MethBDetrendRateTwoOneBefore$~rad$/M$ before the detrend and $\MethBDetrendRateTwoOneAfter$~rad$/M$ after it. For the fastest,
$(5,5)$, it falls from $\MethBDetrendRateFiveFiveBefore$ to $\MethBDetrendRateFiveFiveAfter$~rad$/M$. The detrended variables are therefore
slowly varying, and their spectral representation needs few coefficients.

Both the operations involving the spin-weighted spherical-spheroidal harmonic transform, and the detrending of phases by the fundamental mode require the remnant mass and spin, which are not known a~priori
at evaluation time. They are supplied by a smooth map
$\mathcal{R}:\theta\mapsto(M_f,\chi_f)$, built with the same trend and kernel construction
as the rest of the model and constrained to be invariant under body exchange,
because exchanging the labels of the two bodies does not change the remnant. The same map
is used both to prepare the training data and at evaluation. Therefore, any error in
$\mathcal{R}$ enters as a smooth function of $\theta$, which the coefficient regression
absorbs.

For the remnant fits, the map regresses the
radiated energy (in units of total initial mass of the system) divided by the symmetric mass ratio, instead of $M_f$ itself.
This combination is found to be much flatter across the mass-ratio range than the mass is.
The mass is then obtained from a fit to this radiated energy. The remnant spin is fitted directly.
Both use the trend and kernel machinery described above but not the same settings.
The map is smooth enough in the
intrinsic parameters to take a higher-degree trend: degree \MethBRemnantMassTrendDegree{} for the mass and
\MethBRemnantSpinTrendDegree{} for the spin, against degree \MethBWaveformTrendDegree{} for the waveform coefficients. The remnant is exchange-even. Both fits therefore use the symmetric kernel and the
exchange-even trend of Sec.~\ref{sec:symmetry}, and the fitted map has no
exchange-odd component to suppress.

LOO errors of the map (Sec.~\ref{sec:loo}) are $\MethBRemnantSpinLOORms$ rms in $\chi_f$ and $\MethBRemnantMassLOORms$ rms in $M_f$ (medians and percentiles of the
same LOO test are given in Appendix~\ref{sec:remnant}). The $M_f$ figure excludes SXS:BBH:1124, whose
catalog remnant metadata is an outlier, which is also excluded in the remnant map's training.
The quasinormal frequencies and mixing coefficients
are tabulated once on a grid in $\chi_f$ that straddles zero, with the
retrograde branch continued as above. The training set contains remnants of both
signs.

After applying the two operations above, both segments are then decomposed on shared elements and regressed over the
intrinsic parameters by the construction of Sec.~\ref{sec:regression}, now
with time as the fast coordinate adopted due to the simplicity of time domain
representation of the waveform in the late time perturbative ringdown regime. Every factor introduced above
($M(\chi_f)$, $\omega_{\ell m 0}(\chi_f)$, $M_f$, $s$) is invariant under body
exchange. Therefore, the residuals inherit the exchange parity of the strain modes
exactly and the odd-$m$ symmetry treatment of Sec.~\ref{sec:regression} carries
over unchanged.

The two merger--ringdown segments are not simply joined together at $\tau_s$. The plunge
fit is extended past the merger--ringdown split to $\MethBPlungeFitEnd\,M$ to obtain an
overlapping region with the ringdown element. In this overlapping region, the
reconstruction is a cosine ramp from the plunge variables to the ringdown variables, zero
at $\tau_s$ and one at $\MethBPlungeFitEnd\,M$. The ramp is chosen with vanishing derivative
at both ends, so that the reconstruction is continuously
differentiable across the junction.
\textsc{NRSur7dq4v2} joins its two time-domain surrogates in the same way, with
Planck-window blending masks over a $20\,M$ overlap~\cite{Ravishankar:2026akf}.

\subsection{Inspiral--merger--ringdown reconstruction}

To evaluate the model at a parameter $\theta$, the slow variables are predicted and
the inspiral strain is reconstructed as
\begin{equation}
  h_{\ell m}(\Phi) = c_{\ell m}(\Phi)\,\exp(\mathrm{i}\,m\,s\,\Phi),
  \label{eq:reconstruct}
\end{equation}
with $s$ the recorded phase sign, placed on a physical time axis through the
predicted time--phase map $\tau(\Phi)$. The merger--ringdown modes are evaluated
on their own time grid. The remnant map supplies $(M_f,\chi_f)$ at the query
parameters. The predicted detrended ringdown variables are rewound by the
inverse of
Eq.~\eqref{eq:qnm_detrend}, and are carried back to the spherical basis by
Eq.~\eqref{eq:swsph_m2} and its $m=3$ analogue. The plunge segment is
blended onto the ringdown over $\tau\in[\tau_s, \MethBPlungeFitEnd]\,M$, and the resulting merger-ringdown portion
is cosine-blended onto the inspiral over
$\tau\in[\MethBIMRBlendStart,\MethBIMRBlendEnd]\,M$. Both segments were conditioned in the frame fixed
by the peak alignment of Sec.~\ref{sec:conditioning}, in which $c_{22}$ is
real and positive at the common end of the inspiral--merger portion. Therefore, no further
rotation is applied when they are assembled, and they attach in the frame they
were built in, completing the IMR waveform. Both the
re-wind and the basis transformation are per-mode operations whose only
parameter dependence enters through $(M_f,\chi_f)$. This requires an
interpolation table, and one small matrix product per parameter point, independent of
the number of time samples. The cost of the whole operation is small, after some care is taken to optimize the code, and quantified in
Sec.~\ref{sec:performance}.
 \section{Action-native hybridization}
\label{sec:hybridization}
Numerical simulations are finite in number of cycles (the NR-native length), and evaluating low-mass binaries requires a larger number of cycles in order to extend to the lower frequency of the gravitational wave detectors.
The NR-native length of the model spans the shared $\sim\!\PerfNativeSpanOrbits$-orbit training window
(Sec.~\ref{sec:performance}), which is insufficient for low-mass systems.
For lower starting frequencies, the model carries a
\emph{call-time} PN extension of the NR-native portion of the model. This is triggered by a runtime parameter, on by default for starting frequencies below the model's native starting frequency.
An $f_{\rm low}$ above the native starting frequency trims the pure NR-informed reconstruction,
and only a request below it engages the extension. Hybridization is therefore
a property of the evaluation call, instead of hybridizing the training data first and then surrogating over the hybridized data. The latter method has its own advantages, but as an experimental surrogate construction using applications of new methods,  we first attempt hybridizing the surrogated model.

\begin{figure}[t]
  \centering
  \includegraphics[width=0.92\linewidth]{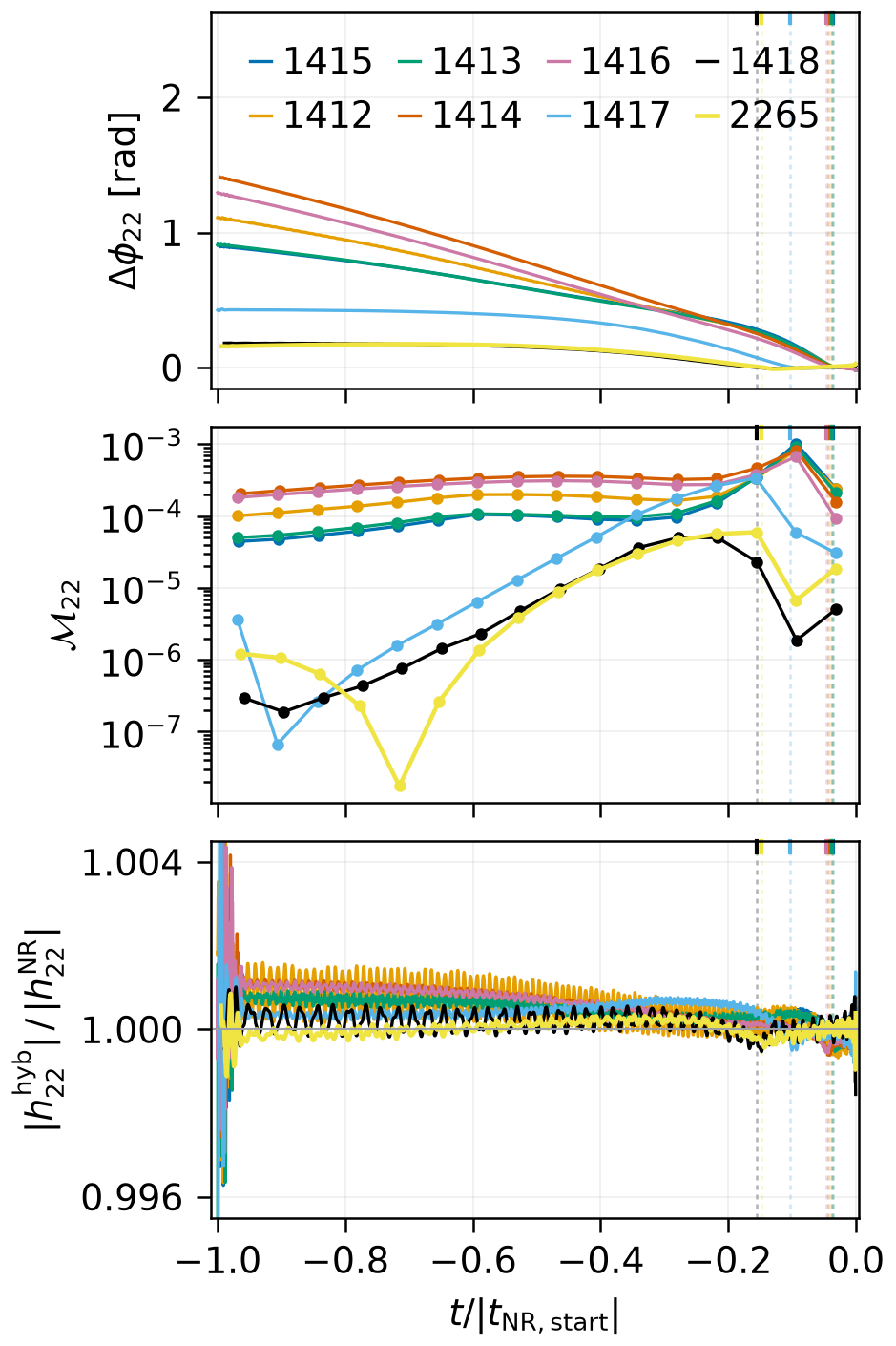}
  \caption{Full-retraining LOO test of the hybridization, extended backward
  in time from the model's NR native length along the PN extension, and scored against
  full-length NR simulation (the truth) over \HybLongSimCountWord{} long simulations. Each panel compares the
  hybrid against NR. Each simulation's time axis is normalized to
  its own full NR length, $x=t/|t_{\rm NR,start}|$, so that the full-length NR-span's start is at
  $-1$ and merger at $0$. The short colored tick at the top of each panel marks
  that simulation's PN-to-model join, i.e.\ the truncated NR-native span's boundary. The NR-native span covers only
  $\HybNativeFracLowPct$--$\HybNativeFracHighPct\%$ of each of these simulations. Most of each curve is
  the PN extension compared with NR.
  The hybridized extension spans $\HybExtOrbitsLow$ to $\HybExtOrbitsHigh$ orbits, depending on
  the simulation.
  \emph{Top:} accumulated $(2,2)$ phase difference, at most $\sim\!\HybExtDephasingMaxRad$~rad at
  the hybridization boundary, closing to zero by merger. That total $\HybExtDephasingLowRad$--$\HybExtDephasingHighRad$~rad is
  accumulated over the $\HybExtOrbitsLow$--$\HybExtOrbitsHigh$ extended orbits, i.e.\ $\HybDephasingPerOrbitLowMrad$--$\HybDephasingPerOrbitHighMrad$~mrad per
  orbit.
  \emph{Middle:} the running $(2,2)$
  mismatch against NR. \emph{Bottom:} the $(2,2)$ amplitude ratio between the hybridized and pure NR waveforms, which is flat to better
  than $\HybAmpRatioDeviationBound$ across the whole span.}
  \label{fig:hybdephasing}
\end{figure}

\paragraph{Hybridization strategy.}
The mechanism for low-frequency extension follows the same action-space description as the model.
Below the native shared NR data span, $\Jphi$ continues along the (uncalibrated) 4PN
trajectory, positioned by \emph{value} continuity. The PN extension is attached (the matching point) by
matching $L_{\rm 4PN}(x_m)$ to the fitted $J(\Phi_m)$. The choice of attachment was selected
by measurement against the alternative choices. Attaching on the endpoint derivative $1/\tau'$,
where the spectral derivative rings, initiated the PN extension of the waveform with a $\sim\!2\times10^{-3}$
frequency error, that the adiabatic factor $\mathrm{d}\Phi/\mathrm{d}\ln
x\sim\HybAdiabaticFactor$ amplified into a $\sim\!5$-radian secular dephasing of the
$(2,2)$ mode over $\sim\!\HybEndpointAddedOrbits$ added orbits, against all three of SpinTaylorT4,
\textsc{NRHybSur3dq8} and \textsc{IMRPhenomXPHM}. Matching instead the value of
$J$ four orbits inside the NR span removes this drift. Whether bare 4PN
suffices at the catalog's attachment frequencies is decided by comparison
against NR, as follows.
We artificially cut out early NR orbits and regenerate them with the hybridization. The PN extension then differs from NR by at most $\HybHoldbackMaxDephasingRad$~rad in $(2,2)$ phase over $\HybHoldbackOrbits$ orbits from $x=\HybHoldbackStartX$, and the difference falls steeply toward
lower attachment frequencies. Fig.~\ref{fig:hybdephasing} shows a related test, with the full-retraining LOO model
(Sec.~\ref{sec:loo}) in place of the simulation's own NR data, over the full NR length of \HybLongSimCountWord{} long simulations, where the extension covers $\HybExtOrbitsLow$ to
$\HybExtOrbitsHigh$ orbits below the NR native boundary. The dephasing accumulates to $\HybExtDephasingLowRad$ to $\HybExtDephasingHighRad$~rad over the
$\HybExtOrbitsLow$ to $\HybExtOrbitsHigh$ extended orbits, i.e., $\HybDephasingPerOrbitLowMrad$ to
$\HybDephasingPerOrbitHighMrad$~mrad per orbit. For comparison, over the same extended orbits the two highest resolutions of these simulations
(Lev2 and Lev3) dephase by $\LevDephasingLongFiveLowRad$ to $\LevDephasingLongFiveHighRad$~rad
($\LevDephasingLongFivePerOrbitLowMrad$ to $\LevDephasingLongFivePerOrbitHighMrad$~mrad per orbit) for the five
near-equal-mass runs, more than the hybrid does against NR, and by $\LevDephasingHighQLowRad$ to
$\LevDephasingHighQHighRad$~rad ($\LevDephasingHighQPerOrbitLowMrad$ to $\LevDephasingHighQPerOrbitHighMrad$~mrad
per orbit) for the three runs at $q=3$ to $4$, less than the hybrid. The next-lower resolution pair
(Lev1 and Lev2) does not consistently bound these differences. Therefore, they indicate the size of the
NR resolution error without being a converged estimate of it. NR's resolution error is a phase difference that accumulates with time, and the mismatch between
two resolutions grows as the square of the compared length~\cite{Mitman:2025tmj}.

From the extended action, the frequency, energy, and clock are derived from
Hamilton's equations and the first law, and a single quadrature supplies the phasing
over the whole extended domain. The radiative sector requires no fitted matching because of the PN normalizers.
Below the boundary, the deviation $R_{\ell m}-1$ decays to zero with a per-mode power law in $v$
whose exponent we measure from its in-band decay. The modes therefore reduce to the PN modes in the
deep low-frequency extension, where PN is asymptotically exact.

Splicing in the action variable $\Jphi$ also has other advantages. The conventional
construction of a hybrid waveform~\cite{MacDonald:2011ne,Varma:2018mmi} works on the
waveform data directly. The PN and
NR descriptions overlap over a
stretch of the early NR orbits, where a matching window is chosen. A time
shift and a phase offset are fitted, ensuring the two descriptions agree over the
window, and the two waveforms are then blended across it with a smooth
weight, applied to the waveform modes themselves~\cite{MacDonald:2011ne} or to each mode's amplitude and
frequency~\cite{Varma:2018mmi}. The window necessarily lies inside the NR span. Therefore, the
first NR orbits of the product are a mixture of NR and PN, and whatever
disagreement is left behind in the alignment is inherited by the early NR orbits of
the product. For that construction, the resulting error is a mismatch
between the hybridized product and the original NR waveform over the NR span,
and in simulations that start with noise, it is up to $\HybSpliceMaxModeMismatch$ on subdominant modes.
The action-native method instead matches one slow, monotonic curve at one point.
Every hybridized quantity in the extension is computed from the stitched $J$,
allowing the continuity of the dynamics at the hybridization boundary to be
inherited, and the NR span is structurally untouched.

Comparing the two constructions on the same
simulation, on the roughest-edge test case (SXS:BBH:1413), the all-mode mismatch between the hybridized and the unhybridized NR
waveform at the hybridization boundary drops from $\HybSpliceAllModeMismatch$ with the conventional joining to
$\HybActionAllModeMismatch$ with the action-native stitch, the $(4,3)$ from
$\HybSpliceFourThreeMismatch$ to $\HybActionFourThreeMismatch$, and the $(2,1)$/$(5,5)$ improve
$\sim\!\HybSpliceGainTwoOneFiveFive\times$. On a clean long simulation (SXS:BBH:1415), the action-native
all-mode mismatch of the waveform at the hybridization boundary is $\HybActionCleanAllModeMismatch$. Waveforms started at different $f_{\rm low}$ agree to a median all-mode mismatch
of order $\HybMultiStartMismatchOrder$.

\paragraph{The conservative dynamics, and the first law as a test.}
A hybridized \AAf{} waveform also carries hybridized \emph{dynamics}: $J$, $E$
and $\omega$ over the full extended domain, exposed by the same
\texttt{dynamics} flag as the NR-native call to the waveform model generator.
Hybridization is currently a capability restricted to \AAf{} variant. The \AdA{} model currently carries no PN extension and
returns the native span only. The PN extension derives
from Hamilton's equations, whose circular-orbit form is the first law, and the approximate helical symmetry holds there. The
residual of the first law at the hybridization boundary therefore serves as an acceptance test, and it is
run on every hybridized waveform at the training points. With the measured (flux-integrated) $E$ describing the NR-span of the waveform,
the residual of the first law identity, $|(\mathrm{d}E/\mathrm{d}J)/\omega-1|$,
tests the join at the hybridization boundary, quantifying the deviation from the helical symmetry in the early inspiral.
The test yields a median $\HybFirstLawNRSideMedian$ on the NR
side, inside the $\HybFirstLawBandPPM$ parts-per-million error band of the per-NR simulation first-law
measurements (Sec.~\ref{sec:action}), and $\HybFirstLawPNSideMedian$ on the analytic PN extension side.
 \section{Validation}
\label{sec:aligned}

This section reports the measured accuracy of the aligned-spin
model. The regression and the binary body exchange-symmetric fit behind it are described
in Sec.~\ref{sec:regression}.

\subsection{Exchange symmetry}
\label{sec:symmetry_measured}

In $u=\log q$, body exchange is the parity
$(u,\chi_a)\rightarrow(-u,-\chi_a)$, and at $q=1$ with equal spins it requires
the odd-$m$ modes and $\partial_u h_{\ell m}$ and $\partial_{\chi_a}h_{\ell m}$
to vanish. Both \NRAAS{} models obey this parity by construction
(Sec.~\ref{sec:symmetry}). Any departures from it come from
floating-point roundoff in the inputs. Evaluating the \AAf{} model at $(q,\chi_{1z},\chi_{2z})$ and at the mirror
configuration $(1/q,\chi_{2z},\chi_{1z})$ gives the same modes to roundoff
($\lesssim10^{-15}$) at $q=1$, with the odd-$m$ modes changing sign. At other
mass ratios, the exchanged coordinates should be exact negatives,
\begin{equation}
  \log\!\left(\tfrac{1}{q}\right) = -\log q, \qquad
  \delta\!\left(\tfrac{1}{q}\right) = \frac{1/q-1}{1/q+1} = -\frac{q-1}{q+1} = -\delta(q),
  \label{eq:mirrorcoords}
\end{equation}
with $\delta$ the mass difference of Eq.~\eqref{eq:massparams}, which enters the fits through $\hat\chi$.
In floating-point arithmetic the two sides of Eq.~\eqref{eq:mirrorcoords} differ in the last digit, and the
fit amplifies this difference according to its conditioning. The resulting relative amplitude
deviations are below $\DiscSwapInteriorMaxDev$ over the interior, and rise to $\DiscSwapWorstDev$ on the
$(3,3)$ at $q=1.5$, $\chi=(+0.85,-0.85)$, near the edge of the odd-$m$ training
domain, where they depend on the machine and the linear-algebra implementation. The
largest such deviation corresponds to a mismatch of $\DiscSwapMaxMismatch$, five orders of magnitude
below the model's fixed-hyperparameter LOO mismatch.
The models we compared against do not obey this parity.
Under the equal-spin exchange symmetry test at $q=1$, \textsc{NRHybSur3dq8} leaves an odd-$m$
residual peak amplitude of $\DiscOddResidualHybLow$ to $\DiscOddResidualHybHigh$ times the peak amplitude of
the $(2,2)$ mode, and \textsc{NRSur7dq4} of $\DiscOddResidualSurSevenLow$ to $\DiscOddResidualSurSevenHigh$
times the $(2,2)$ peak. Swapping
$\chi_{1z}\leftrightarrow\chi_{2z}$ at $q=1$, for $(\chi_{1z},\chi_{2z})=(0.6,-0.2)$, $(0,-0.6)$ and $(0.5,-0.5)$, changes the peak
amplitudes of the $(3,2)$ and $(4,4)$ modes by at most $\DiscSwapThreeTwoHyb$ and $\DiscSwapFourFourHyb$ of the peak for
\textsc{NRHybSur3dq8}, and by $\DiscSwapThreeTwoSurSeven$ and $\DiscSwapFourFourSurSeven$ for
\textsc{NRSur7dq4}. \textsc{NRSur7dq4} (a
precessing model evaluated in the aligned limit, where part of its residual may
come from the model's machinery) is the least symmetric of the three in the
higher modes, by up to $\DiscSwapThreeThreeSurSeven$ in $(3,3)$.

Exchange symmetry also constrains derivatives: $|h_{22}|$ is even in $u=\ln q$, and its derivative must
therefore vanish at $q=1$. We test this with a one-sided finite difference of the peak $(2,2)$ amplitude
in $u$ at zero spins, with steps from $0.04$ down to $0.005$. For \textsc{NRSur7dq4} the derivative values tend to
$\DiscDuHTwoTwoSurSeven$ instead of zero, i.e., the peak amplitude of \textsc{NRSur7dq4} has a kink at equal mass. For \textsc{NRHybSur3dq8} the derivative values decrease
with the step and change sign, from $\DiscDuHTwoTwoHybLargestStep$ to $\DiscDuHTwoTwoHybSmallestStep$, as a
vanishing derivative does under a one-sided difference.

If no symmetries are demanded of the surrogate model during construction, the models inherit the violations of the training data set. The training set's own departure from exchange
symmetry can be measured directly, without involving any model, because the
sign convention of Sec.~\ref{sec:peak}, fixed from the horizon trajectories, places mirror configurations on
opposite branches by construction. For the NR data, on the $\AlNRExchangeFixedSimCount$
configurations that are exchange invariant ($q=1$, equal spins), every odd-$m$ mode
must vanish identically. The residual odd-$m$ amplitude relative to that of the
$(2,2)$ is $\AlNRExchangeFixedOddAmp$ (measured on $\AlNRExchangeFixedOddAmpCount$ mode amplitudes). On
the \AlNRMirrorPairCountWord{} pairs of training simulations at $q=1$ that are mirror images of each
other (the same binary with the spins swapped)~\cite{NRHJSurTrainingSet}, the normalized overlap of each mode between the two members has the sign $(-1)^m$ that
exchange symmetry requires, and the corresponding mismatch without time or phase optimization is
$\AlNRMirrorOverlapDeficit$ (over $\AlNRMirrorOverlapCount$ mode overlaps). Their mode
amplitudes differ by $10^{-4}$ to $2.5\times10^{-2}$ in relative terms.
The 95th percentiles of both measurements lie at the documented
frame-contamination floor of the NR data itself
(Sec.~\ref{sec:aa_accuracy}). Therefore, the exactly-antisymmetric cases agree
with the catalog down to the catalog's own error level.

\subsection{Accuracy of the \AAf{} flavor against NR}
\label{sec:aa_accuracy}

A mismatch depends on the mode set, the alignment freedom, the comparison
window, and any noise weighting.
We choose all four variables and evaluate every model with the same mismatch definition. Every accuracy number quoted in this subsection is a mismatch between the model's modes and
the raw NR modes of the same simulation, computed directly on the mode data.
This mismatch is computed without detector noise weighting or the power spectral density (PSD), and projection onto a sky direction or polarization.
The two waveforms are aligned at the $(2,2)$ amplitude peak,
one relative time shift is searched over $\pm\AlTimeShiftBound\,M$, and the maximization optimizes over one overall complex phase for a
single-mode mismatch, or one orbital-phase rotation plus one overall phase (the reference orbital phase and the polarization angle) for
the all-mode mismatches. The comparison window is the full IMR
waveform, from the start of the NR data to $\AlWindowEndAfterPeak\,M$ after the amplitude
peak. Where a third model is compared against the same reference, the two are
first restricted to a common mode content, by reducing the model with more
modeled modes down to \NRHJ{}'s. The LOO tests are those defined in Sec.~\ref{sec:loo}. The LOO mismatch is evaluated for each
of the $\AlTrainingSetSize$ simulations in turn.
\AAf{} gives a fixed-hyperparameter LOO mismatch of $\AlBlendedLooMismatchTwoTwo$
on the $(2,2)$ mode and $\AlBlendedLooMismatchAllModes$ over all \AlModelModeCountWord{} modes
(Table~\ref{tab:blendloo}).
Trained on all $\AlTrainingSetSize$ simulations, the in-sample values are
$\AlInSampleMismatchTwoTwo$ and $\AlInSampleMismatchAllModes$.
The LOO median is $\AlLooOverInSampleMedianRatioBlendedTwoTwo$ times the in-sample median for the $(2,2)$
mode and $\AlLooOverInSampleMedianRatioBlendedAllModes$ times for all modes.

\begin{table}[t]
  \centering
  \caption{Full-IMR mismatch against NR over the $\AlTrainingSetSize$ training simulations (median with the
  5th--95th percentile span), for the direct fits alone and for the model, which blends the
  direct and ratio fits (Sec.~\ref{sec:radiative}). The LOO is fixed-hyperparameter.}
  \label{tab:blendloo}
  \resizebox{\columnwidth}{!}{%
  \begin{tabular}{lcccc}
    \toprule
    & \multicolumn{2}{c}{$(2,2)$} & \multicolumn{2}{c}{All modes} \\
    Construction & In-sample & LOO & In-sample & LOO \\
    \midrule
    Direct fits only & $\AlDirectInSampleMismatchTwoTwo$ & $\AlLooMismatchTwoTwo$ & $\AlDirectInSampleMismatchAllModes$ & $\AlLooMismatchAllModes$ \\
    Blended & $\AlInSampleMismatchTwoTwo$ & $\AlBlendedLooMismatchTwoTwo$ & $\AlInSampleMismatchAllModes$ & $\AlBlendedLooMismatchAllModes$ \\
    \bottomrule
  \end{tabular}}
\end{table}
 The LOO
error therefore has two contributions of similar size: the in-sample error, set by the
representation and by the training data, and the additional error from the regression across
the parameter space.

The inspiral and the merger--ringdown can also be scored separately, as
diagnostics that localize the error. Computing the same LOO mismatches against NR separately for the part
of the waveform before and after $t=0$ (the peak of the root-sum-square of the
mode amplitudes, Sec.~\ref{sec:peak}), the inspiral--merger part ($t\le0$) has mismatches of
$\AlLooInspiralMismatchTwoTwo$ on the $(2,2)$ and
$\AlLooInspiralMismatchAllModes$ all-mode, and the merger--ringdown part
($t\ge0$) has $\AlLooMergerRingdownMismatchTwoTwo$ and
$\AlLooMergerRingdownMismatchAllModes$. Each waveform segment is normalized by its own power.
The short strong-field merger--ringdown window therefore weighs its subdominant-mode errors much more than the full-IMR number. This shows that the merger--ringdown portion of \AAf{} is a factor of
$\AlLooMergerRingdownOverInspiralFactorTwoTwo$ ($(2,2)$) to $\AlLooMergerRingdownOverInspiralFactorAllModes$ (all modes) less accurate than its inspiral. The modeling
of the merger--ringdown part of a surrogate has always been challenging, and it
is known to be the least accurate portion of a waveform. Ref.~\cite{Finch:2021iip} finds that quasinormal-mode fits to the ringdown of
\textsc{NRSur7dq4} recover the remnant mass and spin one to two orders of
magnitude less accurately than fits to the NR simulations. In
GW230814~\cite{LIGOScientific:2025cmm}, it was found that waveform systematics
could contribute to an apparent deviation from general relativity (GR) in ringdown tests. More
recently, \textsc{NRSur7dq4v2}~\cite{Ravishankar:2026akf} models the inspiral
and the merger--ringdown on separate, overlapping time intervals, with its
largest accuracy gains in the merger--ringdown.

Scored simulation by simulation on the same waveform span,
unweighted mismatch, and LOO setting, \AdA{}, which uses the direct fits alone
(Table~\ref{tab:blendloo}), is marginally more accurate than \AAf{}. Its LOO mismatches are
$\AlLooMismatchTwoTwo$ on the $(2,2)$ and $\AlLooMismatchAllModes$ all-mode. The
per-simulation ratios of the LOO mismatches of \AAf{} to those of
\AdA{} are $\AlLooRatioAAOverReleasedAdATwoTwo$ on the $(2,2)$ and
$\AlLooRatioAAOverReleasedAdAAllModes$ all-mode, and \AAf{} is at or below \AdA{} on
$\AlLooAAAtOrBelowReleasedAdAPctTwoTwo\%$ and $\AlLooAAAtOrBelowReleasedAdAPctAllModes\%$ of the catalog.
For the higher modes the median ratios lie between $\AlLooRatioAAOverReleasedAdAHigherModesMin$ and
$\AlLooRatioAAOverReleasedAdAHigherModesMax$. In-sample, the picture is the same, with median ratios of
$\AlInSampleRatioAAOverReleasedAdAMedianTwoTwo$ on the $(2,2)$, $\AlInSampleRatioAAOverReleasedAdAMedianAllModes$
all-mode and $\AlInSampleRatioAAOverReleasedAdAHigherModesMin$--$\AlInSampleRatioAAOverReleasedAdAHigherModesMax$
for the higher modes.

\paragraph{Worst-case tails}
We find two different causes for the largest full-IMR all-mode LOO
mismatches, against a median and 90\% interval of $\AlBlendedLooMismatchAllModes$.
The two simulations with the largest LOO mismatches, \AlLooWorstSimID{} and \AlLooSecondWorstSimID{}
($q=\AlLooWorstPairMassRatio$ with $\chi_{1z}=\AlLooWorstPairChiOne$ and $\chi_{2z}=\AlLooWorstSimChiTwo$ /
$\AlLooSecondWorstSimChiTwo$; $\AlLooWorstSimMismatch$ and $\AlLooSecondWorstSimMismatch$), also have the largest
in-sample mismatches, $\AlWorstSimInSampleMismatch$ and $\AlSecondWorstSimInSampleMismatch$. This is not a
representation error. Built from their own training coefficients, both waveforms match NR to
$\AlWorstPairOwnCoefficientMismatch$. The error comes from the merger--ringdown fit, which does not reproduce the
higher modes of these two simulations even when they are in the training set. After the peak their
in-sample mismatches are $\AlWorstSimInSampleAfterPeakMismatch$ and $\AlSecondWorstSimInSampleAfterPeakMismatch$,
against a median of $\AlInSampleAfterPeakMedianMismatch$. The next group of poorly fitted simulations, led by
\AlLooThirdWorstSimID{} ($q=\AlLooThirdWorstSimMassRatio$, $\chi=(\AlLooThirdWorstSimChiOne,\AlLooThirdWorstSimChiTwo)$,
$\AlLooThirdWorstSimMismatch$), shows the opposite behavior. Its LOO mismatch is poor, but its in-sample
mismatch is ordinary ($\AlInSampleThirdWorstLooSimMismatch$, within the 90\% interval of the in-sample
distribution). The model reproduces this configuration once its data are in the training set, and
the poor LOO prediction reflects the sparse coverage of this region of the parameter space.

\paragraph{Per-mode accuracy.}
The same LOO runs give full-IMR per-mode mismatches for
\AAf{} against NR (one complex phase optimized per mode), shown next to the NR resolution error, the mismatch
between the two highest resolutions of each simulation, computed the same way:
\begin{center}
{\renewcommand{\arraystretch}{1.45}\setlength{\tabcolsep}{5pt}%
\begin{tabular}{lcc}
  \toprule
  Mode & LOO & NR resolution \\
  \midrule
  $(2,2)$ & \mm{\AlBlendedLooMismatchTwoTwo} & \mm{\AlResolutionMismatchTwoTwo} \\
  $(2,1)$ & \mm{\AlBlendedLooMismatchTwoOne} & \mm{\AlResolutionMismatchTwoOne} \\
  $(3,3)$ & \mm{\AlBlendedLooMismatchThreeThree} & \mm{\AlResolutionMismatchThreeThree} \\
  $(3,2)$ & \mm{\AlBlendedLooMismatchThreeTwo} & \mm{\AlResolutionMismatchThreeTwo} \\
  $(4,4)$ & \mm{\AlBlendedLooMismatchFourFour} & \mm{\AlResolutionMismatchFourFour} \\
  $(4,3)$ & \mm{\AlBlendedLooMismatchFourThree} & \mm{\AlResolutionMismatchFourThree} \\
  $(5,5)$ & \mm{\AlBlendedLooMismatchFiveFive} & \mm{\AlResolutionMismatchFiveFive} \\
  \bottomrule
\end{tabular}}
\end{center}
Higher-mode mismatches must be compared with the resolution error of the NR data~\cite{Knapp:2025ecw},
the mismatch between the two highest resolution levels of each simulation. Where the resolution
error is comparable to the model's LOO mismatch, the NR data limit the comparison.
For the higher modes, this is the case at $q\gtrsim1.5$, where the resolution error is comparable to the
in-sample mismatches. For the $(2,2)$, the resolution error exceeds both the in-sample and the LOO mismatch at all
mass ratios. At
high mass ratio, the data are well resolved: the two recent $q=\AlResolutionModernQEightMassRatio$ simulations in
the training set resolve all modeled modes to $\AlResolutionModernQEightMin$--$\AlResolutionModernQEightMax$. For
$q\gtrsim\AlHomPlateauMassRatioMin$, the LOO mismatches of the $(5,5)$ and $(4,3)$ modes are
$\AlHomPlateauOverResolutionMin$--$\AlHomPlateauOverResolutionMax$ times the local resolution error, and they are
therefore model error.

\begin{figure*}[t]
  \centering
  \includegraphics[width=0.92\linewidth]{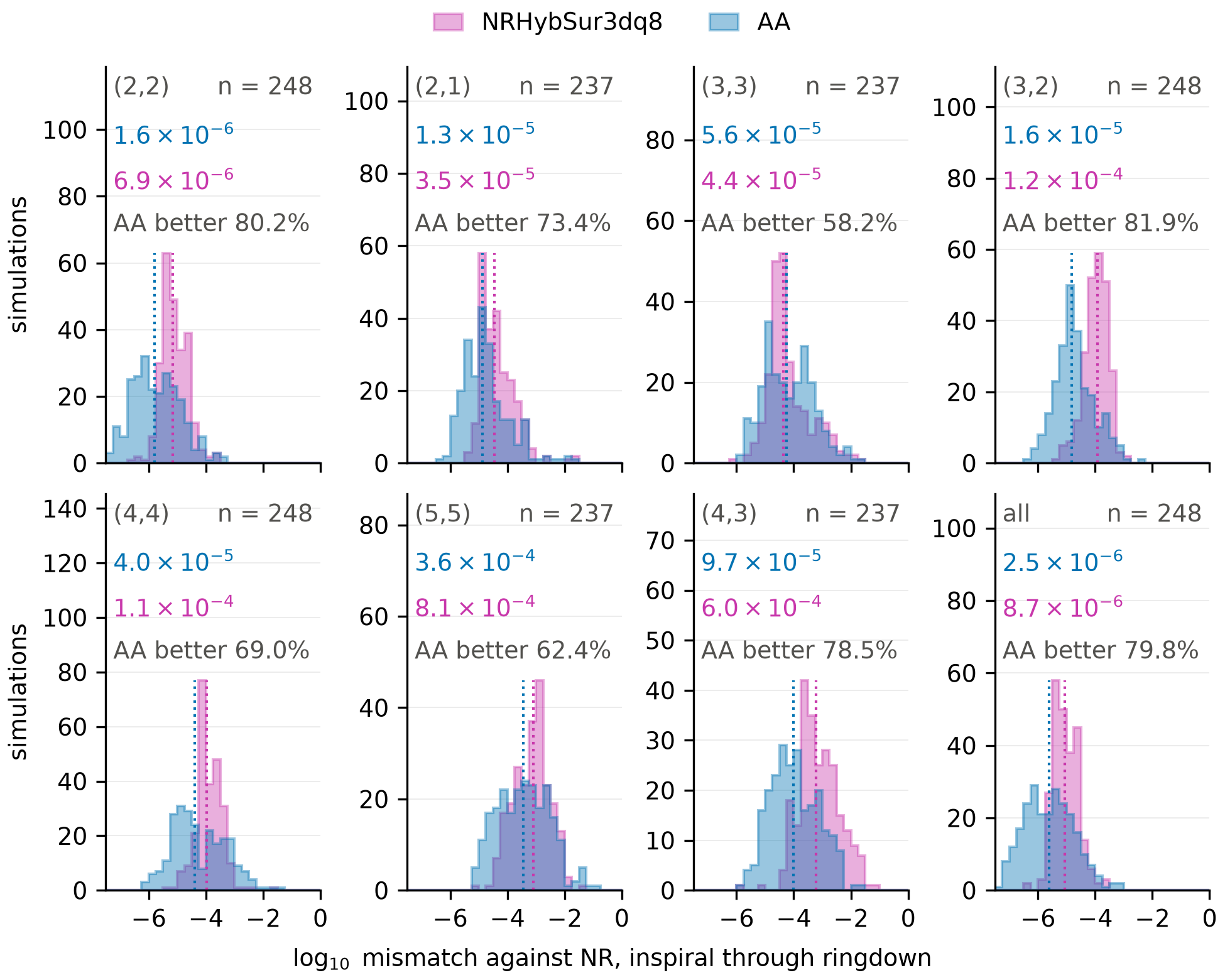}
  \caption{Catalog-wide accuracy of \AAf{} against
  \textsc{NRHybSur3dq8}: distributions of the log10 mismatch
  against each simulation's raw NR modes, one panel per
  modeled mode, along with a combined all-mode panel. These are computed over the $\AlNRHybSurScoredEvenSimCount$ simulations
  both models cover ($\AlNRHybSurScoredOddSimCount$ on the odd-$m$ panels, since the equal-mass,
  equal-spin simulations carry no odd-$m$ power). The mismatch is computed
  using a full IMR span extending to $\AlWindowEndAfterPeak\,M$
  after the amplitude peak, a flat mode-space mismatch with no detector
  weighting, aligned at the $(2,2)$ peak with a $\pm\AlTimeShiftBound\,M$ time shift, and at
  the same \AlModelModeCountWord{}-mode content. Dotted lines and the printed values on each panel
  are the medians. Each panel also prints the fraction of simulations on which
  \AAf{} is more accurate. \AAf{} is fixed-hyperparameter LOO,
  while \textsc{NRHybSur3dq8} is evaluated as published
  and is partially in-sample on this catalog. \AlNRHybSurUncoveredSimCountWord{} simulations with $|\chi|\ge\AlNRHybSurCoverageSpinLimit$ are
  outside \textsc{NRHybSur3dq8}'s coverage and are excluded.}
  \label{fig:modehist}
\end{figure*}

\paragraph{Comparison against \textsc{NRHybSur3dq8}.}
With the same mismatch definition, against LALSimulation
\textsc{NRHybSur3dq8} restricted to
the same \AlModelModeCountWord{} modes, \AAf{} in fixed-hyperparameter LOO is more accurate by a per-simulation factor
of $\AlAAOverNRHybSurFactorTwoTwo$ on the $(2,2)$ and $\AlAAOverNRHybSurFactorAllModes$ all-mode, and
is the more accurate model on $\sim\!\AlAABetterThanNRHybSurPct\%$ of the catalog
(Fig.~\ref{fig:modehist}). On a simulation-to-simulation basis, \AAf{} is more accurate on every
mode. Comparing the panel medians, it has a lower median in every panel
except the $(3,3)$, where \textsc{NRHybSur3dq8}'s median is the better one
($\AlNRHybSurMedianThreeThree$ against $\AlAAMedianThreeThreeNRHybSurSet$). \AAf{} is nevertheless the
more accurate model on $\AlAABetterThanNRHybSurThreeThreePct\%$ of those simulations, because its $(3,3)$ losses
are fewer but larger.

\textsc{NRHybSur3dq8}'s own mismatches against the NR training set of \AAf{} are
$\AlNRHybSurMismatchTwoTwo$ ($(2,2)$) and $\AlNRHybSurMismatchAllModes$
(all-mode). \AlNRHybSurUncoveredSimCountWord{}
simulations with $|\chi|\ge\AlNRHybSurCoverageSpinLimit$ lie outside \textsc{NRHybSur3dq8}'s
coverage and are excluded ($\AlNRHybSurScoredEvenSimCount$ scored on the
even-$m$ and all-mode panels, $\AlNRHybSurScoredOddSimCount$ on the odd-$m$ panels, where the \AlNRHybSurScoredExchangeSymmetricCountWord{}
equal-mass, equal-spin simulations among them carry no odd-$m$ power). The \AAf{} mismatches are fixed-hyperparameter LOO, while \textsc{NRHybSur3dq8} could be partly evaluated at catalog points that overlap its own training set.

\begin{figure}[t]
  \centering
  \includegraphics[width=0.92\linewidth]{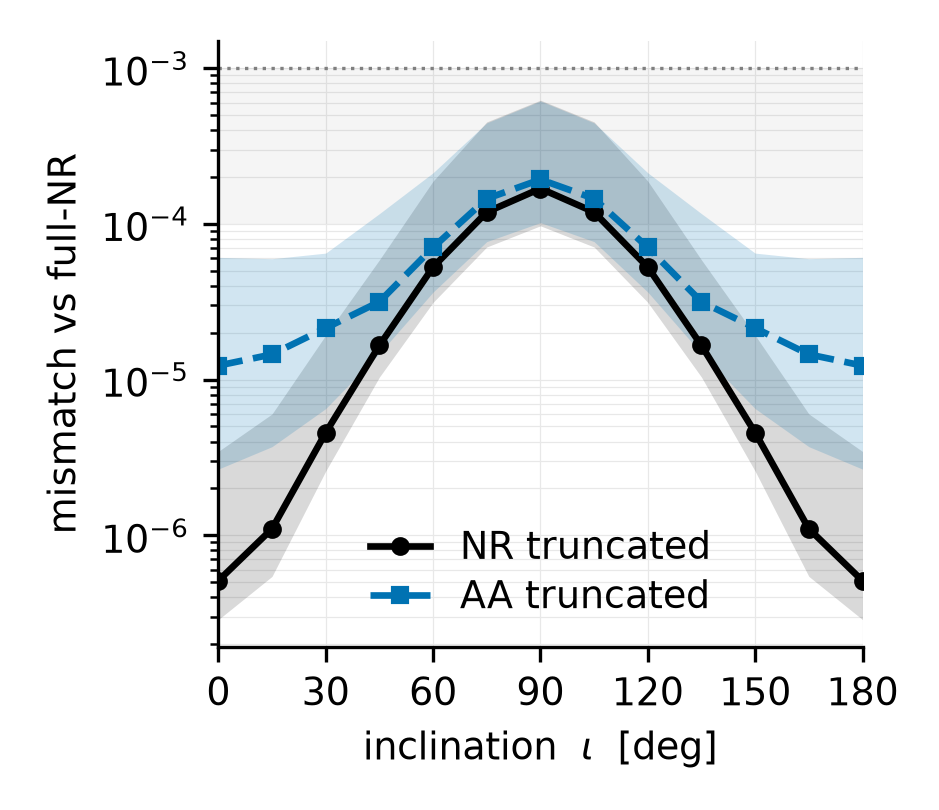}
  \caption{Mode-content completeness: aLIGO-design-PSD-weighted mismatch of
  the \AlModelModeCountWord{}-mode strain against the full-multipole NR waveform versus
  inclination.  Medians with \AlModeCompBandLowPct{}th--\AlModeCompBandHighPct{}th percentile bands over $\AlModeCompSimCount$
  simulations spanning the training range ($M=\AlModeCompTotalMass\,M_\odot$) are plotted. ``NR
  truncated`` is NR's own \AlModelModeCountWord{} modeled modes against full NR, the
  model-independent truncation floor. ``AA
  truncated'' is the model's full-retraining LOO \AlModelModeCountWord{}-mode strain,
  retrained without the scored simulation, against NR on
  the same window.}
  \label{fig:modecomplete}
\end{figure}

\paragraph{Mode-content completeness.}
The model includes \AlModelModeCountWord{} multipoles (\AlModelModeList), while the NR catalog contains more. The choice of the modes was made using a study involving the faithfulness of the model across inclination of the binary against the NR data. The error from \emph{omitting} modes
is distinct from the per-mode fit error, and adds to the waveform systematics. It is a property of the modeled-mode
choice, it is shared by any model carrying the same set, and it grows toward
edge-on inclinations as the $(2,2)$ is geometrically suppressed. We quantify it with the
aLIGO-design-PSD-weighted mismatch of the \AlModelModeCountWord{}-mode strain against the
\emph{full} NR waveform (all available multipoles, including the $m=0$
memory), maximized over phase, time shift, and orbital phase, versus
inclination $\iota$, over $\AlModeCompSimCount$ simulations spanning the training range at
$M=\AlModeCompTotalMass\,M_\odot$ (Fig.~\ref{fig:modecomplete}. Medians are shown, with \AlModeCompBandLowPct{}th--\AlModeCompBandHighPct{}th
percentile bands in the figure). Edge-on, where the omitted power is the
largest, the error due to the \AlModelModeCountWord{}-mode truncation  (a mode-truncation floor computed from the NR data) has a median of $\AlModeCompEdgeOnSevenModeFloor$,
a factor $\AlModeCompFiveToSevenFloorGain$ below the floor $\AlModeCompEdgeOnFiveModeFloor$ for a five-mode choice. That factor is
the gain from carrying the additional $(4,3)$ and $(5,5)$ in the seven-mode model. The model's full-retraining LOO
\AlModelModeCountWord{}-mode strain, with the inspiral retrained without the scored simulation, is estimated to be
$\AlModeCompEdgeOnLooModel$ at edge-on, $\AlModeCompEdgeOnModelOverFloor\times$ above the mode-truncation floor. There, the
accuracy is set by the mode truncation, instead of by the fit. Face-on, the
situation is reversed. The truncation floor falls to $\sim\!\AlModeCompFaceOnSevenModeFloor$, while the
LOO curve asymptotes to $\AlModeCompLooFitLevel$. At this inclination, almost no power is in the
omitted modes. Therefore the LOO curve reflects the fit error. Above $\iota\approx\AlModeCompCrossingInclination^{\circ}$, the error from the omitted modes is larger
than the fit error, mostly from $(2,0)$ and $(6,6)$ modes.

\paragraph{First-law consistency of the predicted dynamics.}
The first-law residual $|(\mathrm{d}E/\mathrm{d}J)/\omega-1|$ of
Sec.~\ref{sec:hybridization} can also be evaluated at points away from the NR
training data. At such points,
the model predicts $J(\Phi)$, $E(\Phi)$ and the clock through independent
regression fits. The exact dynamics satisfies
$\mathrm{d}E/\mathrm{d}J=\omega$ identically. Any violation by the
predicted curves is therefore due to the regression. Over $\HybClosureRandomPoints$
random parameter points, the residual evaluated on the predicted surrogate energy and action variables has a
median of $\HybClosurePredictedMedian$, about $\HybClosureExcessRatio$ times the $\HybFirstLawNRSideMedian$ obtained on the
stitched NR variables at the hybridization boundary. This excess is attributable to the error of the Gaussian-process
regression away from the training data. This first-law check can run at
any point of the parameter space and serves as an internal consistency check of the predicted dynamics.
 \subsection{Distinguishability from numerical relativity}
\label{sec:distinguishability}

Mismatch quantifies faithfulness against a reference waveform, and by itself it is not an observational
statement. We therefore convert the accuracy into a threshold \emph{distinguishability}
signal-to-noise ratio (SNR). In the single-parameter form of Refs.~\cite{Lindblom:2008cm,Baird:2012cu}, two
waveforms are indistinguishable in a detector below a critical signal-to-noise
ratio
\begin{equation}
  \rho_{\rm crit} = \frac{1}{\sqrt{2\,\mathrm{MM}}},
  \label{eq:rhocrit}
\end{equation}
with $\mathrm{MM}$ being the mismatch.
A larger $\rho_{\rm crit}$ means the model is faithful up to a higher SNR.
Every row of Table~\ref{tab:rhocrit} is the all-mode measurement of
Sec.~\ref{sec:aa_accuracy}. Therefore, the \NRHJ{} models and the external
surrogates are compared with the same mismatch definition.

\begin{table}[t]
  \centering
  \caption{Distinguishability SNR $\rho_{\rm crit}=1/\sqrt{2\,\mathrm{MM}}$
  from the all-mode mismatch against each simulation's raw NR modes, over
  the full IMR waveform, for the aligned-spin set.
  Medians with 5th--95th percentile ranges are shown. Higher SNR is better.
  The \AAf{} and \AdA{} LOO rows are fixed-hyperparameter LOO, with the inspiral and
  merger--ringdown regressions refit per held-out training point.
  \textsc{NRHybSur3dq8} was built from this same SXS catalog, and its rows are partly
  in-sample. The rows below the middle rule are a separate measurement, restricted to the
  time interval and parameter range that \textsc{NRSur7dq4} covers ($\ValNRSurBoxSimCount$
  simulations and \ValBoxModeCountWord{} modes, described in the text), and are not comparable with
  the rows above it.}
  \label{tab:rhocrit}
  {\renewcommand{\arraystretch}{1.3}%
  \begin{tabular}{lccc}
    \toprule
    Model & $n$ & $\rho_{\rm crit}$ & Median MM \\
    \midrule
    \AAf{} (in-sample)     & \ValTrainingSimCount{} & $\ValRhoAAInSample$ & \mm{\ValMedianMMAAInSample} \\
    \AAf{} (LOO)           & \ValTrainingSimCount{} & $\ValRhoAALOO$ & \mm{\ValMedianMMAALOO} \\
    \AdA{} (LOO)           & \ValTrainingSimCount{} & $\ValRhoAdAReleasedLOO$ & \mm{\ValMedianMMAdAReleasedLOO} \\
    \textsc{NRHybSur3dq8}  & \ValRhoHybSurN{} & $\ValRhoHybSur$  & \mm{\ValMedianMMHybSur} \\
    \midrule
    \multicolumn{4}{l}{Range of \textsc{NRSur7dq4}, \ValBoxModeCountWord{} modes, $\ValNRSurBoxSimCount$ simulations} \\
    \AAf{} (in-sample)     & \ValNRSurBoxSimCount{} & $\ValRhoBoxAAInSample$ & \mm{\ValMedianMMBoxAAInSample} \\
    \AAf{} (LOO)           & \ValNRSurBoxSimCount{} & $\ValRhoBoxAALOO$ & \mm{\ValMedianMMBoxAALOO} \\
    \AdA{} (in-sample)     & \ValNRSurBoxSimCount{} & $\ValRhoBoxAdAInSample$ & \mm{\ValMedianMMBoxAdAInSample} \\
    \textsc{NRHybSur3dq8}  & \ValNRSurBoxSimCount{} & $\ValRhoBoxHybSur$  & \mm{\ValMedianMMBoxHybSur} \\
    \textsc{NRSur7dq4}     & \ValNRSurBoxSimCount{} & $\ValRhoBoxNRSur$    & \mm{\ValMedianMMBoxNRSur} \\
    \textsc{NRSur7dq4v2}~\cite{Ravishankar:2026akf} & \ValNRSurBoxSimCount{} & $\ValRhoBoxNRSurVTwo$ & \mm{\ValMedianMMBoxNRSurVTwo} \\
    \textsc{NRSur7dq4} neural emulator & \ValNRSurBoxSimCount{} & $\ValRhoBoxNRSurNN$ & \mm{\ValMedianMMBoxNRSurNN} \\
    \bottomrule
  \end{tabular}}
\end{table}

\AAf{} has
$\rho_{\rm crit}=\ValRhoAALOO$ in fixed-hyperparameter LOO, against an
in-sample value of $\ValRhoAAInSample$, and exceeds
\textsc{NRHybSur3dq8} ($\ValRhoHybSur$, partly in-sample) with the same mismatch definition.
The model's fixed-hyperparameter LOO error is therefore not detectable in current single-event
detections (SNR $\sim\!10$--$40$, GW150914 $\sim\!24$). Error at this level
matters only in a regime where every current model, \textsc{NRSur7dq4}
included, requires denser NR coverage.

\textsc{NRSur7dq4} and \textsc{NRSur7dq4v2} (evaluated with
\texttt{gwsurrogate}~\cite{Field:2025isp}) have no entry in the upper part of the table because they do not cover
its time interval and parameter range. Of the $\ValTrainingSimCount$ simulations, $\ValNRSurBoxOutsideCount$ lie outside their native validity
domain ($q\le\ValNRSurBoxQMax$, $|\chi|\le\ValNRSurBoxChiMax$), and their waveforms start $\ValNRSurStartBeforePeak\,M$ before the peak
and end $\ValNRSurEndAfterPeak\,M$ after it, shorter than the table's window at both ends. A mismatch is not comparable across windows of different length. The mismatch grows as the
square of the phase error~\cite{Owen:1995tm,Owen:1998dk,Lindblom:2008cm}, and in NR waveforms the
phase error between two resolutions accumulates with time. For a phase error growing linearly in
time, the mismatch between two resolutions therefore grows as the square of the compared
length~\cite{Mitman:2025tmj}. The mismatches in the lower part of Table~\ref{tab:rhocrit} are
computed over \textsc{NRSur7dq4}'s shorter window and are therefore smaller than they would be over
the full window. The
lower part scores every model on the interval and range \textsc{NRSur7dq4}
covers, at the $\ValNRSurBoxSimCount$ simulations inside its box, with every model and the NR
data restricted to the \ValBoxModeCountWord{} modes it carries. On that window the ordering is
unchanged. The LOO \AAf{} model has a median mismatch of
$\ValMedianMMBoxAALOO$, \textsc{NRHybSur3dq8} $\ValMedianMMBoxHybSur$,
\textsc{NRSur7dq4} $\ValMedianMMBoxNRSur$, and the \textsc{NRSur7dq4} neural
emulator~\cite{Purrer:2026wgp}, evaluated as released, $\ValMedianMMBoxNRSurNN$. The
LOO \AAf{} model is closer to NR than \textsc{NRSur7dq4} at every
one of the $\ValNRSurBoxSimCount$ simulations and closer than \textsc{NRHybSur3dq8} at $\ValBoxAALOOCloserThanHybSurPct\%$
of them, and the ordering holds separately for $q<2$, $2\le q<3$ and
$q\ge3$. Only the \AAf{} LOO row is out of sample. The \AAf{} and \AdA{} in-sample rows are in-sample, and the external
surrogates are partly in-sample. Ending the window $\ValBoxShortWindowEndAfterPeak\,M$ after the peak instead of $\ValBoxWindowEndAfterPeak\,M$
changes no median by more than $10\%$.
 \section{Cross-model campaign and comparisons with NR}
\label{sec:campaign}

As a further check, the \NRHJ{} model is compared against \CampTargetCountWord{} independent
waveform models (the NR surrogates \textsc{NRHybSur3dq8} and \textsc{NRSur7dq4}, the
phenomenological IMRPhenomTPHM and IMRPhenomXPHM, and the effective-one-body
SEOBNRv5PHM) over a large Monte-Carlo set of random aligned-spin configurations,
drawn with a distinct random seed per model ($\CampDrawsTotalMillion$~million points over the \CampTargetCountWord{}
targets, run against the \AAf{} model, with both models in each pair restricted to the same modes). For each draw, the coherent H1/L1 network
faithfulness~\cite{Harry:2017weg} uses a single set of sky position,
polarization angle, inclination, coalescence phase, and coalescence time shared
by both detectors, with the per-detector likelihoods summed before normalization.
The sky position, polarization angle and inclination are held at the signal's
own values, and the faithfulness is maximized over coalescence phase and time
only, by an exact fast-Fourier-transform time--phase maximization, with the
alignment window scaled to the total mass. This network mismatch is
noise-weighted with power spectral densities estimated from the public LIGO
Hanford and Livingston data around GW250114.
The five models are mutually independent, which allows the campaign to localize the
source of any large disagreement. A mismatch shared across all \CampTargetCountWord{} targets points at this model as being erroneous, while a mismatch against a single external target waveform model points to the other.

We find that \NRAAS{} passes the combined model-versus-model test against the other waveform models.
Against IMRPhenomTPHM, over $\CampTPHMDraws$ draws, the network mismatch has a
median of $\CampMedianMMTPHM$ and a $90$th percentile of $\CampPNinetyMMTPHM$.
The corresponding values against \textsc{NRSur7dq4} are $\CampMedianMMNRSur$ and
$\CampPNinetyMMNRSur$, and against SEOBNRv5PHM $\CampMedianMMSEOB$ and
$\CampPNinetyMMSEOB$.

Where the campaign's own points were compared against NR (below), the
disagreement mostly belonged to the targets.

In a second comparison campaign, we compare the mismatches of the \AAf{},
\textsc{NRHybSur3dq8} and SEOBNRv5PHM models against one another at identical parameters, with
every model restricted to the same \CampTriangleModeCountWord{} modes, $(2,2)$, $(2,1)$, $(3,3)$,
$(3,2)$ and $(4,4)$. Over $\CampTrianglePoints$ random aligned-spin configurations
($q\le\CampBoxQMax$, $|\chi_{1z}|,|\chi_{2z}|\le\CampBoxChiMax$, $M$ between $\CampBoxMtotMin$ and
$\CampBoxMtotMax\,M_\odot$), each evaluated at \CampTriangleInclinationCountWord{} inclinations, the network mismatch
between \AAf{} and \textsc{NRHybSur3dq8} has a median of $\CampTriangleMedianAAHyb$ and a
$90$th percentile of $\CampTrianglePNinetyAAHyb$, and exceeds $\CampTriangleThreshold$ in $\CampTriangleAAHybAboveThresholdRows$ of the
$\CampTriangleRows$ evaluations. The median rises from $\CampTriangleFaceOnAAHyb$ face-on to
$\CampTriangleEdgeOnAAHyb$ edge-on, where the higher modes contribute most. Therefore, the two surrogates also agree closely in the higher modes. Against SEOBNRv5PHM, the two surrogates have nearly
the same mismatches, with medians of $\CampTriangleMedianSEOBAA$ for \AAf{} and
$\CampTriangleMedianSEOBHybSur$ for \textsc{NRHybSur3dq8}. They therefore agree with each other an order of magnitude more closely than either agrees with SEOBNRv5PHM. At the
parameters of the $\CampAnchoredSimCount$ training simulations with $q\le\CampBoxQMax$ and
$|\chi_{1z}|,|\chi_{2z}|\le\CampBoxChiMax$, again at \CampTriangleInclinationCountWord{} inclinations, the same \CampTriangleModeCountWord{}-mode
mismatch against NR has a median of $\CampAnchoredMedianNRAA$ for \AAf{} and
$\CampAnchoredMedianNRHybSur$ for \textsc{NRHybSur3dq8}, and both are below $\CampTriangleThreshold$ in $\CampAnchoredBothBelowThresholdPct\%$
of these evaluations. Both surrogates were trained on many of these simulations, \AAf{} on all of them.
Therefore, these values give the scale of each surrogate's distance
from NR, and do not necessarily rank the two.

At $\CampNRCheckAboveThresholdPoints$ of the $\CampNRCheckPopulationMillion$~million campaign points against the comparison
models other than IMRPhenomXPHM ($\CampNRCheckAboveThresholdPct\%$), the mismatch between \AAf{} and the comparison model
exceeds $\CampNRCheckThreshold$. To find which of the two is in error, each
such point is first checked against a second, external NR surrogate,
\textsc{NRHybSur3dq8} or, for $q\le4$, \textsc{NRSur7dq4}. At $\CampNRCheckPoints$ of the points,
an external surrogate agrees with the comparison model at least ten times more
closely than \AAf{} does. This points to \AAf{} as the source of the error. To
test this, we compute the mismatch of \AAf{}, \AdA{}, the comparison model and
\textsc{NRHybSur3dq8} against the nearest available NR simulation in the training
catalog, at that simulation's parameters and with the campaign's extrinsic parameters. At those points, \AAf{}
yields a median in-sample mismatch of
$\CampNRCheckMedianAA$ against NR. The campaign mismatches between \AAf{} and the comparison model at
these points are larger by a factor of $\CampNRCheckCampaignOverNRMedian$ at the median (90th percentile
$\CampNRCheckCampaignOverNRPNinety$). \AAf{} is closer to NR than the comparison model on $\CampNRCheckAACloserThanTargetPct\%$ of
the points compared against NR, and than \textsc{NRHybSur3dq8} on $\CampNRCheckAACloserThanHybSurPct\%$. The paired
\AAf{}/\AdA{} ratio is $\CampNRCheckAAOverAdAMedian$ at the median. Therefore, at these points the large campaign mismatches come from the comparison
models and not from \AAf{}.

A second
comparison against NR takes $\CampTPHMSetPoints$ points from the
IMRPhenomTPHM part of the campaign, with the intrinsic parameters chosen to be those of the nearest available training
simulation, at a median distance of $\CampTPHMSetNearestSimDistance$ in $(\log q,\chi_{1z},\chi_{2z})$ coordinate space,
and the extrinsic parameters of the campaign kept the same. This comparison gives a median network mismatch against NR of
$\CampTPHMSetMMTPHM$ for IMRPhenomTPHM, against
$\CampTPHMSetMMAA$ for \AAf{} and $\CampTPHMSetMMAdA$ for \AdA{} (both in-sample). We therefore find that IMRPhenomTPHM is farther away from NR than both surrogates at $\CampTPHMSetTPHMWorsePct\%$ of the points. The IMRPhenomTPHM mismatch is also found to increase by a factor of $\CampTPHMSetEdgeOverFaceTPHM$
from face-on to edge-on at these points, while both surrogates stay relatively flat
(Appendix~\ref{app:iota}), showing the variation of faithfulness against observed mode content. Under NR, \AAf{} and \AdA{} are
nearly indistinguishable at these points: the per-point mismatch ratio has a median
of $\CampTPHMSetAAOverAdAMedian$, a $\CampTPHMSetAAOverAdAEffectPct\%$ level effect.

\paragraph{Mode by mode, over the whole training catalog.}
Comparing \AAf{} with \AdA{} one mode at a time, over all $\ValTrainingSimCount$ training simulations, leads to the same conclusion. Table~\ref{tab:aaadamodes} gives the comparison with both model variants
(which use a common merger--ringdown model). Therefore, it measures the in-sample accuracy of the inspiral alone.
The two flavors are practically indistinguishable,
with \AdA{} slightly more accurate in the interior of the domain, and \AAf{} at the edges of parameter space.
Pooled over the $\CampModePairCount$ (simulation, mode)
pairs, the median ratio \AdA{}$/$\AAf{} is $\CampModePooledRatio$ and \AAf{} is closer
to NR on $\CampModePooledAABetterPct\%$ of them. For the individual modes, the median ratio lies between
$\CampModeRatioMin$ and $\CampModeRatioMax$. Both models are fitted on all $\ValTrainingSimCount$
simulations. Therefore, this comparison is in-sample.

\begin{table}[t]
  \centering
  \caption{Paired \AAf{}/\AdA{} faithfulness against NR, one mode at a time,
  over all $\ValTrainingSimCount$ training simulations. Both
  flavors are composed with the same merger--ringdown model and
  scored on the campaign's noise-weighted
  two-detector (H1, L1) network mismatch. ``Ratio'' is the median of the
  per-simulation ratio \AdA{}$/$\AAf{}, and a value above one means \AAf{} is
  the closer to NR of the two. ``\AAf{} better'' is the fraction of simulations in which \AAf{} is closer to NR. Rows with $n=\ValOddModeSimCount$ exclude the simulations in which that mode vanishes. Both columns are in-sample.}
  \label{tab:aaadamodes}
  \begin{tabular}{lccccc}
    \toprule
    Mode & $n$ & \AAf{} med & \AdA{} med & Ratio & \AAf{} better \\
    \midrule
    $(2,2)$ & \ValTrainingSimCount{} & \mm{\CampModeAAMedianMMTwoTwo} & \mm{\CampModeAdAMedianMMTwoTwo} &
      \mm{\CampModeRatioTwoTwo} & $\CampModeAABetterPctTwoTwo\%$ \\
    $(2,1)$ & \ValOddModeSimCount{} & \mm{\CampModeAAMedianMMTwoOne} & \mm{\CampModeAdAMedianMMTwoOne} &
      \mm{\CampModeRatioTwoOne} & $\CampModeAABetterPctTwoOne\%$ \\
    $(3,3)$ & \ValOddModeSimCount{} & \mm{\CampModeAAMedianMMThreeThree} & \mm{\CampModeAdAMedianMMThreeThree} &
      \mm{\CampModeRatioThreeThree} & $\CampModeAABetterPctThreeThree\%$ \\
    $(3,2)$ & \ValTrainingSimCount{} & \mm{\CampModeAAMedianMMThreeTwo} & \mm{\CampModeAdAMedianMMThreeTwo} &
      \mm{\CampModeRatioThreeTwo} & $\CampModeAABetterPctThreeTwo\%$ \\
    $(4,4)$ & \ValTrainingSimCount{} & \mm{\CampModeAAMedianMMFourFour} & \mm{\CampModeAdAMedianMMFourFour} &
      \mm{\CampModeRatioFourFour} & $\CampModeAABetterPctFourFour\%$ \\
    $(4,3)$ & \ValOddModeSimCount{} & \mm{\CampModeAAMedianMMFourThree} & \mm{\CampModeAdAMedianMMFourThree} &
      \mm{\CampModeRatioFourThree} & $\CampModeAABetterPctFourThree\%$ \\
    $(5,5)$ & \ValOddModeSimCount{} & \mm{\CampModeAAMedianMMFiveFive} & \mm{\CampModeAdAMedianMMFiveFive} &
      \mm{\CampModeRatioFiveFive} & $\CampModeAABetterPctFiveFive\%$ \\
    \bottomrule
  \end{tabular}
\end{table}

The dependence of each model's faithfulness on the viewing inclination, and
the one configuration where \textsc{NRHybSur3dq8} remains the more faithful model
face-on, are examined further in Appendix~\ref{app:iota}.

 \section{Analytic gradients and parameter estimation}
\label{sec:gradients}

\begin{figure}[tbp]
  \centering
  \includegraphics[width=0.92\linewidth]{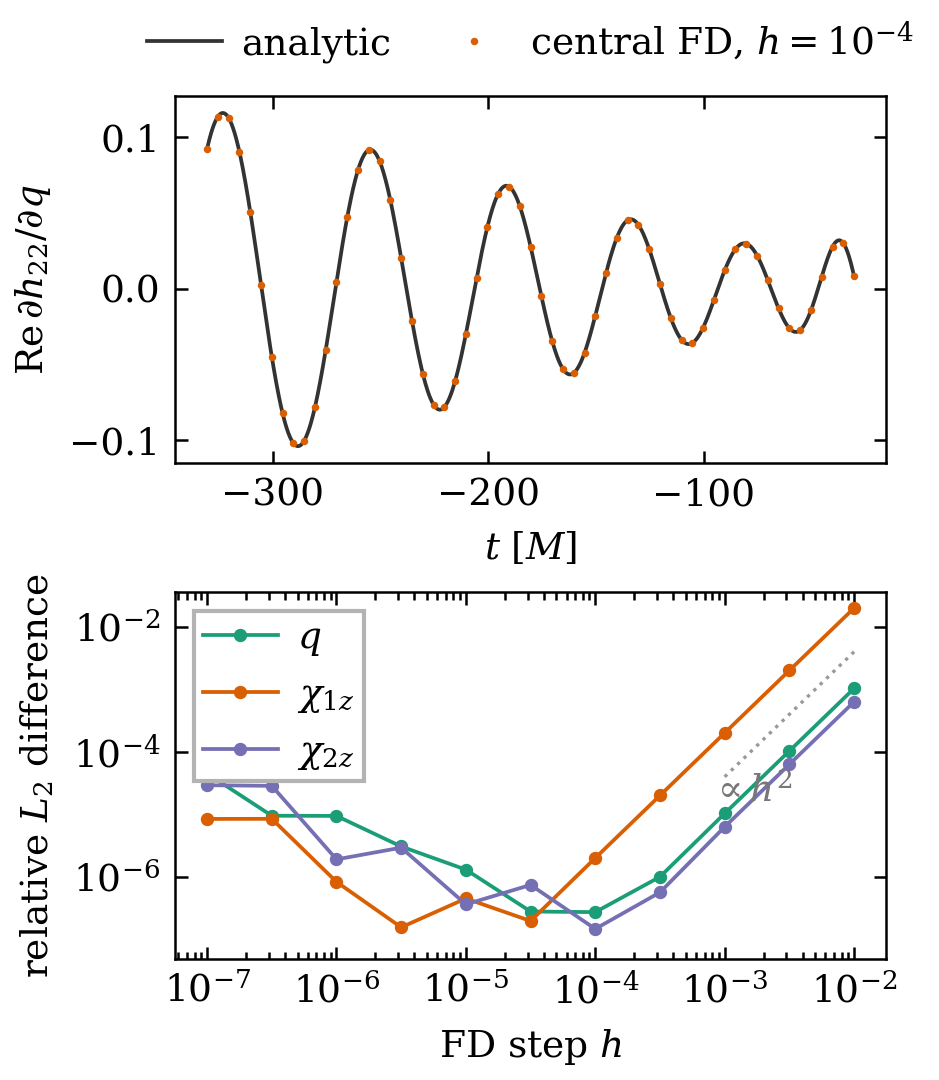}
  \caption{Analytic derivatives of the \AAf{}'s $(2,2)$ mode with
  respect to the intrinsic parameters, at
  $(q,\chi_{1z},\chi_{2z})=\GradFigPoint$. \emph{Top:}
  $\operatorname{Re}\,\partial h_{22}/\partial q$ over the last $\GradFigSpanM\,M$ of the
  NR native span, the analytic (line) against central finite differences
  of the model's own evaluation at step $h = \Delta q=\GradFigStep$ (dots). \emph{Bottom:}
  relative $L_2$ difference between the analytic derivative and the central
  finite difference over different step sizes, for all three parameters. The left branch rises as the finite difference's roundoff error and the right
  branch as its $h^{2}$ truncation error. At the best step the two derivatives
  agree to $\sim\!2\times10^{-7}$.}
  \label{fig:gradcheck}
\end{figure}

The surrogate exposes \emph{analytic} parameter derivatives of the strain,
built by the chain rule through every stage of the model, with the derivatives
along the fixed parameter directions exact up to the spectral representation
error. The Chebyshev
trend and the kernel of Eq.~\eqref{eq:gpr} are differentiated in the fit's
internal coordinates and carried back to the physical parameters through
the coordinate maps of Sec.~\ref{sec:regression}. The phase derivative
follows by the inverse-function theorem,
\begin{equation}
  \frac{\partial\Phi}{\partial\theta} =
  -\frac{\partial\tau/\partial\theta}{\partial\tau/\partial\Phi},
  \label{eq:phasederiv}
\end{equation}
and the
fast orbital phase contributes its explicit $\mathrm{i}\,m\,s$ term at fixed
peak-aligned time. Every analytic derivative is validated against a finite-difference
result. Central differences at two step sizes are combined so that
their leading finite-difference truncation error cancels (Richardson extrapolation). The
size of that correction also measures the accuracy of the finite difference
itself at each point. The automated check demands agreement
between the analytic and finite-difference derivatives only down to the
level the finite difference can resolve, and it fails on anything worse. Figure~\ref{fig:gradcheck}
shows the derivative itself and its validation on the \AAf{} model.
The gradients are fused into a Kokkos
kernel.
The full-IMR reconstruction,
including transport through the inspiral--merger/merger--ringdown blend window, is likewise differentiable. The
chain is analytic \emph {end-to-end} for the \AAf{} flavor's native and
hybridized paths alike (the hybridized path is currently in NumPy only). One trained model serves both waveforms and gradients. The
PN extension's derivatives are transported through the $\Jphi$ matching point by the
implicit-function theorem. The
partial derivatives of the PN normalizers, and of the dynamics series are generated with SymPy~\cite{Meurer:2017yhf}, by running the PN code on symbolic arguments, with the result written out as
generated code. The derivative code is produced from the same code as the values.

\begin{figure}[tbp]
  \centering
  \includegraphics[width=\linewidth]{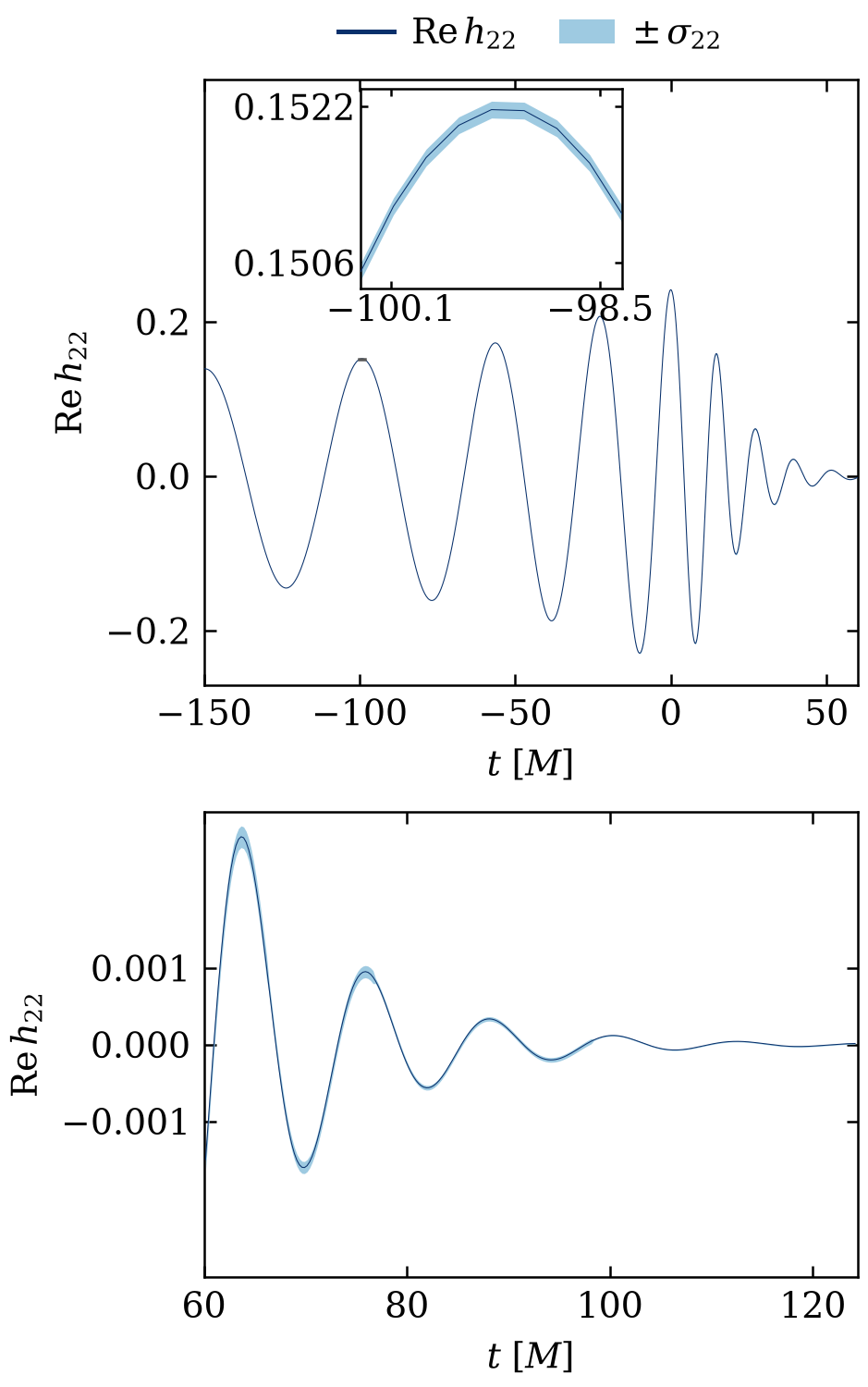}
  \caption{The \AAf{} $(2,2)$ waveform with its own one-sigma
  amplitude uncertainty band, at $q=\GradErrBandQ$, $\chi_{1z}=\GradErrBandChiOne$,
  $\chi_{2z}=\GradErrBandChiTwo$, with time in units of $M$ from the amplitude peak.
  The lower frame is a zoomed-in version of the part from $t=\GradErrBandRingdownFrameStartM\,M$
  to the last served sample with the vertical scale expanded.
  The band is drawn to scale in both frames. The inset in the top panel
  magnifies a $\GradErrBandInsetWidthM\,M$
  window around the inspiral extremum near $t=\GradErrBandExtremumTimeM\,M$, marked by the small
  rectangle, where the band is $\GradErrBandInsetBandPercent\%$ of the amplitude. The band grows
  relative to the signal as the ringdown decays, rising to $\GradErrBandFinalBandFraction$ of the
  mode amplitude at the final sample. It is an uncertainty of amplitude at fixed time. The
  clock uncertainty, median $\sigma_\tau=\GradErrBandClockSigmaMedianM\,M$, though available, is not drawn.}
  \label{fig:errorband}
\end{figure}

\begin{figure}[tbp]
  \centering
  \includegraphics[width=\linewidth]{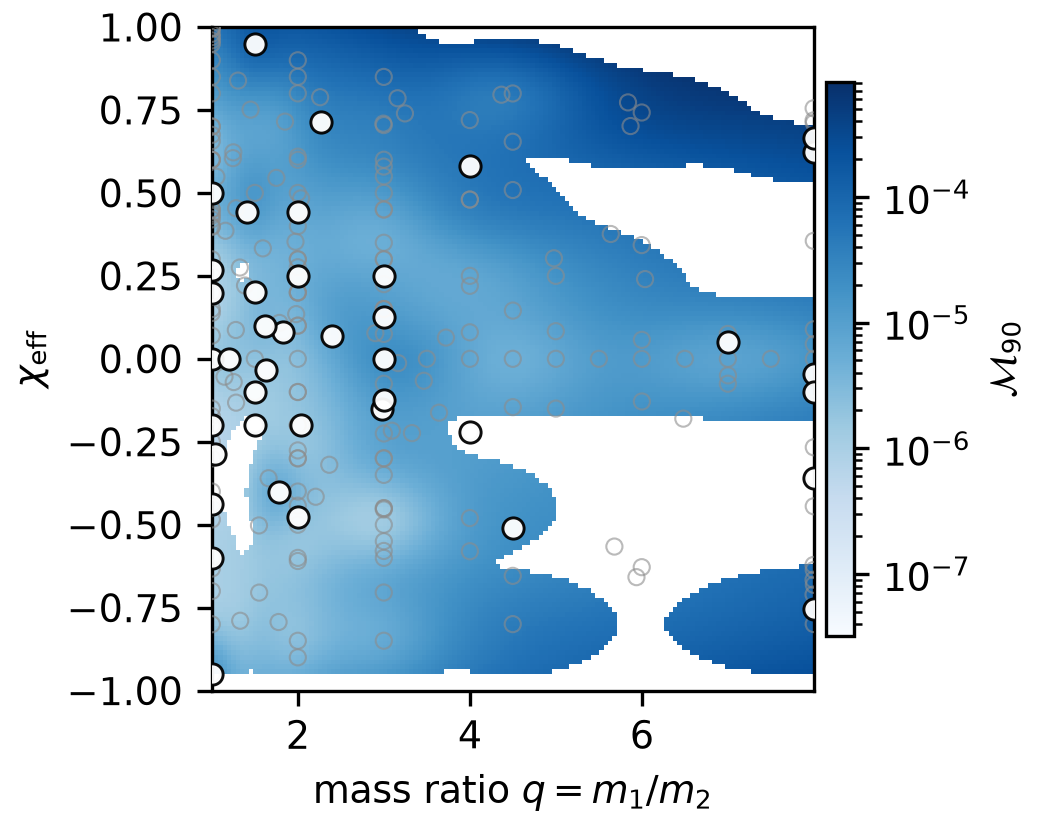}
  \caption{A surrogate model for the LOO mismatch. Predicted $\GradEnvelopePercentile$th-percentile mismatch envelope of the \AAf{} model,
  $\mathcal{M}_{90}=10^{\mu+z_{90}\sigma}$ with $z_{90}=\GradEnvelopeZScore$, over the
  $(q,\chi_{\rm eff})$ plane at $\chi_a=0$, where $\chi_{\rm
  eff}=(q\chi_{1z}+\chi_{2z})/(1+q)$ and $\chi_a=(\chi_{1z}-\chi_{2z})/2$. The
  envelope is a GPR fitted to the logarithm of the fixed-hyperparameter LOO all-mode mismatches of the training
  simulations, and calibrated by a LOO of that fit.
  Open gray circles are the training simulations at their projections. Filled
  white circles with black edges mark points whose
  measured mismatch exceeds their predicted $\mathcal{M}_{90}$. Cells where the
  posterior reproduces the prior carry no information, and are
  left blank instead of shaded. Its variation over the plane is discussed in Sec.~\ref{sec:gradients}.}
  \label{fig:envelope_map}
\end{figure}

\paragraph{Predictive uncertainty.}
The GPR predicts, alongside every coefficient, a
variance for that coefficient. Propagating those variances through the
waveform reconstruction gives an error forecast for the waveform. At any
parameter point it predicts the waveform, along with an estimate of how
wrong that waveform is likely to be. This forecast is the \emph{one-sigma strain error band}
(Fig.~\ref{fig:errorband}): per mode and per sample, a one-standard-deviation amplitude
uncertainty in the mode's own units, shown as a band around the waveform. A second estimate, used for the
parameter-space maps, is a \emph{ninetieth-percentile mismatch envelope} (Fig.~\ref{fig:envelope_map}), obtained
from a separate GPR fitted to the fixed-hyperparameter LOO mismatches of the training simulations, a single number
per parameter point rather than a band in time. We check the calibration of the error forecast in two ways
on a separate $\GradNumTrainingSims$-fold fixed-hyperparameter LOO of the inspiral regressions, where the actual error of each surrogated variable (its held-out prediction minus its value from the NR
simulation) is known. The first check measures how often
the actual error falls outside the predicted error bars. For each surrogated variable, we count how far away the NR data lies from the surrogate's prediction, and where in the predicted error band it lies. We compute how often the error lands within one, two, and three predicted standard deviations of the model. A perfectly calibrated complex-Gaussian error model would give $\GradCalRefOneSigmaPercent\%$,
$\GradCalRefTwoSigmaPercent\%$, and $\GradCalRefThreeSigmaPercent\%$. The measured rates run $\GradCalCoverOneSigmaLowPercent$--$\GradCalCoverOneSigmaHighPercent\%$, $\GradCalCoverTwoSigmaLowPercent$--$\GradCalCoverTwoSigmaHighPercent\%$,
and $\GradCalCoverThreeSigmaLowPercent$--$\GradCalCoverThreeSigmaHighPercent\%$ across the $\GradNumSurrogatedVariables$ surrogated variables. The second check measures whether
the overall size of the error envelope is correct. Per surrogated variable, the median over the training set of the ratio of the actual error magnitude (its LOO error) to the predicted one (from the mode itself) ranges from $\GradCalRatioMedianLow$--$\GradCalRatioMedianHigh$. It is below one on every surrogated variable except the
$(2,2)$, where it equals one. This shows that the forecast errs on the
safer side by overstating the error.
However, this has a limitation. Across the
well-covered interior of the parameter space, the predicted variance is nearly
constant. There the forecast gives the right overall size of the error, but it
cannot rank the points by their error. It distinguishes points
only toward the sparsely sampled edges of the parameter space, where the variance
grows.
Figure~\ref{fig:envelope_map} shows this envelope over the
$(q,\chi_{\rm eff})$ plane, with $\chi_{\rm eff}=(q\chi_{1z}+\chi_{2z})/(1+q)$. The predicted mismatch is smallest near equal mass and at moderate spins,
where the training set is dense, and grows toward high mass ratio and large
$|\chi_{\rm eff}|$, where it is sparse.

\paragraph{Physical-units parameter-estimation demonstration.}

\begin{figure*}[tbp]
  \centering
  \includegraphics[width=\linewidth]{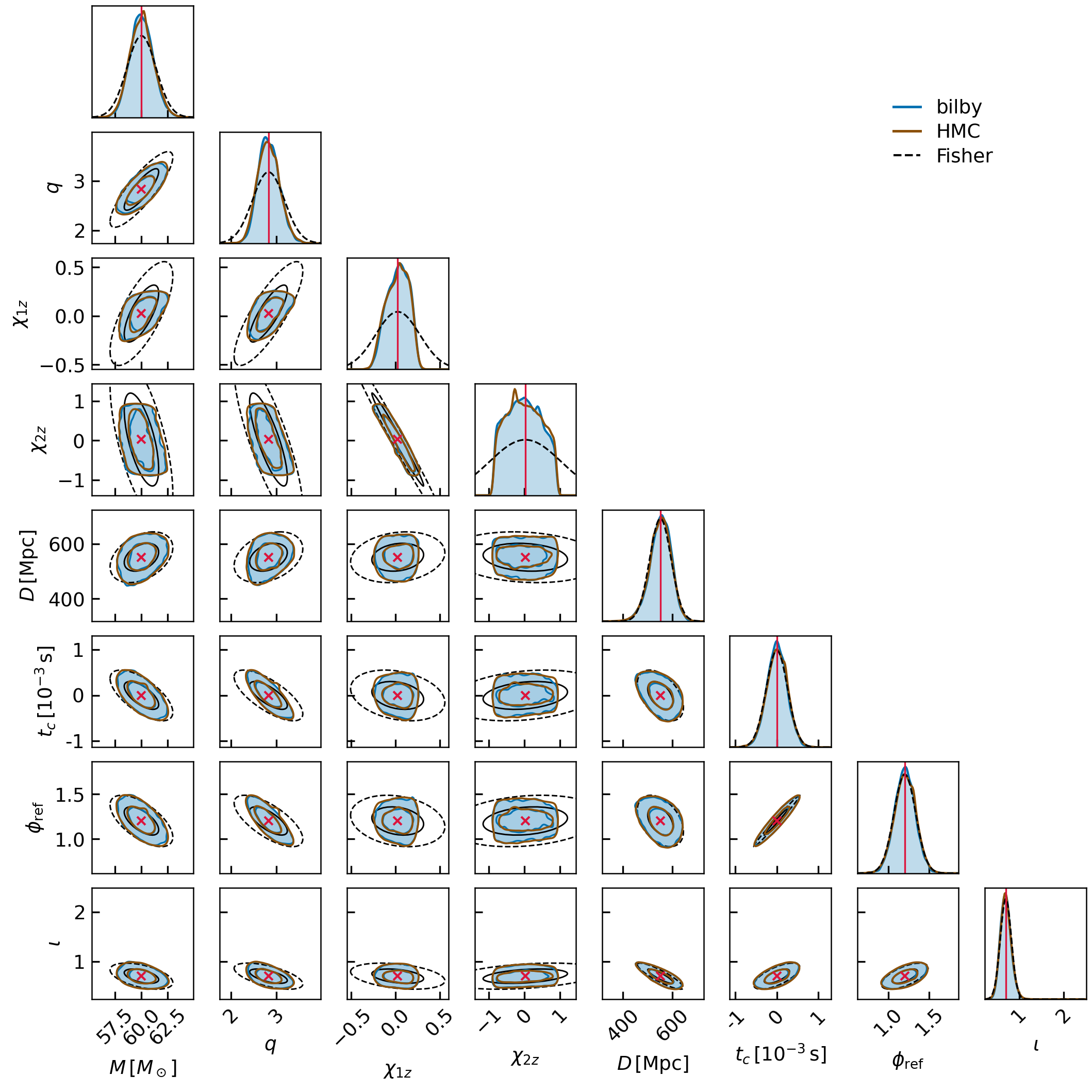}
  \caption{Zero-noise \emph{self}-injection recovery:
  \textsc{bilby}/\textsc{dynesty} (blue) and the gradient-HMC chains (brown)
  overlaid on the Fisher ellipses (black). All three runs are drawn at two enclosed probabilities, $50$ and $90\%$. The Fisher ellipses
  are solid at $50$ and dashed at $90$. The sampler regions are highest-density
  regions of the sampled posterior and the Fisher ellipses are the
  corresponding regions of the Gaussian forecast. The two therefore coincide only
  where the posterior is Gaussian. The crimson cross marks the
  injected values. The agreement between the samplers and the Fisher forecast is quantified in the text. The injected data are generated by the same model that
  is then recovered. These runs do not incorporate the surrogate modeling error envelopes.}
  \label{fig:bilby}
\end{figure*}

We test the model's gradients on one end-to-end parameter-estimation problem. The gradient of the physical strain with respect to the \GradNumPhysicalParamsWord{} parameters ($M$, $q$,
$\chi_{1z}$, $\chi_{2z}$, luminosity distance $D_L$, coalescence time $t_c$, $\phi_{\rm ref}$, $\iota$) is used in three independent checks. The first is a Fisher-matrix forecast of the posterior widths. The second is a simple Hamiltonian Monte Carlo (HMC) sampler developed by the author that uses the Fisher matrix as its mass matrix. The third samples the same likelihood with \textsc{bilby} and the \textsc{dynesty} nested sampler, which uses no gradients. The signal is injected in zero noise, with a total mass of $\GradPEInjMassMsun\,M_\odot$, a network SNR of $\GradPEInjSNR$, and intrinsic parameters at the center of the training range.
The Fisher forecast gives $\sigma_M=\GradFisherSigmaMass\,M_\odot$, $\sigma_q=\GradFisherSigmaQ$, $\sigma_{\chi_{1z}}=\GradFisherSigmaChiOne$, $\sigma_{\chi_{2z}}=\GradFisherSigmaChiTwo$ (the secondary spin is essentially unconstrained at this SNR), $\sigma_{D_L}=\GradFisherSigmaDistMpc$~Mpc, $\sigma_{t_c}=\GradFisherSigmaTcMs$~ms, $\sigma_{\phi_{\rm ref}}=\GradFisherSigmaPhiRef$ and $\sigma_\iota=\GradFisherSigmaIota$, with the expected strong correlations between $t_c$ and $\phi_{\rm ref}$, between $D_L$ and $\iota$, and between $\chi_{1z}$ and $\chi_{2z}$. Both samplers recover the injection. For every parameter, the distance from the posterior mean to the injected value, measured in units of the posterior standard deviation, is at most $\GradHMCMaxBiasSigma$ for the HMC run and $\GradNestedMaxBiasSigma$ for the nested-sampling run. The HMC run accepted $\GradHMCAcceptPercent\%$ of its proposed trajectories and kept $\GradHMCKeptSamples$ samples. Fig.~\ref{fig:bilby} overlays both posteriors on the Fisher ellipses. The two posteriors agree with each other, and with the Fisher forecast to well within one standard deviation, and the nested-sampling marginal widths are comparable to the Fisher widths ($D_L$: $\GradNestedSigmaDistMpc$ vs $\GradFisherSigmaDistMpc$~Mpc, $\iota$: $\GradNestedSigmaIota$ vs $\GradFisherSigmaIotaThreeFig$).

\subsection{Surrogate error in parameter estimation}
\label{sec:pe_error}
For parameter estimation, the relevant question is whether the surrogate's
modeling error biases the inference more than the statistical noise does. This is the relevant regime for current high-SNR
observations, and for future detectors.
A template error $\delta h$ shifts the maximum-likelihood parameters by
$\Delta\theta^i_{\rm sys}\simeq (\Gamma^{-1})^{ij}\,(\partial_j h\,|\,\delta h)$
\citep{Cutler:2007mi}, with $\Gamma$ the Fisher matrix built from the
analytic strain gradients of Sec.~\ref{sec:gradients} against the aLIGO design
sensitivity. The Cauchy--Schwarz inequality bounds the bias in every parameter
by the statistical width $\sigma_i=\sqrt{(\Gamma^{-1})^{ii}}$,
\begin{equation}
  \frac{|\Delta\theta^i_{\rm sys}|}{\sigma_i}\;\le\;\rho\,\sqrt{2\,\mathcal{MM}}
  \;=\;\frac{\rho}{\rho_{\rm crit}},
  \qquad \rho_{\rm crit}\equiv\frac{1}{\sqrt{2\,\mathcal{MM}}},
\end{equation}
uniformly over the \GradNumPhysicalParamsWord{} physical parameters $(M,q,\chi_{1z},\chi_{2z},D_L,t_c,
\phi_{\rm ref},\iota)$. The surrogate's systematic bias therefore stays below
the statistical error for every parameter up to a SNR
$\rho_{\rm crit}$, evaluated here from the exhaustive \AAf{} all-mode
through-ringdown LOO mismatch (all $\GradNumTrainingSims$ simulations, fixed-hyperparameter LOO).

\begin{figure}[tbp]
  \centering
  \includegraphics[width=0.92\linewidth]{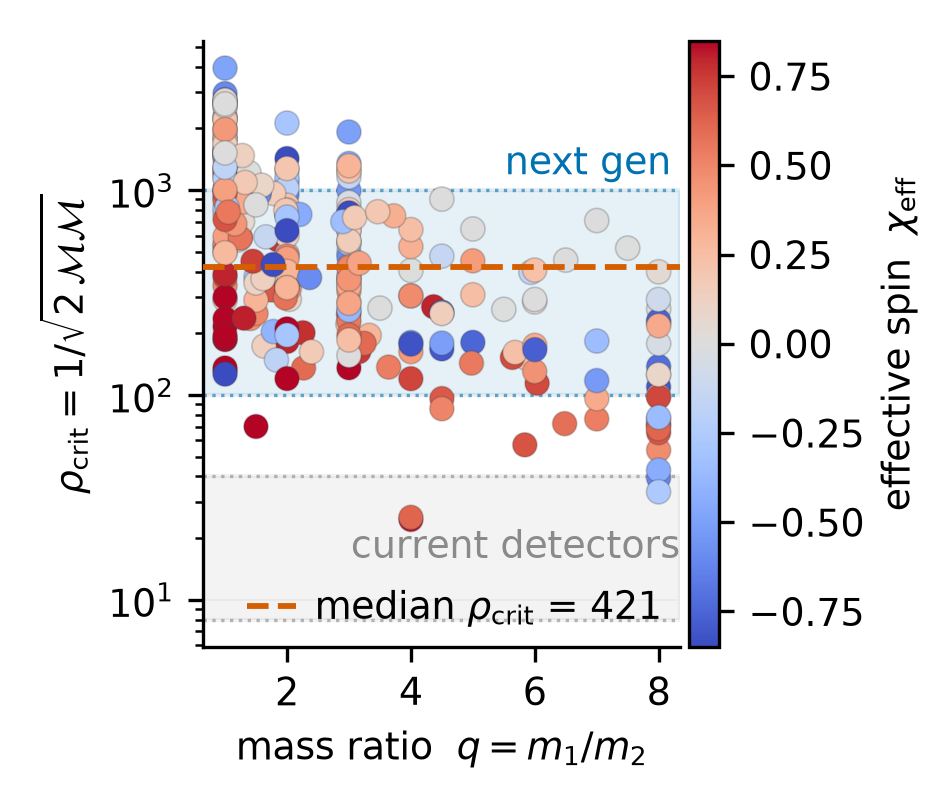}
  \caption{Indistinguishability SNR $\rho_{\rm
  crit}=1/\sqrt{2\,\mathcal{MM}}$ from the all-mode through-ringdown \AAf{} fixed-hyperparameter LOO mismatch over all $\GradNumTrainingSims$ simulations, versus mass ratio, colored
  by effective spin $\chi_{\rm eff}$. The surrogate's modeling error is
  undetectable in parameter estimation below $\rho_{\rm crit}$. The
  current-detector and $3$G signal-to-noise bands are marked, and the dashed
  line is the median, $\rho_{\rm crit}=\GradRhoCritMedian$. The same $\rho_{\rm crit}$ is the
  SNR below which the Cutler--Vallisneri systematic bias stays
  under the statistical $1\sigma$ error in every parameter.}
  \label{fig:pe_indist}
\end{figure}

Across the training set $\rho_{\rm crit}=\GradRhoCritCI$ (median with
the 5th--95th percentile span, from the all-mode fixed-hyperparameter
LOO distribution), falling to $\sim\!\GradRhoCritMin$ only at the worst points of the tail at the edge
of the training range (Sec.~\ref{sec:aa_accuracy}; Fig.~\ref{fig:pe_indist}).
At the current-detector scale ($\rho\lesssim40$) the surrogate is
indistinguishable from NR and its bias is a small fraction of the statistical
error everywhere in the training range. At $\rho=30$ the bias-to-noise ratio is
$\lesssim\!\GradBiasToNoiseAtThirtyPercent\%$ for a median simulation. Only for $\rho\gtrsim\!\GradRhoCritMedianRounded$, at the loud
events expected in third-generation detectors, does the modeling error approach
the statistical floor. In that regime, every current model requires denser NR
coverage, and a better fit alone does not help. These numbers are measured with the unweighted
mode-space mismatch. The median detector-network noise-weighted all-mode LOO mismatch is $\GradNoiseWeightedHarshness$
times the unweighted one, which scales $\rho_{\rm crit}$ by $\GradNoiseWeightedRhoScale$.

 \section{Performance}
\label{sec:performance}

\paragraph{Single-call (CPU) throughput.}
We quote two configurations, the pure NR span and a low-frequency hybridized call that contains both the NR-native and the PN spans,
at the same parameters ($q=\PerfBenchMassRatio$, $\chi_{1z}=\PerfBenchSpinOne$, $\chi_{2z}=\PerfBenchSpinTwo$,
$M=\PerfBenchTotalMass\,M_\odot$, $\Delta t = 1/\PerfSampleRate$~s). Each timing is the warm median over
interleaved (across model variants, to prevent cached evaluations) rounds in fresh single-threaded, physical-core pinned processes (through OMP\_PROC\_BIND).
The full native
NR span ($f_{\rm min}=0$, here $\PerfNativeDurationSec$~s from $\PerfNativeStartFreqHz$~Hz) costs $\PerfNativeAACompiledMs$~ms for the \AAf{}
compiled evaluator and $\PerfNativeAdAMs$~ms for the \AdA{} model on a single core
of AMD EPYC 7742 (2.3 GHz).
LALSimulation \textsc{NRHybSur3dq8}, restricted to the \PerfModeCountWord{} modes we
produce, and forced to the same span (number of orbits), costs $\PerfNativeNRHybSurLowMs$--$\PerfNativeNRHybSurHighMs$~ms on two nodes with the same AMD EPYC 7742 processors.
A $\PerfHybCallFminHz$~Hz call at the same parameters engages the
call-time hybridization and costs $\PerfHybCallAACompiledMs$~ms (against \textsc{NRHybSur3dq8}'s
$\PerfHybCallNRHybSurMs$~ms, $\PerfHybCallSpeedup\times$). We can infer the speedup due to the compiled Kokkos backend by comparing it with the development numpy backend.

The pure NumPy backend, which is the original development
implementation, runs at $\PerfNativeAANumpyMs$~ms on the native span and $\PerfHybCallAANumpyMs$~ms on the
$\PerfHybCallFminHz$~Hz hybridized call.
Analytic parameter gradients on the compiled evaluator cost $\PerfNativeAAGradMs$~ms per call for
\AAf{} and $\PerfNativeAdAGradMs$~ms for \AdA{} on the native span. The \AAf{} NumPy reference
path costs $\PerfNativeAAGradNumpyMs$~ms.

The speed advantage of the compiled backend shows on the long, hybridized cases. On the $q=\PerfLongMassRatio$,
$M=\PerfLongTotalMass\,M_\odot$ reference case ($f_{\rm min}=\PerfLongFminHz$~Hz, $\sim\!\PerfLongDurationSec$~s of
waveform at $\Delta t=1/\PerfSampleRate$~s), the compiled hybrid call costs $\PerfLongAACompiledMs$~ms
against $\PerfLongNRHybSurMs$~ms for \textsc{NRHybSur3dq8} at the same \PerfModeCountWord-mode content
and thread count, $\sim\!\PerfLongSpeedup\times$ faster. The NumPy hybrid configuration costs
$\PerfLongAANumpyMs$~ms (Fig.~\ref{fig:speedbatched}).

\begin{figure}[t]
  \centering
  \includegraphics[width=0.92\linewidth]{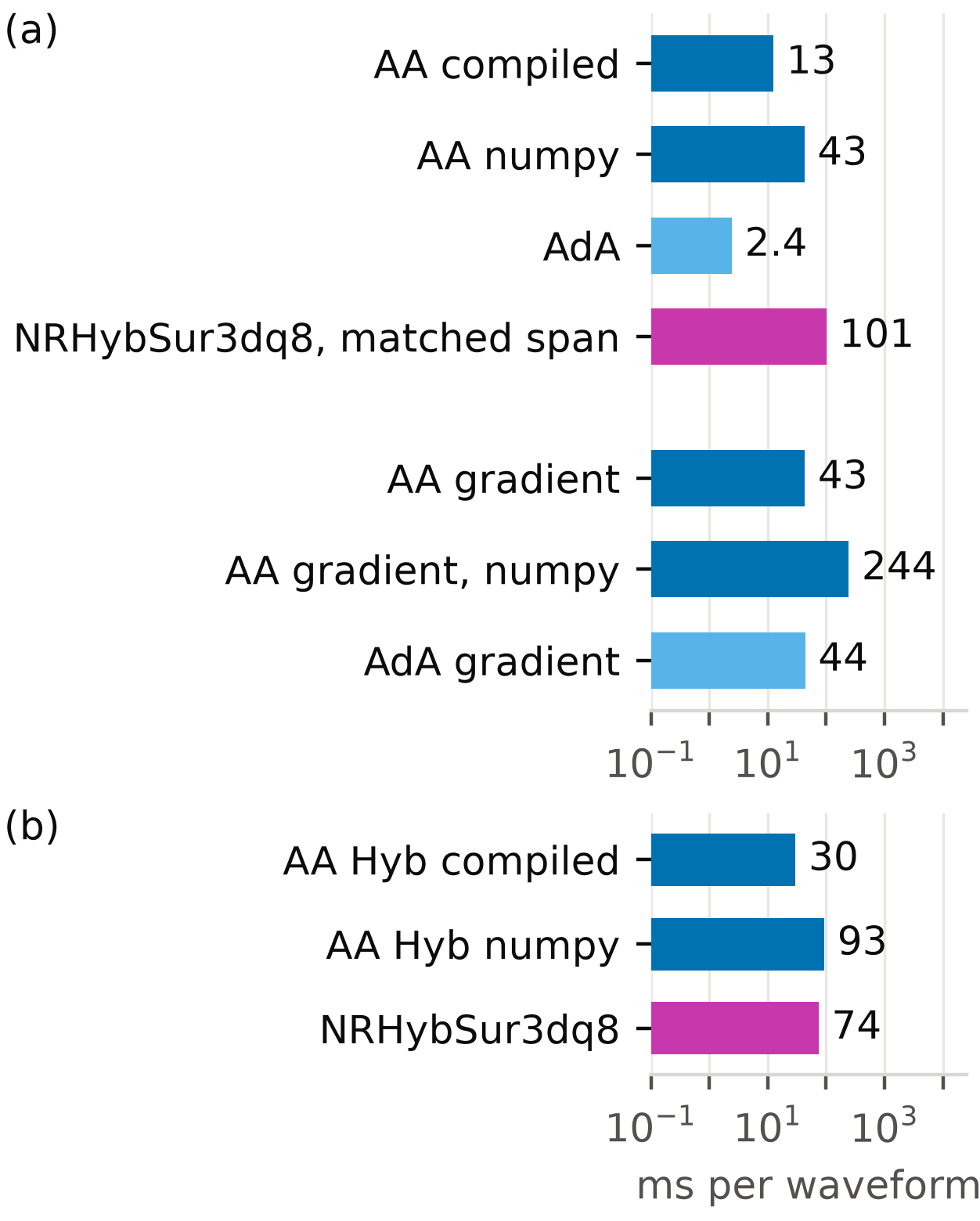}
  \caption{Single-call wall times of the \NRAAS{} family (compiled and
  NumPy backends) against LALSimulation \textsc{NRHybSur3dq8} at matched
  sample rate and \PerfModeCountWord-mode content. Every time is a warm-start median over
  interleaved rounds (alternating between different model flavor calls) in fresh
  pinned single-threaded processes, in milliseconds per waveform.
  Panel~(a) is the pure native NR span ($q=\PerfBenchMassRatio$, $M=\PerfBenchTotalMass\,M_\odot$,
  $f_{\rm min}=0$), with the \textsc{NRHybSur3dq8} row forced to the same
  span. Panel~(b) is a $\PerfHybCallFminHz$~Hz call at the same parameters, which engages the
  call-time hybridization. The scale is shared with
  Fig.~\ref{fig:speedbatched}.}
  \label{fig:speed}
\end{figure}

\begin{figure}[t]
  \centering
  \includegraphics[width=0.92\linewidth]{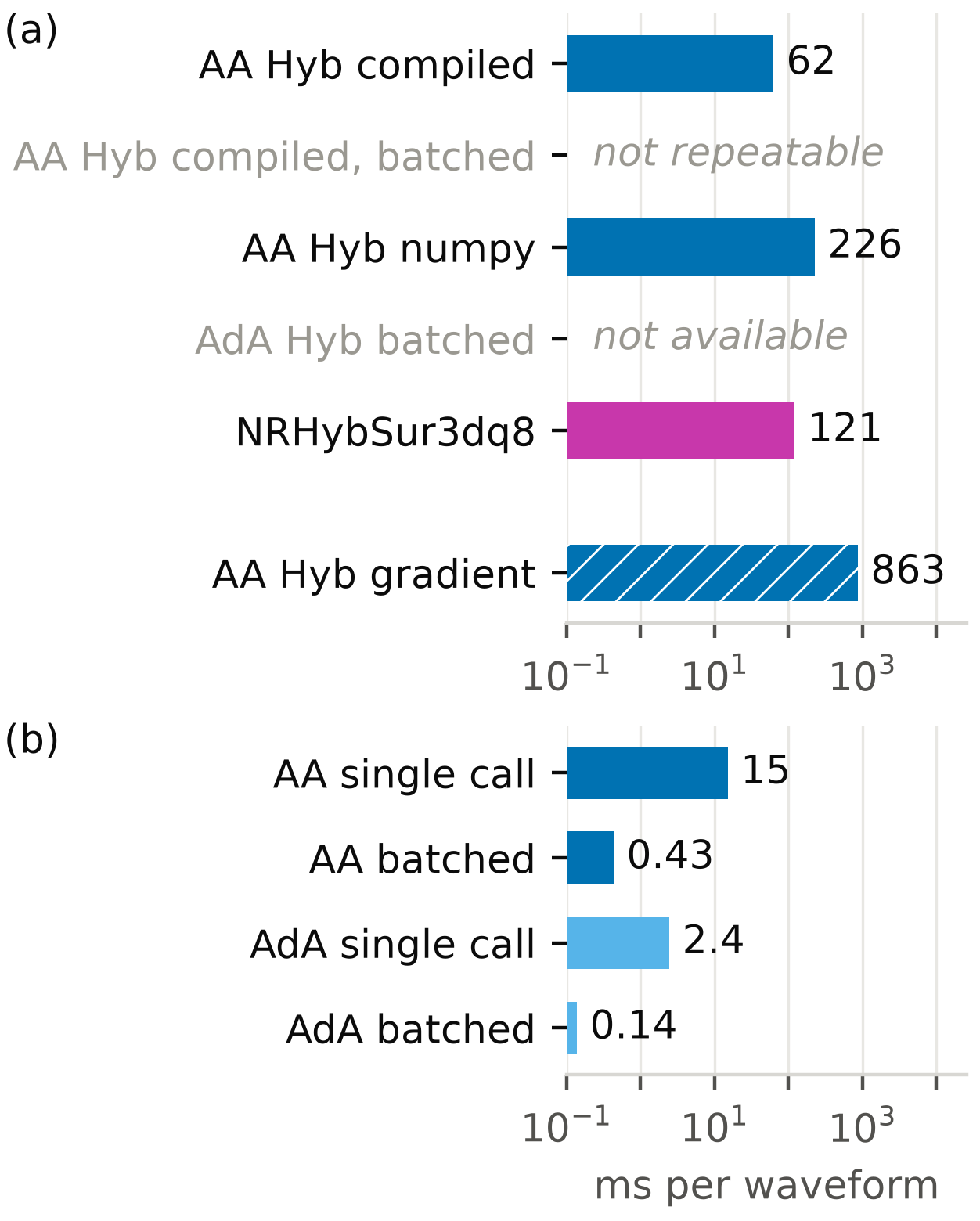}
  \caption{The long hybridized case and the batched evaluation times,
  with the same measurement conventions and x-scale as
  Fig.~\ref{fig:speed}. Panel~(a) is
  the long hybridized case ($q=\PerfLongMassRatio$, $M=\PerfLongTotalMass\,M_\odot$, $f_{\rm min}=\PerfLongFminHz$~Hz,
  $\sim\!\PerfLongDurationSec$~s of waveform). Its batched row is not shown because the measurement spread over many observations was high. The
  hatched gradient bar is an upper bound (no compiled path exists for
  the hybrid gradient). Panel~(b) is batched throughput on a $\PerfBatchedThreads$-thread
  exclusive node, measured on the trimmed $\PerfBatchedFminHz$~Hz configuration at
  $M=\PerfBenchTotalMass\,M_\odot$ and shown beside its own single-call reference bars. Each
  batched bar is the per-waveform times over the largest batch size, whose per-waveform
  time repeats across interleaved rounds (alternating between models) to within $\PerfDriftLimitPct\%$. The batch size is $\PerfBatchedAABatch$ for both models.}
  \label{fig:speedbatched}
\end{figure}

\paragraph{Batched (Kokkos) evaluator.}
This package is manually tuned for maximizing performance and parallel efficiency.
A pybind11/Kokkos evaluator (CUDA and OpenMP execution spaces) contracts the GPR
to coefficients via stacked general matrix--matrix multiplications, and fuses the $\exp(\mathrm{i}\,m\,s\,\Phi)$
rotation in the kernel. On the \AAf{} model, batching the standard case on a
$\PerfBatchedThreads$-thread CPU node (exclusive, warm medians across interleaved rounds) takes
the per-waveform cost from $\PerfBatchedBatchOneMs$~ms at batch $1$ to $\PerfBatchedAAMs$~ms at batch $\PerfBatchedAABatch$
(Fig.~\ref{fig:speedbatched}),
a $\PerfBatchedAmortization\times$ speedup relative to single-call evaluation measured on the same
grid (the trimmed $\PerfBatchedFminHz$~Hz configuration, $\PerfBatchedSingleCallMs$~ms single-threaded). The
batch-$1$ point already performs better than a single call, for two reasons. The
batched evaluator runs on all $\PerfBatchedThreads$ threads and holds its initialization state across
calls, while the single-call row is single-threaded and repeats its setup
on every call. The
$\PerfBatchedAmortization\times$ compares batched throughput against the single-threaded
path. Within the $\PerfBatchedThreads$-thread path, batching alone accounts for $\PerfBatchedBatchingGain\times$
of it. The \AdA{} model takes $\PerfBatchedAdAMs$~ms per waveform at batch $\PerfBatchedAdABatch$, and the timings for every batch
size up to $\PerfBatchedAdABatch$ vary by at most $\PerfDriftLimitPct\%$ across rounds.

\paragraph{Sampling-rate independence.}
\AAf{} evaluation is pointwise in time. The model stores continuous
functions (piecewise Chebyshev in $\Phi$, paired with the time map), and each
requested sample is an exact evaluation of those functions at that time.
Arbitrary \texttt{times=} grids are supported. The model has no internal time
grid, integration step, or resampling. Therefore, the requested sampling rate
changes only which points are evaluated.
Coarse and fine grids are measured to be bitwise identical at shared samples.
Therefore, no part of the model error is sourced by the sampling rate, and the in-sample and LOO mismatches of Sec.~\ref{sec:aa_accuracy} account for the entire model error budget
at any sampling rate.

\paragraph{Mass range and coverage.}
The common phase domain, which is used for the spectral representation, spans about $\PerfNativeSpanOrbits$ orbits before merger. It is the
intersection of the available phase span of the peak-aligned waveforms of the training set. This means that the shortest simulation currently sets the earliest usable phase, and the longer simulations will have to shorten their early inspiral. This shortening is a median of $\PerfShorteningMedianOrbits$ orbits and up to $\PerfShorteningMaxOrbits$ for the
longest of the simulations available in the training set. The lightest usable total mass at a fixed lower frequency is the mass
at which the starting $(2,2)$ frequency equals $f_{\rm low}$. At $f_{\rm
low}=\PerfMassRangeFlowHz$~Hz, this minimum total mass ranges from $\PerfMinMassLow$ to $\PerfMinMassHigh\,M_\odot$ across the
aligned-spin parameter range, with a median of $\PerfMinMassMedian\,M_\odot$. The high-spin, high-$q$ corner
sets the upper bounds on the mass, because those binaries inspiral slowly and complete $\PerfNativeSpanOrbits$
orbits before merger at a higher frequency. For comparison,
\textsc{NRSur7dq4}~\cite{Varma:2019csw} fixes a $\PerfNRSurWindowM\,M$ time window of
$\PerfNRSurOrbitsLow$ to $\PerfNRSurOrbitsHigh$ orbits over its $q\le4$ range, which gives minimum masses of $\PerfNRSurMinMassLow$ to $\PerfNRSurMinMassHigh\,M_\odot$ at
the same lower frequency (Table~\ref{tab:massrange}). These minimum masses
apply only to the \emph{native} NR span. The call-time action hybridization of
Sec.~\ref{sec:hybridization} extends any evaluation to an arbitrarily low
starting frequency, removing the minimum-mass restriction at the cost of
using PN instead of NR waveforms in the extension. The NR span is not
modified. This differs from training-data hybridization, as in
\textsc{NRHybSur3dq8}~\cite{Varma:2018mmi}. As this method models the waveform in the phase domain, it may be amenable to reusing the entire
lengths of all NR waveforms.

\begin{table}[t]
\centering
\caption{Waveform duration and the lightest usable total mass at $f_{\rm
low}=\PerfMassRangeFlowHz$~Hz for the non-hybridized model. \NRHJ{}'s coverage is the intersection of the training set's
peak-aligned phase ranges, while \textsc{NRSur7dq4} fixes a $\PerfNRSurWindowM\,M$ time
window.}
\label{tab:massrange}
\begin{tabular}{lccc}
\toprule
Model & $q$ range & orbits & $M_{\min}/M_\odot$ \\
\midrule
\NRHJ{} (this work) & $\le 8$ & $\sim\!\PerfNativeSpanOrbits$ & $\PerfMinMassLow$--$\PerfMinMassHigh$ (med.\ $\PerfMinMassMedian$) \\
\textsc{NRSur7dq4}  & $\le 4$ & $\PerfNRSurOrbitsLow$--$\PerfNRSurOrbitsHigh$ & $\PerfNRSurMinMassLow$--$\PerfNRSurMinMassHigh$ \\
\bottomrule
\end{tabular}
\end{table}
 \section{Software and availability}
\label{sec:software}

\subsection{Implementation}
\label{sec:implementation}

\NRHJ{} is developed in \textsc{Python} and implemented in \textsc{C++}, with a \textsc{Python} user interface.
The \textsc{C++} code is compiled into a shared library and loaded at runtime. The Python code paths
are the primary development target, and the \textsc{C++} code was added later to accelerate and optimize,
and to enable performance portability to high-performance computing environments. Owing to this Python-based development
approach, which has its own validation criteria, the validation of the \textsc{C++} code is done by comparing the outputs
of the \textsc{C++} and \textsc{Python} code paths, and ensuring that they are identical up to machine precision.
The \textsc{C++} code is also validated in-sample against the original training data, and the results are consistent with the validation
of the \textsc{Python} code path.

Kokkos is used to provide a performance-portable implementation of the \textsc{C++} code, enabling execution on CPUs and GPUs.
Various hand-coded optimizations are implemented, including fused multiply add (FMA), the use of vectorization and dense linear algebra, and minimum interpolation schemes.

SymPy~\cite{Meurer:2017yhf} has been used in generating the code for the analytic PN expressions. The
binding energy and the fluxes are transcribed from PostNewtonian.jl~\cite{PostNewtonianjl} and checked against that
package, and the angular momentum derived from them is checked against the closed-form expression of
Ref.~\cite{Blanchet:2024mnz}. The
mode-amplitude normalizers of Sec.~\ref{sec:radiative} are generated
from the Mathematica ancillary file of Ref.~\cite{Henry:2022ccf}, and checked against an independent evaluation
of that file. The 4PN $(2,2)$ term is taken from Ref.~\cite{Blanchet:2023sbv}. Symbolic derivatives
of the PN expressions are also generated using the same validated expression in SymPy, and validated against
finite-difference derivatives. These derivatives are used to construct the model, and also for dispatching
the gradients of waveforms. Spectral derivatives, transforms and spectral decay indicators are provided by the \textsc{spectools} library, developed by the author. No JAX / automatic-differentiation / XLA libraries are used in the code,
and all derivatives are computed analytically, handled case by case
for potential ill-conditioning, and validated against finite-difference derivatives.

The \textsc{C++} code is compiled with \textsc{CMake}, and the \textsc{Python} interface uses \textsc{pybind11}.
The package has been released on pypi with various installation modes. This includes compilation from source,
deployment on GPUs, and choice of optimization flags. By default, the compiled backend is recommended using the
`--no-binary` argument to pip. In the release, although we recommend compiling the package from source so that
the binary is adapted to the hardware architecture, we also supply precompiled binaries for Windows, Mac, and Linux systems
 compiled at baseline hardware capabilities, which would not deliver the speeds of the compiled backend.

\subsection{Public interface}
\label{sec:interface}

The package provides the model family through one Python object. It returns the
modes in geometric units on any requested time grid (\texttt{times=}). Its method
\texttt{physical()} returns the waveform in physical units: masses in $M_\odot$, time
step in s, $f_{\rm min}$ and $f_{\rm ref}$ in Hz, and distance in Mpc, with the mode
conventions of the LIGO Algorithm Library (LAL). The polarizations $h_+$ and $h_\times$
are formed with spin-weighted spherical harmonics that agree with LAL to
$\SoftSWSHLALTolerance$. It also provides timing benchmarks for each variant, the
\texttt{dynamics} output $(J,E,\omega)$, and the remnant map
(Appendix~\ref{sec:remnant}). Public Python APIs are available
for the evaluation of waveforms and of their gradients, each with an
error envelope, for the evaluation of the conservative dynamics of the inspiral with
their own envelopes, and for the evaluation of the remnant mass and spin.
Gradients of the remnant map are computed internally, where the
merger--ringdown gradient chain needs them, but are not part of the public
interface.

The evaluator adopts the NRSur/LAL conventions. $f_{\rm low}=0$ (or
\texttt{None}) corresponds to the full NR native span, as in \texttt{ChooseTDModes}.
Setting $f_{\rm ref}=0$ places the phase reference at $f_{\rm min}$ (at the first sample
when $f_{\rm min}$ is itself $0$). Aligned spins are constants of the
evolution. Therefore, $f_{\rm ref}$ sets only the phase convention ($\arg
h_{22}=-2\phi_{\rm ref}$ at $f_{\rm ref}$). The hybridization of
Sec.~\ref{sec:hybridization} is controlled by the option \texttt{hybridization}, which
selects a code path for a given $f_{\rm low}$. With \texttt{"extend\_if\_below\_reach"}
(the default), a request above the native starting frequency is trimmed from the NR-based waveform,
and the PN extension is used only below it. With \texttt{"never"}, the model returns only NR-informed content. If $f_{\rm low}$ is below
the native starting frequency, it raises an error that states the lowest $f_{\rm low}$ it can
return without the PN extension.
\texttt{"require"} raises an error when no PN extension can be made.

\subsection{Availability}
\label{sec:availability}

The evaluation code is released under a modified PolyForm Noncommercial 1.0.0
licence as the Python package
\href{https://pypi.org/project/nrhjsurrogate/}{\texttt{nrhjsurrogate}}~\cite{nrhjsurrogate},
installable from the Python Package Index with
\texttt{pip install nrhjsurrogate}. The documentation is at
\href{https://psu-edu.github.io/nrhjsurrogate/}{\texttt{psu-edu.github.io/nrhjsurrogate}}.
The code that trains a model is not part of the package.
The package is published as a source distribution. Therefore, an
installation can build the \textsc{Kokkos} evaluator against the user's own
compiler, fetching and building the pinned \textsc{Kokkos} release if none is locally available. An installation on a machine with no suitable compiler still
succeeds and uses either the complete \textsc{NumPy} backend, or a precompiled binary.

The model files are distributed separately from the code, and are an order of magnitude larger than the
package. They are deposited at
\href{https://zenodo.org/records/22262339}{\textsc{Zenodo}} under
\href{https://doi.org/10.5281/zenodo.22262339}{10.5281/zenodo.22262339}, with the
GPL-3.0-or-later licence. The record
holds the two generations described here: the
\NRHJ{}\_\AAf{}, \NRHJ{}\_\AdA{} and the shared merger--ringdown model
\NRHJ{}\_MR, together with the preceding generation of the same three. The
package resolves a model file by name and will fetch it from that record on
first use, or load it from a specified directory. Each file's SHA-256 is
pinned inside the package and checked against the download. A corrupted or
substituted model file is refused. The command
\texttt{nrhjsur-verify} checks a local installation against reference values shipped with the
package, and is the check users can run before using the model.
 \section{Discussion: limitations and extensions}
\label{sec:discussion}

\begin{figure*}[tbp]
  \centering
  \includegraphics[width=\linewidth]{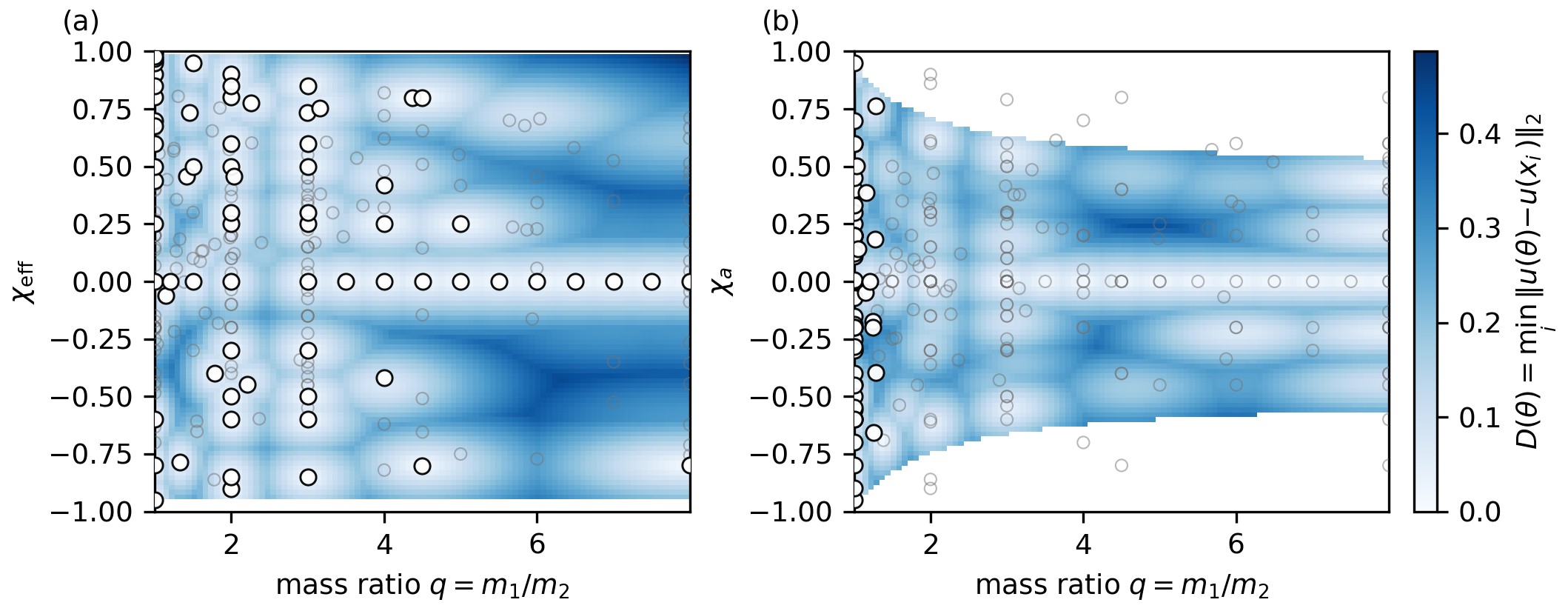}
  \caption{Map of how far each point in the parameter space lies from the
  nearest training simulation, on the $(q,\chi_{\rm eff})$ and $(q,\chi_a)$
  planes. Distance is measured in the (mass-ratio, spin) parameters,
  $D(\theta)=\min_i\|u(\theta)-u(x_i)\|_2$, with $u$ the $(\log q,\chi_{1z},\chi_{2z})$
  coordinates each scaled linearly to $[-1,1]$ over the training range. The $\GradNumTrainingSims$ training simulations are overlaid. Those with black edges lie on the displayed plane, and the grey ones are projected onto it. Dark regions indicate gaps in training coverage, and the darkest point marks where the next simulation would most reduce model error.}
  \label{fig:coverage_map}
\end{figure*}

The error envelope, a GPR fitted to the fixed-hyperparameter LOO mismatches, predicts that the model would benefit from denser catalog
coverage at high mass ratio and at the sparse spin corners
(Fig.~\ref{fig:envelope_map}). The intrinsic modeling-error metric can
therefore guide where new simulations should be placed to improve the model's
accuracy the most. The distance to the nearest
training simulation in the model's parameter-space metric
(Fig.~\ref{fig:coverage_map}) makes these coverage voids explicit, and the analytic gradients of
Sec.~\ref{sec:gradients} allow the placement to be driven by the metric of the intrinsic
waveform manifold, and predicted uncertainty directly.

\begin{figure*}[tbp]
  \centering
  \includegraphics[width=\linewidth]{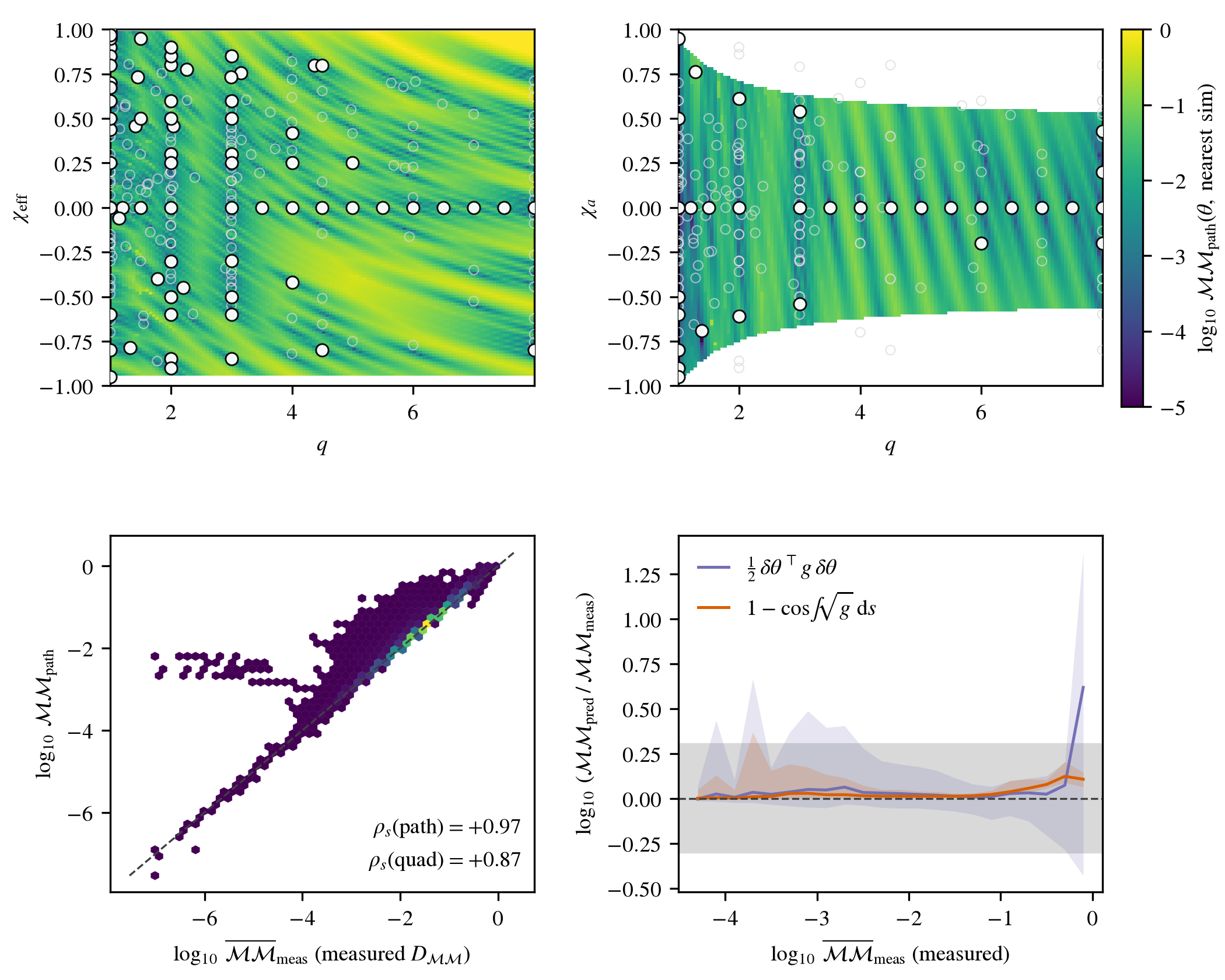}
  \caption{Equivalent-mismatch distance predicted from the model's own
  parameter-space metric, and its validation against direct measurement with NR. \textbf{Top:}
  the predicted mismatch $\mathcal{MM}_{\rm path}$ of Eq.~\eqref{eq:pathmm}
  between each grid node and its nearest training simulation, integrated along
  the straight coordinate path in $(\log q,\chi_{1z},\chi_{2z})$, on the two
  planes of Fig.~\ref{fig:coverage_map}. Unlike the measured field~\eqref{eqn:mf}, this is defined between arbitrary configurations and needs no further waveform
  evaluation. \textbf{Bottom left:} prediction against the measured
  $D_{\mathcal{MM}}$ at every node. The dashed line is equality, and the two rank
  correlations compare the path integral with the single-point quadratic form
  $\tfrac12\delta\theta^{\top}g\,\delta\theta$. \textbf{Bottom right:} the ratio
  of predicted to measured mismatch, binned along the measured axis. The bands
  are the 25-75\% spread within each bin, and the gray strip marks a factor of
  two ($\times 2$ and $\times 0.5$ on the log scale).
  The quadratic form tracks the measurement up to $\mathcal{MM}\sim\DiscMetricQuadDepartMM$ and then
  grows without bound, while the measured mismatch cannot exceed one (its largest value here is $\DiscMetricMeasuredMMMax$). The
  path integrated mismatch remains flat across the full range. All projected metric values at
  $\DiscMetricNumPositiveDefinite$ grid points are positive-definite and all $\DiscMetricNumPathIntegrals$ path
  integrals converged. The metric is derived from the model itself. Therefore, where the model is inaccurate, its metric is also inaccurate. The high-aligned-spin corner is both the most sparsely covered region, and the region where this limitation is largest. The $\GradNumTrainingSims$ training simulations
  are overlaid on the field panels: white markers with black edges lie on the
  displayed plane ($|\chi_a|<0.06$ left, $|\chi_{\rm eff}|<0.06$ right),
  hollow gray markers are projections.}
  \label{fig:wf_metric_equiv_mismatch}
\end{figure*}

Waveform-space coverage can also be measured directly, by evaluating the model on a
grid over the two planes of Fig.~\ref{fig:coverage_map}, the $(q,\chi_{\rm eff})$
and $(q,\chi_a)$ slices, and recording, at each node, the smallest all-mode mismatch
against the raw NR waveforms of the training set,
\begin{equation}
D_{\mathcal{MM}}(\theta)=\min_i\,\mathcal{MM}[\mathrm{AA}(\theta),
\mathrm{NR}(x_i)]     \label{eqn:mf}
\end{equation}
($\DiscCoverageGridSide\times\DiscCoverageGridSide$ per plane, minimum over all $\GradNumTrainingSims$
simulations). That mismatch field can be defined at any location where the waveform model can be evaluated.
However, we take a better approach, using the model's own intrinsic parameter-space
metric. Following the template-metric construction of
Refs.~\cite{Owen:1995tm,Balasubramanian:1995bm}, the mismatch between two nearby
configurations $\theta$ and $\theta+\delta\theta$, computed with the same
inner product $\langle\cdot|\cdot\rangle$ as the measured
$D_{\mathcal{MM}}$ field, is quadratic at leading order,
\begin{align}
  \mathcal{MM}(\theta,\theta+\delta\theta)&=
  \tfrac{1}{2}\,g_{ij}(\theta)\,\delta\theta^{i}\delta\theta^{j}
  +\mathcal{O}(\delta\theta^{3}),
  \label{eq:wfmetric}\\
  g_{ij}&=\left.\langle\partial_{i}h|\partial_{j}h\rangle/
  \langle h|h\rangle\right|_{\perp},\nonumber
\end{align}
where $\partial_{i}$ is the analytic derivative of the surrogate strain
with respect to the parameters
$\theta=(\log q,\chi_{1z},\chi_{2z})$ (Sec.~\ref{sec:gradients}) and
$\perp$ denotes the projection that removes the relative time- and
phase-shift directions from the parameter derivatives. Once $g_{ij}$ is
tabulated, the mismatch between \emph{any} two nearby configurations
can be computed without further waveform evaluations. The metric defines a line
element on the waveform manifold, and a finite separation accumulates
along a path $\gamma$ as an angle,
\begin{equation}
  \mathrm{d}s^{2}=g_{ij}\,\mathrm{d}\theta^{i}\,\mathrm{d}\theta^{j},
  \qquad
  \mathcal{MM}_{\rm path}=1-\cos\!\int_{\gamma}\!\mathrm{d}s,
  \label{eq:pathmm}
\end{equation}
which is bounded and reduces to Eq.~\eqref{eq:wfmetric} at small
separation. Geodesics on a curved manifold are generally not straight lines. Therefore, the statement that
$\gamma$ is the straight coordinate path in $(\log q,\chi_{1z},\chi_{2z})$ is
part of the definition. The metric is
strongly anisotropic (median eigenvalues $\DiscMetricEigMedianSoft$, $\DiscMetricEigMedianMid$ and $\DiscMetricEigMedianStiff$,
condition number $\DiscMetricConditionMedian$). Therefore, a scalar volume element $\sqrt{\det g}$
describes it poorly. At $q\gtrsim6$ its soft (least sensitive) direction is almost exactly the
secondary spin, a known degeneracy that the metric recovers independently.
Integrating Eq.~\eqref{eq:pathmm} along the straight coordinate path to
each node's nearest training simulation reproduces the measured field~\eqref{eqn:mf} at
a Spearman $r_s=\DiscMetricSpearmanPath$ (Fig.~\ref{fig:wf_metric_equiv_mismatch}), against $r_s=\DiscMetricSpearmanQuad$ for
the single-point quadratic form of Eq.~\eqref{eq:wfmetric},
which grows without bound and departs from the measurement above
$\mathcal{MM}\sim\DiscMetricQuadDepartMM$ while the bounded form does not. Since any path is at
least as long as the geodesic, the straight-line integral is a conservative
upper bound on the true separation.

The accuracy of the model near equal mass is also limited by systematics of
the NR training data: the resolution error of the higher modes
(Sec.~\ref{sec:aa_accuracy}), the frame contamination of the catalog, and the
exchange asymmetry of mirror configurations (Sec.~\ref{sec:symmetry_measured}).
The NR training data are themselves asymmetric at the level measured in
Sec.~\ref{sec:symmetry_measured}. Therefore, a model that reproduced them exactly would be
equally asymmetric. \textsc{NRHybSur3dq8} and \textsc{NRSur7dq4}, which are constructed
differently from \NRHJ{}, show asymmetries of the same order
(Sec.~\ref{sec:symmetry_measured}). We attribute the asymmetry to the simulations and their
post-processing. It limits the accuracy that any regression can achieve for the $(3,2)$
and $(4,4)$ modes near equal mass.

The disagreement between exchange partners also provides
a model-independent error estimate for the training data, obtained without any
convergence series. We report the asymmetry here as a limitation, and expect to
remedy it in
future work. Symmetrizing the training set (averaging exchange partners, or
equivalently mirroring them into $\log q<0$ before fitting) restores the
symmetry and averages two independent numerical realizations of the
same physics. It should therefore improve the accuracy instead of trading against it. We have not
adopted it here because it requires retraining and its cost in fixed-hyperparameter LOO mismatch has not been measured.

The next step is to extend this representation to precessing and
eccentric simulations, which will require extending the parameter space and
training set.
 \section{Conclusions}
\label{sec:conclusions}

We have presented \NRAAS, a family of NR
surrogates, currently for non-eccentric aligned-spin binary black holes, to be extended to more generic systems. The \AAf{} model uses a fit inspired by the action-angle representation of binary dynamics, using the azimuthal action $\Jphi(\Phi)$, the conservative numerical Hamiltonian $E(\Phi)$, the clock $\tau(\Phi)$, and the NR-to-PN
radiative ratios, on shared Chebyshev elements. They are regressed
across the scaled $(\log q,\chi_s,\chi_a)$ coordinates with a Gaussian process. From these
fits we derive fast, accurate hybridized waveforms. The first
law~\cite{LeTiec:2015kgg,Fujita:2016igj} and PN mode structure yield a central
\AAf{} model accurate, against held-out NR over the full
IMR span, to a median fixed-hyperparameter LOO $(2,2)$ mismatch of $\AlBlendedLooMismatchTwoTwo$ and an
all-mode $\AlBlendedLooMismatchAllModes$ (medians with \FrontSpanLowerPercentile--\FrontSpanUpperPercentile{} percentile
spans).
The model evaluates a full
IMR waveform in $\FrontSingleCallMs$~ms in a single CPU call and
$\FrontBatchedMsPerWaveform$~ms per waveform in batched evaluation. GPU capabilities are provided through a Kokkos-enabled compiled backend. Hybridization to arbitrarily low
starting frequency is a call-time flag that extends the action itself along 4PN,
leaving the NR span structurally untouched. The inspiral/merger--ringdown split is derived
by an adiabaticity-breakdown diagnostic, and the
hybridized dynamics $(J,E,\omega)$ ship as a product alongside the
strain. Beyond accuracy and speed, the model supplies analytic parameter
derivatives fused into the batched evaluator, a calibrated predictive
uncertainty, and a validated remnant map, which together enable differentiable,
uncertainty-aware parameter estimation in physical units.

We also show how the nearly analytic derivatives of the waveforms, along with the intrinsic metric of the waveform model manifold, can be used to identify gaps in the training data. This will be valuable in future NR simulation campaigns to build accurate waveform models for next-generation gravitational wave detections.

With the robustness of the aligned spin non-eccentric model established, future work will build on the methods presented here, extending it to eccentric and precessing systems.

\section{Use of AI and LLMs}

The conceptual ideas behind this work, the details of the implementation, and the outline of the code design were conceived without the use of artificial intelligence (AI) and large language models (LLMs). The numerical methods and tools used here were developed by the author previously without the use of AI/LLMs for other purposes, and a preliminary implementation of the code was authored in Python. AI-assisted coding with Claude (Anthropic) was used only for later tuning of the GPR fits, the Kokkos implementation, writing plotting scripts, and automating testing and release of the software.

LLMs were used to validate the equations, figures, and tables and their labels, and to edit the drafted manuscript.

\section{Acknowledgments}

The author thanks Jan Steinhoff, B.S.Sathyaprakash, and Viviana Caceres for comments and feedback.

This research is supported by National Science Foundation grants Nos. PHY-2608388, AST-2307147, AST-2606673, PHY-2608011, PHY-2308886, PHY-
2309064, and PHY-250294. This work was carried out in part at ICTS-TIFR with support of the Department of Atomic Energy, Government of India, under project no. RTI4001.

\appendix
\section{The remnant map}
\label{sec:remnant}

The exchange-invariant remnant map $\rho:\theta\mapsto(M_f,\chi_f)$ that is used to construct
 the merger--ringdown element (Sec.~\ref{sec:mr}) is also exposed as a public
\texttt{remnant()} API, and we validate it standalone against the SXS catalog
metadata over the $\GradNumTrainingSims$ production simulations. Those points are the
map's training set. Therefore, we quote the errors of the fixed-hyperparameter LOO refit of the map over all $\GradNumTrainingSims$ simulations as a conservative measure. This yields absolute errors of $\RemMfAbsErrLOOCI$ in
$M_f/M$ and $\RemChifAbsErrLOOCI$ in $\chi_f$. Remnant properties have also been modelled by dedicated
surrogates~\cite{Varma:2018aht,Boschini:2023ryi,Thomas:2025rje}.

\begin{figure}[tbp]
  \centering
  \includegraphics[width=0.92\linewidth]{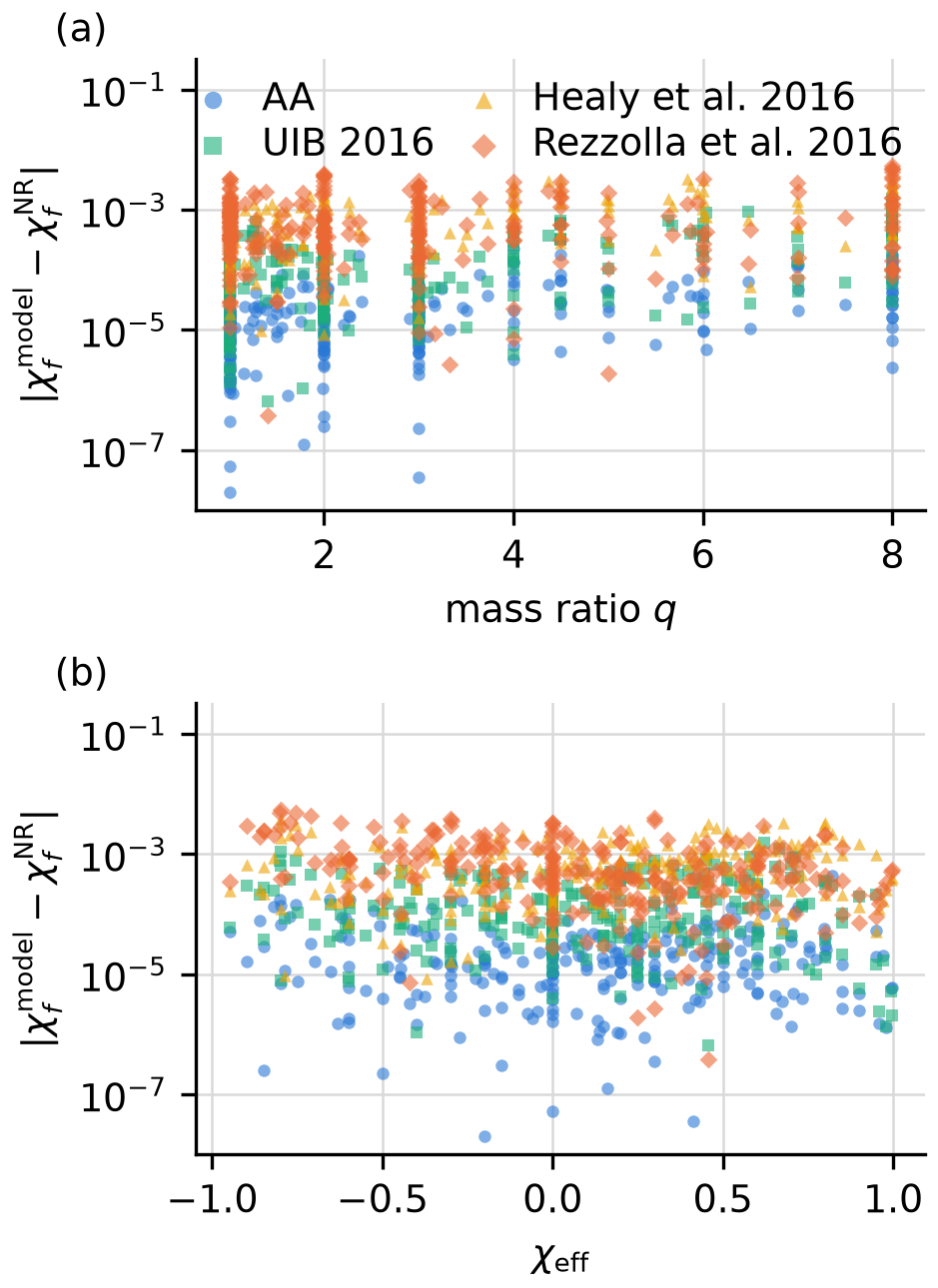}
  \caption{Validation of the remnant map over the $\GradNumTrainingSims$ production
  simulations, $\chi_f$ sector. Absolute $\chi_f$ error of the
  LOO map (\AAf{}) and of three published aligned-spin fits (UIB
  2016~\cite{Jimenez-Forteza:2016oae}, Healy et al.\ 2016~\cite{Healy:2016lce}, Rezzolla et
  al.\ 2016~\cite{Hofmann:2016yih}), evaluated at the same $\GradNumTrainingSims$ parameter
  points, against the SXS metadata, as a function of mass ratio~(a) and of $\chi_{\rm eff}$~(b).}
  \label{fig:remnant}
\end{figure}

Figure~\ref{fig:remnant} shows the LOO $\chi_f$ errors against the SXS metadata beside three
published aligned-spin remnant fits evaluated at the same parameters. One catalog point, SXS:BBH:1124 ($q=1$, $\chi=(+0.998,+0.998)$), is
missed in $M_f$ by $\sim\!\RemOutlierMfMiss$ by \emph{every} predictor, ours and
all three published fits, while all of them match its $\chi_f$. The most likely
explanation is that the metadata value is the outlier there.
The reference is the catalog's remnant metadata. The map predicts the remnant mass
and spin only. It does not model the remnant \emph{kick}.

 \section{Inclination dependence of the cross-model comparison studies}
\label{app:iota}
\enlargethispage{2\baselineskip}

\begin{figure*}[tbp]
  \centering
  \includegraphics[width=\linewidth]{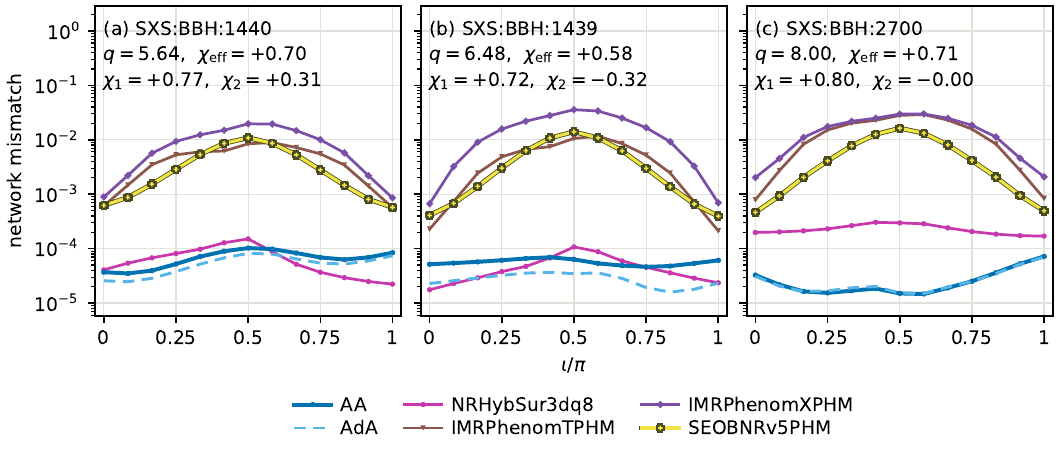}
  \caption{All-mode coherent-network mismatch against NR versus inclination at three
  high-$q$, high-spin configurations, $\IotaNumInclinations$ inclinations each. The variation of
  the mismatch between the face-on and edge-on extremes is quantified in the
  text. The figure
  shows the variation of the models' accuracy with inclination. The three simulations shown,
  SXS:BBH:2700, SXS:BBH:1440 and SXS:BBH:1439, are among those where \AAf{} is least accurate in fixed-hyperparameter LOO, and they are in its training set.
  Therefore, the ordering of the models here does not hold across the catalog.
  The surrogates (\AAf{}, \AdA{} and \textsc{NRHybSur3dq8}) are restricted to
  the modes of \NRHJ{}. IMRPhenomXPHM-SpinTaylor, IMRPhenomTPHM and
  SEOBNRv5PHM are used through their polarizations $h_+$ and $h_\times$, which include
  all the modes each model has, and are not restricted to the same mode set. Any mode present
  in the NR waveform that such a model does not produce contributes its full power to the
  mismatch. Therefore, part of their mismatch comes from missing modes rather than
  modeling error. This part grows towards an edge-on configuration, where the subdominant modes carry
  more power. Total mass is set to $M=\IotaTotalMassMsun\,M_\odot$.}
  \label{fig:iota}
\end{figure*}

\paragraph{Inclination faithfulness.}
Computing the faithfulness with the sky position and polarization angle held
fixed~\cite{Harry:2017weg}, and varying the inclination at fixed intrinsic
parameters, shows how the disagreement of a model with NR depends on the
inclination of the source. This reflects how the modelling errors in the
individual modes, which are weighted differently at each inclination, appear in
the observed signal. We do this for three training simulations, SXS:BBH:2700,
SXS:BBH:1440 and SXS:BBH:1439, chosen from \AAf{}'s worst performing cases
(Fig.~\ref{fig:iota}). They rank $\IotaRankWorstFirst$th, $\IotaRankWorstSecond$th and $\IotaRankWorstThird$th worst of the $\IotaNumCommonSims$
catalog simulations by \AAf{}'s fixed-hyperparameter LOO all-mode mismatch, i.e.\ all three are in its
worst quartile. All three are also inside its training set. The results in this
appendix show the in-sample accuracy of the \AAf{} model near its worst
performing cases.
Over $\IotaNumInclinations$ inclinations at these three high-$q$, high-spin configurations, \AAf{}'s network mismatch
against NR varies by $\IotaVariationAALow$--$\IotaVariationAAHigh\times$, and
\textsc{NRHybSur3dq8}'s by $\IotaVariationHybLow$--$\IotaVariationHybHigh\times$. The other approximants vary far more over the
same inclinations. IMRPhenomXPHM-SpinTaylor varies by $\IotaVariationXPHMLow$--$\IotaVariationXPHMHigh\times$, IMRPhenomTPHM
by $\IotaVariationTPHMLow$--$\IotaVariationTPHMHigh\times$, and SEOBNRv5PHM by $\IotaVariationSEOBLow$--$\IotaVariationSEOBHigh\times$. This has implications for
the model-versus-model campaign of Sec.~\ref{sec:campaign},
which compares \NRHJ{} against other models.
A cross-check against a model whose own error grows by more
than an order of magnitude toward edge-on measures the error of that model, and not of the
surrogate.

At these $\IotaNumPoints$ points (three simulations at $\IotaNumInclinations$ inclinations each, in-sample for \AAf{}), neither
\AAf{} nor \textsc{NRHybSur3dq8} is consistently more accurate, even though the
three simulations were chosen from \AAf{}'s worst performing cases. \AAf{} is the
more accurate model at $\IotaAAWinsPoints$ of the $\IotaNumPoints$ points ($\IotaAAWinsPercent\%$: all $\IotaAAWinsTwoSevenHundred$ on SXS:BBH:2700, $\IotaAAWinsOneFourFourZero$ of $\IotaNumInclinations$ on
SXS:BBH:1440, $\IotaAAWinsOneFourThreeNine$ of $\IotaNumInclinations$ on SXS:BBH:1439) and \textsc{NRHybSur3dq8} at the other
$\IotaHybWinsPoints$ ($\IotaHybWinsPercent\%$), with the pointwise ratio of mismatch against NR between that of \AAf{} and \textsc{NRHybSur3dq8}
spanning $\IotaRatioAAHybLow$--$\IotaRatioAAHybHigh$. Catalog-wide, in fixed-hyperparameter LOO, \AAf{} is the better model on $\IotaCatalogAAWinPercent\%$ of those $\IotaNumCommonSims$
simulations (Sec.~\ref{sec:aa_accuracy}).

\paragraph{Face-on comparison.}
Face-on, where only the $m=2$ modes radiate, \textsc{NRHybSur3dq8} is more
faithful at SXS:BBH:1439 ($\IotaFaceOnHybOneFourThreeNineMantissa$ versus $\IotaFaceOnAAOneFourThreeNine$), while \AAf{}
matches it at SXS:BBH:1440 ($\IotaFaceOnAAOneFourFourZeroMantissa$ versus $\IotaFaceOnHybOneFourFourZero$) and is better by
$\IotaFaceOnHybOverAATwoSevenHundred\times$ at SXS:BBH:2700 ($\IotaFaceOnAATwoSevenHundred$ versus $\IotaFaceOnHybTwoSevenHundred$).
This deficit is invisible in the all-mode and inclined comparisons.

\bibliographystyle{apsrev4-2}
\bibliography{refs}

\end{document}